\documentclass[onecolumn, draftclsnofoot,11pt]{IEEEtran}
\IEEEoverridecommandlockouts
\global\long\def\EE{\mathbb{E}}
\global\long\def\PP{\mathbb{P}}

\global\long\def\11{\mathbbm{1}}
 
\def\ulineX{\underline{X}}
\def\ulinex{\underline{x}}

\def\ulineW{\underline{W}}
\def\ulinew{\underline{w}}

\def\ulineCalW{\underline{\mathcal{W}}}

\def\uliney{\underline{y}}

\def\ulineX{\underline{X}}

\def\CalA{\mathcal{A}}
\def\CalB{\mathcal{B}}
\def\CalC{\mathcal{C}}
\def\CalJ{\mathcal{J}}
\def\CalP{\mathcal{P}}
\def\CalQ{\mathcal{Q}}
\def\CalR{\mathcal{R}}
\def\CalW{\mathcal{W}}
\def\CalU{\mathcal{U}}
\def\CalV{\mathcal{V}}
\def\CalT{\mathcal{T}}
\def\CalX{\mathcal{X}}
\def\CalY{\mathcal{Y}}

\def\3To1BC{$3-$to$-1$}

\def\define{:{=}~}

\def\Expectation{\mathbb{E}}

\newif\ifProofForORDBC

\def\parsec{\par\noindent}
\def\med{\medskip\parsec}

\def\PMFIndicator{\mathbbm{1}_{\left\{\mathtt{PMF}(\CalC_1,\CalC_2)\right\}}}

\newcommand{\e}{\epsilon}

\usepackage{hyperref}
\hypersetup{
colorlinks=true, %
 pdfstartview={FitH},
    linkcolor=red,
    citecolor=blue, %DarkGreen,
    urlcolor={blue!80!black}
}

\usepackage{stackengine}
\def\deq{\mathrel{\ensurestackMath{\stackon[1pt]{=}{\scriptstyle\Delta}}}}
\def\define{\mathrel{\ensurestackMath{\stackon[1pt]{=}{\scriptstyle\Delta}}}}

\newcounter{relctr} %% <- counter for relations
\everydisplay\expandafter{\the\everydisplay\setcounter{relctr}{0}} %% <- reset every eq
\newcommand\labelrel[2]{%
  \begingroup
    \refstepcounter{relctr}%
    \stackrel{\textnormal{(\alph{relctr})}}{\mathstrut{#1}}%
    \originallabel{#2}%
  \endgroup
}
\AtBeginDocument{\let\originallabel\label}

\newif\ifJournal
\usepackage{amssymb}
\usepackage{amsmath}
\usepackage{mathrsfs}
\usepackage{ulem}
\usepackage{epsf,epsfig}
\usepackage{cite}
\usepackage{color}
\usepackage{dsfont}
\usepackage{bbm}
\usepackage{amsthm}
\usepackage{setspace}
\usepackage{float}

\newtheorem{theorem}{Theorem}

\newcommand{\comment}[1]{}

\begin{document}

\sloppy
\newtheorem{remark}{\it Remark}
\newtheorem{thm}{Theorem}
\newtheorem{corollary}{Corollary}
\newtheorem{definition}{Definition}
\newtheorem{lemma}{Lemma}
\newtheorem{example}{Example}
\newtheorem{prop}{Proposition}
%\vspace{-0.2in}
\title{\huge Source Coding for Synthesizing Correlated Randomness}
%\vspace{-0.2in}

% \author{\IEEEauthorblockN{Touheed Anwar Atif, Arun Padakandla and S. Sandeep
%     Pradhan}\\
% \IEEEauthorblockA{Department of Electrical Engineering and Computer Science,\\
% University of Michigan, Ann Arbor, MI 48109, USA.\\
% Email: \tt\small touheed@umich.edu, arunpr@umich.edu pradhanv@umich.edu}
% }

\author{\IEEEauthorblockN{Touheed Anwar Atif\IEEEauthorrefmark{1},
Arun Padakandla\IEEEauthorrefmark{2} and  S. Sandeep Pradhan\IEEEauthorrefmark{1} \\}
\IEEEauthorblockA{Department of Electrical Engineering and Computer Science,\\
\IEEEauthorrefmark{1}University of Michigan, Ann Arbor, MI 48109, USA.\\
\IEEEauthorrefmark{2}University of Tennessee, Knoxville, USA\\
Email: \tt touheed@umich.edu, arunpr@utk.edu, pradhanv@umich.edu}

\thanks{This work was presented in part at IEEE International Symposium on Information Theory (ISIT) 2020.}}

% \author{

% \IEEEauthorblockN{Touheed Anwar Atif}
% \IEEEauthorblockA{University of Michigan, USA \\
% Email: touheed@umich.edu  
% }
% \and
% \IEEEauthorblockN{Arun Padakandla}
% \IEEEauthorblockA{University of Tennessee, USA \\
% Email: arunpr@utk.edu
% }\and
% \IEEEauthorblockN{S. Sandeep Pradhan}
% \IEEEauthorblockA{University of Michigan, USA \\
% Email: pradhanv@umich.edu
% }}

\maketitle
\vspace{-0.5in}
\setlength{\baselineskip}{17.4pt}
\begin{abstract}
We consider a scenario wherein two parties Alice and Bob are provided $X_{1}^{n}$ and $X_{2}^{n}$  -- samples that are IID from a PMF $P_{X_1 X_2}$. Alice and Bob can communicate to Charles over (noiseless) communication links of rate $R_1$ and $R_2$ respectively. Their goal is to enable Charles generate samples $Y^{n}$ such that the triple $(X_{1}^{n},X_{2}^{n},Y^{n})$ has a PMF that is close, in total variation, to $\prod P_{X_1 X_2 Y}$. In addition, the three parties may posses pairwise shared common randomness at rates $C_1$ and $C_2$. We address the problem of characterizing the set of rate quadruples $(R_1,R_2,C_1,C_2)$ for which the above goal can be accomplished. We provide a set of sufficient conditions, i.e. an inner bound to the achievable rate region, and necessary conditions, i.e. an outer bound to the rate region for this three party setup. We provide a joint-typicality based random coding argument involving  encoding and decoding operations to perform soft covering and a pertinent relaxation of the PMF requirement for the encoders.
\end{abstract}

{\hypersetup{
colorlinks=true, %
 pdfstartview={FitH},
    linkcolor=black,
    citecolor=blue, %DarkGreen,
    %urlcolor={blue!80!black}
}
\singlespacing
\tableofcontents}
	\setcounter{tocdepth}{1}
\addtocontents{toc}{\protect\setcounter{tocdepth}{1}}

\section{Introduction}
The task of generating correlated randomness at different terminals in a 
network has applications in several communication \cite{7997946, DBLP:conf/icalp/GhaziS17} and computing \cite{2013TOC_IshKusEtAl, 2015CRYPTO_IshMajSahGen, 10.1007/978-3-540-85174-5_32} scenarios. The presence of distributed correlated randomness also serves as a primitive in several cryptographic protocols \cite{2004TIT12_CsiNar}. In this article, we study the problem of characterizing fundamental 
information-theoretic limits of generating such correlated randomness in network scenarios.

\begin{figure}[htb]
 \centering
\includegraphics[width=3.2in]{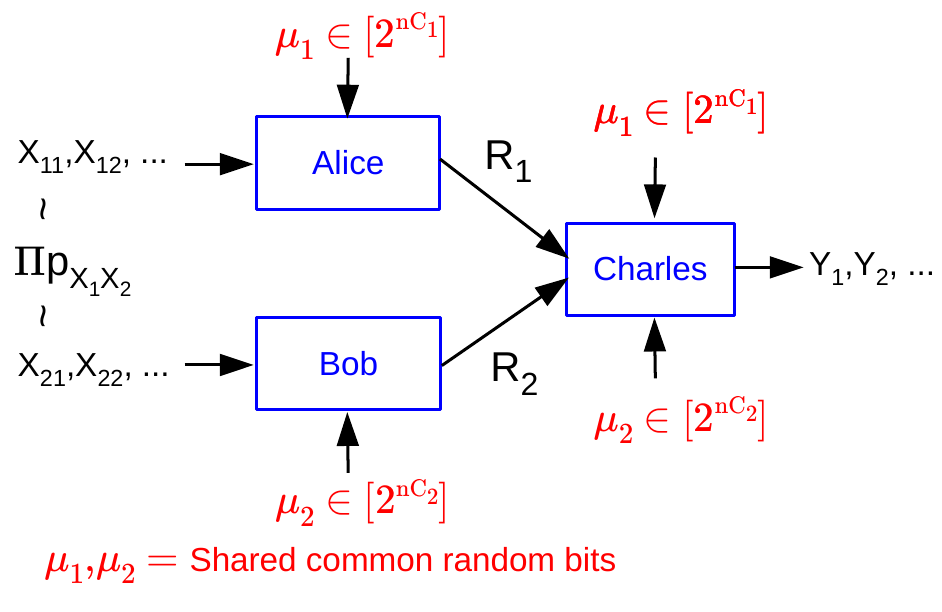}
\vspace{-0.15in}
\caption{Illustration of distributed agents performing source coding for synthesizing correlated randomness.}
\vspace{-0.07in}
\label{Fig:NetworkSoftCovering}

\end{figure}

We consider the scenario depicted in Fig~\ref{Fig:NetworkSoftCovering}. Three distributed parties -
Alice, Bob and Charles - have to generate samples that are independent 
and identically distributed (IID) with a target probability mass function (PMF) 
$P_{X_{1}X_{2}Y}$. Alice and Bob are provided with samples that are IID 
$P_{X_{1}X_{2}}$ - the corresponding marginal of the target PMF 
$P_{X_{1}X_{2}Y}$. They have access to unlimited private randomness and share 
noiseless communication links of rates $R_{1},R_{2}$ with Charles. In addition, the three parties share common randomness at rate $C$. For what rate triples $(R_{1},R_{2}, C)$ can Alice and Bob enable Charles to generate the required samples? In this article, we undertake a Shannon-theoretic study and characterize inner \cite{arxiv_CurrentPaper} and outer bounds on the aforementioned set of rate triples.

The roots of this line of study - distributed terminals generating IID copies of correlated random variables - can be traced back to the work of Wyner \cite{197503TIT_Wyn}. Wyner \cite{197503TIT_Wyn} considered the scenario of distributed parties generating IID samples distributed with PMF $P_{XY}$, when fed with a common information stream. In characterizing this rate, Wyner discovered a measure, commonly referred to as \textit{Wyner's common information}, that quantifies the amount of common information between two correlated random variables. A renewed interest in this study led Cuff \cite{201311TIT_Cuf} to study the scenario depicted in Fig.~\ref{Fig:NetworkSoftCovering} with just two terminals corresponding to Alice, Charles, and Bob being absent. Cuff \cite{201311TIT_Cuf} characterized the entire set of rate pairs $(R,C)$ and showed that Wyner's common information forms one vertex of this region. Cuff's work also shares an interesting connection with an analogous problem in quantum information theory. Prior to \cite{201311TIT_Cuf}, Winter \cite{winter2004extrinsic} considered the problem of simulating quantum measurements with limited common randomness. This work was generalized in \cite{wilde2012information} where the authors characterized a complete trade-off between communication and common randomness rates.
Building on this, \cite{201907ISIT_HeiAtiPra} studied a distributed scenario consisting of three distributed parties and derived inner bounds. 

Motivated by applications in security \cite{10.1007/978-3-030-56880-1_14}, cryptography \cite{10.1145/62212.62215}, need for co-ordinated control among distributed terminals \cite{201708TIT_Tre}, among others, this line of study has received considerable attention lately \cite{song2016likelihood,cuff2016soft }. The works of Wyner \cite{197503TIT_Wyn}, Cuff \cite{201311TIT_Cuf} and others \cite{yassaee2015channel} naturally lead us to consider the scenario depicted in Fig.~\ref{Fig:NetworkSoftCovering}. In contrast to these works, our scenario requires two distributed terminals, observing correlated information, to co-ordinate their communication to a central decoder. This poses certain technical challenges in the design and analysis of the encoders and the decoders, thereby leaving the information-theoretic study of our scenario unresolved. As we describe in the sequel, our work overcomes these challenge via (i) a novel design of the encoders and decoder, and (ii) identification of appropriate mathematical tools for performance analysis and rate region characterization.

The key challenge here is to ensure that Bob's simulated samples $Y^{n}$ are \textit{correlated simultaneously} with $X_{1}^{n}$ and $X_{2}^{n}$ in a single-letter fashion. In particular, it maybe noted that the conventional side-information approach of treating one of the sources, say $X_{2}^{n}$, as side-information and adopting the proof of channel synthesis with side-information \cite{yassaee2015channel} does not work. The reason for this is the need for simultaneous correlation as mentioned above. Indeed, it maybe noted that, while the channel synthesis with side-information problem \cite{yassaee2015channel} has been addressed and solved several years ago, the problem of distributed channel synthesis has remained open.

\begin{figure}
 \centering
\includegraphics[width=3.4in]{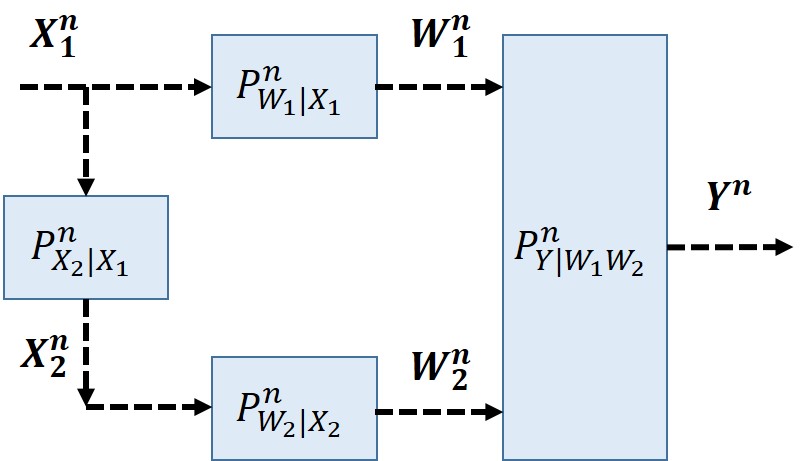}\vspace{-0.1in}
\caption{Figure demonstrating the generation of random variables $X_1,X_2,W_1,W_2,Y$ from the joint PMF $P_{\ulineX\ulineW Y}$ while incorporating the Markov chains specified in the theorem statement.}
\vspace{-0.27in}
\label{Fig:setup}
\end{figure}

We propose a novel approach to addressing this problem. We first prove an inner bound that appears smaller at first sight. Specifically, we prove achievability of one corner point of the achievable rate region wherein the lower bound on one of the rates is higher. This larger lower bound enables us simulate the generated samples to be correlated with a larger sub-collection of auxiliary random variables. We then leverage this for lowering the lower bound on the other rate components. By then using convexification, we prove that by swapping the order and performing time-sharing, we can enlarge the inner bound to what one might conjecture to be a natural inner bound via binning. The reader will find Figs.~\ref{Fig:setup}, \ref{Fig:AliceEncoding}, and  \ref{Fig:BobEncoding} illustrate the new steps in our proof technique.

We also emphasize that while the stated inner bound might appear natural for a reader familiar with the problem of distributed source coding \cite{197805PhDThe_Tun}, the problem of distributed channel synthesis is different and involves more constraints. Indeed, in this problem, it is required that the generated random variables appear to have a single-letter distribution as specified, not just that they meet certain distortion criterion. This difference is clearly emphasized in the rate region obtained for the conventional channel synthesis problem studied by Cuff \cite{201311TIT_Cuf} for which we are aware of optimality. Observe that, as against to a single lower bound on the rate that we obtain in the Shannon's source coding problem, Cuff's problem yields two lower bounds, proving that the distributed channel synthesis problem is more involved, and perhaps hinting at the need for a new proof technique that we have developed in this article. We also note that we have the opportunity to employ a more sophisticated Chernoff-Hoeffding concentration inequality due to Ahlswede Winter \cite{ahlswede2002strong} - a tool not regularly employed in proof of coding theorems.

\begin{figure}[htb]
    \centering
    \includegraphics[width=2.9in]{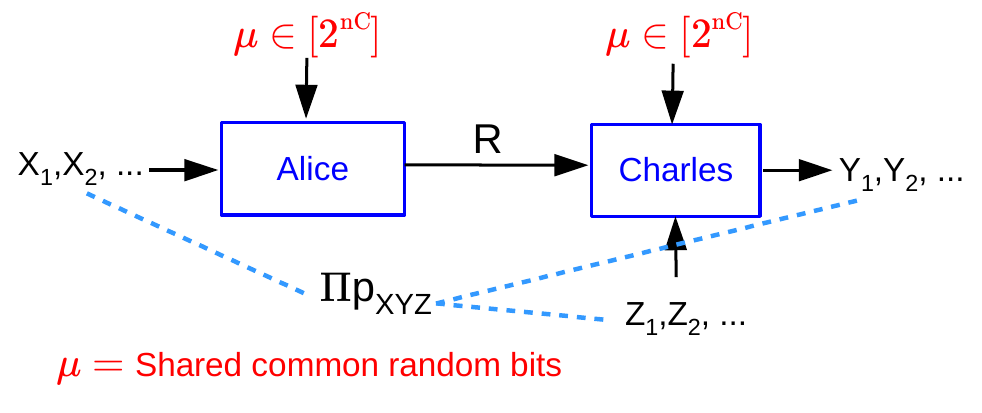}
    \vspace{-0.12in}
\caption{Synthesizing correlated randomness with side information available at the decoder.}
\vspace{-0.25in}
\label{Fig:SideInfoSoftCovering}
\end{figure}

Lastly, we highlight another novelty of our findings. In addressing the scenario in Fig.~\ref{Fig:NetworkSoftCovering}, it is natural to try and build on Cuff's \cite{201311TIT_Cuf} findings - relying on the use of a likelihood encoder that maps the observed 
sequence and common random bits into a codebook of sufficient rate. 
Essentially, the encoder performs a MAP decoding of the observed sequence into 
the chosen codebook. While this choice greatly simplifies the analysis, it 
permits little room for generalization.  Our experience in network information theory suggests that encoding and decoding via joint-typicality can be naturally generalized to diverse multi-terminal scenarios. Motivated by this, we propose joint-typicality based 
encoding and decoding to perform soft covering \cite{arxiv_CurrentPaper}. As a reader will note, the transition from a likelihood encoder to a joint-typicality based encoder results in challenges in analysis due to the hard constraints that the encoders are valid PMFs. 
Toward this, we develop a novel construction of random encoders, by relaxing the PMF requirement. This relaxation plays a central role in generalizing the results to the distributed case. The mathematical tools we have adopted to overcome these challenges maybe viewed as part of our technical contribution. In view of the general applicability of typicality-based coding schemes, we regard the typicality-based soft covering we propose as an important step. Furthermore, we leverage ideas from the outer bounds for the distributed source coding problem \cite{aaron2008ImprovedOuterBounds} to characterize an outer bound for this problem. This article therefore contains a complete suite of results for the distributed channel synthesis problem, thereby 1) filling our knowledge gaps in regards to our scenario and 2) deriving bounds for this problem that is on par with our knowledge for the distributed source coding problem. Elaborating on the last point, we note that with infinite common randomness, our rate regions reduce to those that are currently the best known for the distributed source coding problem. 

A preliminary version of this work appeared in \cite{arxiv_CurrentPaper}. Subsequently,  building on this work, the authors in \cite{kurri2021multiple} considered a side information and three-way common information generalization of the problem considered in \cite{arxiv_CurrentPaper}, and derived inner and outer bounds. 

The paper is organized as follows. After setting up notation and stating the problem (Sec.~\ref{Sec:Prelims}), we provide our main results, the inner and outer bounds to the achievable rate-region of a distributed problem, in Sec.~\ref{Sec:DistribuetdSoftCovering}. Before providing a complete proof of the above inner bound, we consider the two-terminal side-information scenario (Fig.~\ref{Fig:SideInfoSoftCovering}) in Sec.~\ref{Sec:SoftCoverWithSideInfo}, wherein the decoder is provided with side-information. This provides us with an ideal pedagogical step to present our typicality based encoder, decoder and indicate the mathematical challenges in its information-theoretic analysis. Unlike \cite{yassaee2015channel}, we propose a joint-typicality based encoder and decoder and provide a complete proof of achievability of the full rate region. Building on the tools developed therein, we present the proof of our main results in Sec.~\ref{Sec:ProofOfDistTheorem}.

\section{Preliminaries and Problem Statement}
\label{Sec:Prelims}
We supplement standard information theory notation with the following. For a 
PMF $P_{X}$, we let $P^{n}_{X}=\prod_{i=1}^{n}P_{X}$. 
Given a sequence $x^n \in \CalX^n$, let 
$P_{x^n}$ denote its empirical distribution.
For any distribution $P_X$ on $\CalX$, define the $\delta$-typical set $T_{\delta}(X)$ as
\[
T_{\delta}(X)\define\left\{x^n \in \CalX^n: \|P_{x^n}-P_X \|_{\infty} \leq
\frac{\delta}{|\CalX|}, \ \ 
P_{x^n} \ll P_X \right\}.
\]
For any distribution $P_{XY}$ on $\CalX \times \CalY$, define the $\delta$-jointly typical set $T_{\delta}(X,Y)$ as
\[
T_{\delta}(X,Y)=T_{\delta}(P_{XY})\define\left\{(x^n,y^n) \in \CalX^n \times \CalY^n: \|P_{x^n,y^n}-P_{XY} \|_{\infty} \leq
\frac{\e}{|\CalX||\CalY|}, \ \ 
P_{x^n,y^n} \ll P_{XY} \right\},
\]
where $P_{x^n,y^n}$ is the empirical joint distribution of 
two sequences $(x^n,y^n)$.
Note that if $(x^n,y^n) \in 
T_{\delta}{(P_{XY})}$, then 
$x^n \in T_{\delta}{(P_X)}$, and 
$y^n \in T_{\delta}{(P_Y)}$. 
For any conditional distribution $P_{Y|X}:\CalX \rightarrow \CalY$, and any $x^n \in \CalX^n$, define the 
$\delta$-conditional typical set $T_{\delta}(Y|x^n)$ as
\[
T_{\delta}(Y|x^n)\deq\left\{y^n \in \CalY^n: \left\|  P_{x^n,y^n} -P_{Y|X}P_{x^n} \right\|_{\infty} \leq
\frac{\delta}{|\CalY|}, P_{x^n,y^n} \ll P_{x^n}P_{Y|X} \ \ 
\right\}.
\] For an integer $n\geq 1$, 
$[n]\define \{1,\cdots,n\}$. The total variation between PMFs $P_X$ and $Q_X$ 
defined over $\mathcal{X}$ is denoted $\|P_X-Q_X\|_1 = 
\frac{1}{2}\sum_{x \in \mathcal{X}} |P_X(x) - Q_X(x)| = \sup_{\CalA \subset \CalX} |P_X(\CalA ) - Q_X(\CalA )|.$
\begin{comment}{
\med\textbf{Channel Synthesis Problem}:
\label{Sec:SoftCoverWithSideInfo}
% \begin{definition}(Channel Synthesis) Given a PMF $P_{XY}$, a channel synthesis protocol comprising of an encoder and decoder mapping is said to $\epsilon$ faithfully synthesize a channel if 
% \begin{align}
%     \|P^n_{XY} - \tilde{P}_{X^nY^n}\|_1 \leq \epsilon
% \end{align}
% where $\tilde{P}_{X^nY^n}$ denotes the induced PMF by the action of the encoder and decoder mapping.
% \end{definition}
\med\textbf{Problem Statement}: 
}\end{comment}
\begin{definition}
\label{Defn:AlternateSideInfo}
Given a PMF $P_{XYZ}$ on $\mathcal{X}\times\mathcal{Y}\times \mathcal{Z}$, a rate pair $(R,C)$ is said to be achievable, if $\forall \epsilon > 0$ and all sufficiently large $n$, there exists a collection of 
$2^{nC}$ randomized encoders $E^{(\mu)}:\mathcal{X}^{n} \rightarrow [\Theta]$
for $\mu \in [2^{nC}]$ and a corresponding collection of $2^{nC}$ randomized 
decoders $D^{(\mu)}:\mathcal{Z}^{n} \times [\Theta] \rightarrow \mathcal{Y}^{n}$ 
for $\mu \in [2^{nC}]$ such that $\| P^{n}_{XYZ}- 
P_{X^{n}Y^{n}Z^{n}}\|_{1} \leq \epsilon$, $\frac{1}{n}\log_2 \Theta \leq 
R+\epsilon$, where for all $x^n,y^n,z^n \in \mathcal{X}^n\times\mathcal{Y}^n\times \mathcal{Z}^n $
\begin{align}
P_{X^{n}Y^{n}Z^{n}}(x^{n},y^{n},z^{n}) \deq \sum_{\substack{ \mu \in 
[2^{nC}] \\ }}2^{-nC}\sum_{m \in[ \Theta]}
P^{n}_{XZ}(x^{n},z^{n})P^{(\mu)}_{M|X^{n}}(m|x^{n})P^{(\mu)}_{Y^{n} 
|Z^{n},M}(y^{n}|z^{n},m), \nonumber
\end{align}
$P^{(\mu)}_{M|X^{n}}, P^{(\mu)}_{Y^{n}|Z^{n}M}$ are the PMFs induced by  
encoder and decoder respectively, corresponding to shared random message 
$\mu$, with $M$ being the random variable corresponding to the message transmitted. We let $\mathcal{R}_s(P_{XYZ})$ denote the set of achievable rate pairs.
\end{definition}
Cuff \cite[Thm.~II.1]{201311TIT_Cuf} provides a single-letter characterization for 
$\mathcal{R}_s(P_{XY})$ when  $\mathcal{Z}=\phi$ is empty. A single-letter  characterization of $\mathcal{R}_s(P_{XY})$ in the general case was provided in   \cite{yassaee2015channel}.
% Our first main result (Thm.~\ref{Thm:SideInformation}) is a characterization of $\mathcal{R}_s(P_{XYZ})$. 
Building on this, we address the network scenario 
(Fig.~\ref{Fig:NetworkSoftCovering}) for which we state the problem below. In the following, we let $\ulineX = (X_1,X_2),  \ulinex^n = (x^n_1,x^n_2)$.

\begin{definition}
\label{Defn:Distributed}
Given a PMF $P_{X_{1}X_{2}Y}$ on $\mathcal{X}_1\times\mathcal{X}_2\times\mathcal{Y}$ , a rate quadruple  $(R_1,R_2,C_1, C_2)$ is said to be achievable, 
if $\forall\epsilon\!>\!0$ and all sufficiently large $n$, 
there exists $2^{nC_1}\times 2^{nC_2}$ randomized encoder pairs $(E_{1}^{(\mu_1)}, E_{2}^{(\mu_2)} )$, where  $E_{j}^{(\mu_j)} : 
\mathcal{X}_{j}^{n} 
\rightarrow [\Theta_{j}] : \mu_j \in [2^{nC_j}],  j \in [2],$ and a 
corresponding collection of $2^{nC} $ randomized decoders 
$D^{(\mu)}:[\Theta_{1}]\times [\Theta_{2}] \rightarrow 
\mathcal{Y}^{n}$ for $\mu \in [2^{nC}]$, where $C\deq C_1+C_2$ and $\mu \deq (\mu_1,\mu_2)$, such that $\| P^{n}_{\ulineX Y}- P_{\ulineX^{n} Y^{n}}\|_{1} \leq \epsilon$, $\frac{1}{n}\log_2 \Theta_{j} \leq R_{j}+\epsilon : j \in [2]$, where for all $\ulinex^n,\uliney^n \in \mathcal{\ulineX}^n\times\mathcal{Y}^n $
\begin{eqnarray}
P_{\ulineX^{n}Y^{n}}(\ulinex^{n},y^{n}) \deq \sum_{\mu \in 
[2^{nC}]}\!\!2^{-nC} \!\!\!\sum_{\substack{ (m_{1},m_{2}) 
\in \\ [ \Theta_{1}]\times [\Theta_{2}] }}\!\!
P^{n}_{\ulineX }(\ulinex^{n})P^{(\mu_1)}_{M_{1}|X_{1}^{n}}(m_{1}
|x_ { 1 } ^ { n } )P^ { (\mu_2)}_{ M_ { 2 } |X_{2}^{n}}(m_{2}|x_{2}^{n})P^{(\mu)}_ 
{ Y^{ n } |M_{1},M_{2}}(y^{n}|m_{1},m_{2}),
\nonumber
\end{eqnarray}
$P^{(\mu_j)}_{M_{j}|X_{j}^{n}} : j \in [2], P^{(\mu)}_{Y^{n}|M_{1},M_{2}}$ are the 
PMFs induced by the two randomized encoders and decoder, respectively,
corresponding to common random index $(\mu_1,\mu_2)$. We 
let $\mathcal{R}_{d}(P_{\ulineX Y})$ denote the set of achievable rate triples.
\end{definition}
Our main results are the characterization of an inner bound and an outer bound to $\mathcal{R}_{d}(P_{\ulineX Y})$ 
which are provided in Theorem~\ref{Thm:Distributed} and Theorem~\ref{Thm:DistributedOuterBound}, respectively.

\section{Distributed Soft Covering - Main Results}
\label{Sec:DistribuetdSoftCovering}
In this section, we provide an  inner bound and an outer bound to the 
achievable rate-region for the distributed setting (Fig. \ref{Fig:NetworkSoftCovering}). Our first result in this regard is the following inner bound to $\mathcal{R}_{d}(P_{\ulineX Y})$. In the following, we let $\ulineX = (X_1,X_2), \ulineW = (W_1,W_2), \ulinex = (x_1,x_2)$ and $\ulinew = (w_1,w_2)$.
% \begin{theorem}
% \label{Thm:Distributed}
% Given a PMF $P_{X_1X_2Y}$, $(R_1,R_2,C) \in \mathcal{R}_d(P_{\ulineX Y})$ if there exists a PMF $P_{W_1 W_2 \ulineX Y}$ such that (i) $P_{\ulineX Y}(\ulinex,y) = \sum_{\ulinew \in \ulineCalW}P_{\ulineW \uline{X} Y}(\ulinew,\ulinex,y)$ for all $(\ulinex,y)$, (ii) $W_1-X_1-X_2-W_2$ and $\ulineX-\ulineW-Y$ are Markov chains, and
% % \begin{align*}
% %     R_j \geq &I(X_j;U_j) - I(U_1,U_2) : j \in [2]\\
% %     R_1+R_2 &\geq I(X_1;U_1) + I(X_2;U_2) - I(U_1;U_2)\\
% %     R_j + C &\geq I(X_j,Y;U_j)-I(U_1;U_2)  : j \in [2]\\
% %     % R_2 + C &\geq I(X_2,Y;U)-I(U;V)\\
% %     R_1+R_2 +C &\geq I(X_1,Y;U_1) + I(X_2,Y;U_2)-I(U_1;U_2).
% % \end{align*}
% \begin{align}
%     {R}_1 & \geq I(X_1;W_1|Q) - I(W_1;W_2|Q) \nonumber \\
%     {R}_2 & \geq I(X_2;W_2|Q) - I(W_1;W_2|Q) \nonumber \\
%     {R}_1 + R_2 & \geq I(X_1;W_1|Q) + I(X_2;W_2|Q)-  I(W_1;W_2|Q) \nonumber \\
%     R_1+R_2+C & \geq I(X_1X_2W_2Y;W_1|Q) +  I(X_1X_2Y;W_2|Q) - I(W_1;W_2|Q)
% \end{align}
% \end{theorem}
\begin{theorem}
\label{Thm:Distributed}
Given a PMF $P_{X_1X_2Y}$, let $\mathcal{P}(P_{X_1X_2Y})$ denote the collection of all PMFs $P_{QW_1W_2\ulineX Y}$ defined on $\mathcal{Q}\times\mathcal{W}_1\times\mathcal{W}_2\times\mathcal{\ulineX}\times\mathcal{Y}$ such that (i) $P_{\ulineX Y}(\ulinex,y) = \sum_{(q,\ulinew) \in \mathcal{Q} \times\ulineCalW}P_{Q\ulineW \uline{X} Y}(q,\ulinew,\ulinex,y)$ for all $(\ulinex,y)$, (ii) $\sum_{\ulinew \in \uline{\mathcal{W}}}P_{Q\ulineW \uline{X} Y}(q,\ulinew,\ulinex,y) = P_Q(q)P_{\uline{X} Y}(\ulinex,y) $ for all $(q,\ulinex,y)$ (iii) $W_1-QX_1-QX_2-W_2$ and $\ulineX-Q\ulineW-Y$ are Markov chains, (iv) $|\mathcal{W}_1|\leq |\mathcal{X}_1|$, $|\mathcal{W}_2|\leq |\mathcal{X}_2|$, and 
$|\mathcal{Q}|\leq 7$.
Further, let $\beta(P_{Q\ulineW\ulineX Y})$ denote the set of rates and common randomness quadruple $(R_1,R_2,C_1,C_2) \in [0,\infty)^4$ that satisfy
\begin{align}\label{eq:dist_rate_region_theorem}
    {R}_1 & \geq I(X_1;W_1|Q) - I(W_1;W_2|Q) \nonumber \\
    {R}_2 & \geq I(X_2;W_2|Q) - I(W_1;W_2|Q) \nonumber \\
    {R}_1 + R_2 & \geq I(X_1;W_1|Q) + I(X_2;W_2|Q)-  I(W_1;W_2|Q) \nonumber \\
    {R}_1 + C_1 & \geq I(X_1X_2Y;W_1|Q) -  I(W_1;W_2|Q), \nonumber \\
    {R}_2 + C_2 & \geq I(X_1X_2Y;W_2|Q) -  I(W_1;W_2|Q), \nonumber \\
     R_1+R_2+C_1 & \geq I(X_1X_2Y;W_1|Q) + I(X_2;W_2|Q) - I(W_1;W_2|Q) \nonumber \\
      R_1+R_2+C_2 & \geq I(X_1X_2Y;W_2|Q) + I(X_1;W_1|Q) - I(W_1;W_2|Q) \nonumber \\
    R_1+R_2+C_1+C_2 & \geq I(X_1X_2Y;W_1W_2|Q) 
\end{align} where the mutual information terms are evaluated with the PMF $P_{QW_1W_2\ulineX Y}$. We have
\begin{eqnarray}  
\mathcal{R}_I(P_{\underline{X}Y}) \define \text{Closure}
    \left(\bigcup_{P_{Q\ulineW\ulineX Y} \in \mathcal{P}(P_{X_1X_2Y})}\beta(P_{Q\ulineW\ulineX Y})\right) \subseteq \mathcal{R}_d(P_{\ulineX Y}).\end{eqnarray}
In other words, $(R_1,R_2,C_1,C_2)$ is achievable if $(R_1,R_2,C_1,C_2) \in \mathcal{R}_I(P_{\underline{X}Y})$.
\end{theorem}
\begin{remark}
Before providing a proof to the above theorem, we briefly discuss two corner points of the rate region with
respect to the common randomness available. Firstly, consider the regime when both $C_1$ and $C_2$ are unlimited. This implies that only the first three constraints are active and hence the inner bound to the achievable rate-region reduces to the Berger-Tung inner bound \cite{berger}. Secondly, consider the case when only one of the $C_1,$ and $C_2$, say $C_2$, is unlimited. In the first glance, one may think that the rate $R_1$ is only constraint by the first and the sum rate ($R_1 + R_2$)  constraint. However, a careful observation yields an additional constraint $R_1 + R_2 + C_1$ limiting the rate of $R_1$. The insight to this is the joint distributed simulation task that the problem addresses. It suggests that if $R_2$ and $C_2$ are at their minimum then $R_1$ has to provide for any additional rate that is needed in simulating the joint distribution.
\end{remark}

\begin{proof}
The proof of this theorem is provided in Section \ref{sec:ProofOfDistTheoremInnerBound}.
\end{proof}

We consider an example to illustrate the significance of the inner bound. 
\begin{example}
Consider a distributed setup as shown in Fig. \ref{Fig:NetworkSoftCovering}. Let the input alphabets of the two encoders, $\CalX_1$ and $\CalX_2$, and the output alphabet $\CalY$ be given by the binary set $\{0,1\}$.
Let the joint distribution $P_{X_1X_2Y} = P_{X_1X_2}P_{Y|X_1X_2}$ be defined as
\begin{align*}
    P_{X_1X_2}(0,0) = P_{X_1X_2}(1,1) = \frac{(1-p)}{2} \quad \text{and} \quad P_{X_1X_2}(0,1) = P_{X_1X_2}(1,0) = \frac{p}{2},
\end{align*}
and 
\begin{align*}
    P_{Y|X_1X_2}(0|0,0) = P_{Y|X_1X_2}(0|1,1) = 1-\delta \quad \text{and} \quad 
    P_{Y|X_1X_2}(0|0,1) = P_{Y|X_1X_2}(0|1,0) = \delta,
\end{align*}
for $p = \delta = 0.2.$ The trade-off between the achievable sum communication rate and sum common randomness rate is numerically computed and is depicted in Fig. \ref{Fig:rateregion}. The figure demonstrates the usefulness of common randomness in decreasing the sum communication rates. However, below a certain threshold, no amount of common randomness can be used toward decreasing the communication rates further.

\begin{figure}[h]
    \centering
    \includegraphics[width=4.5in]{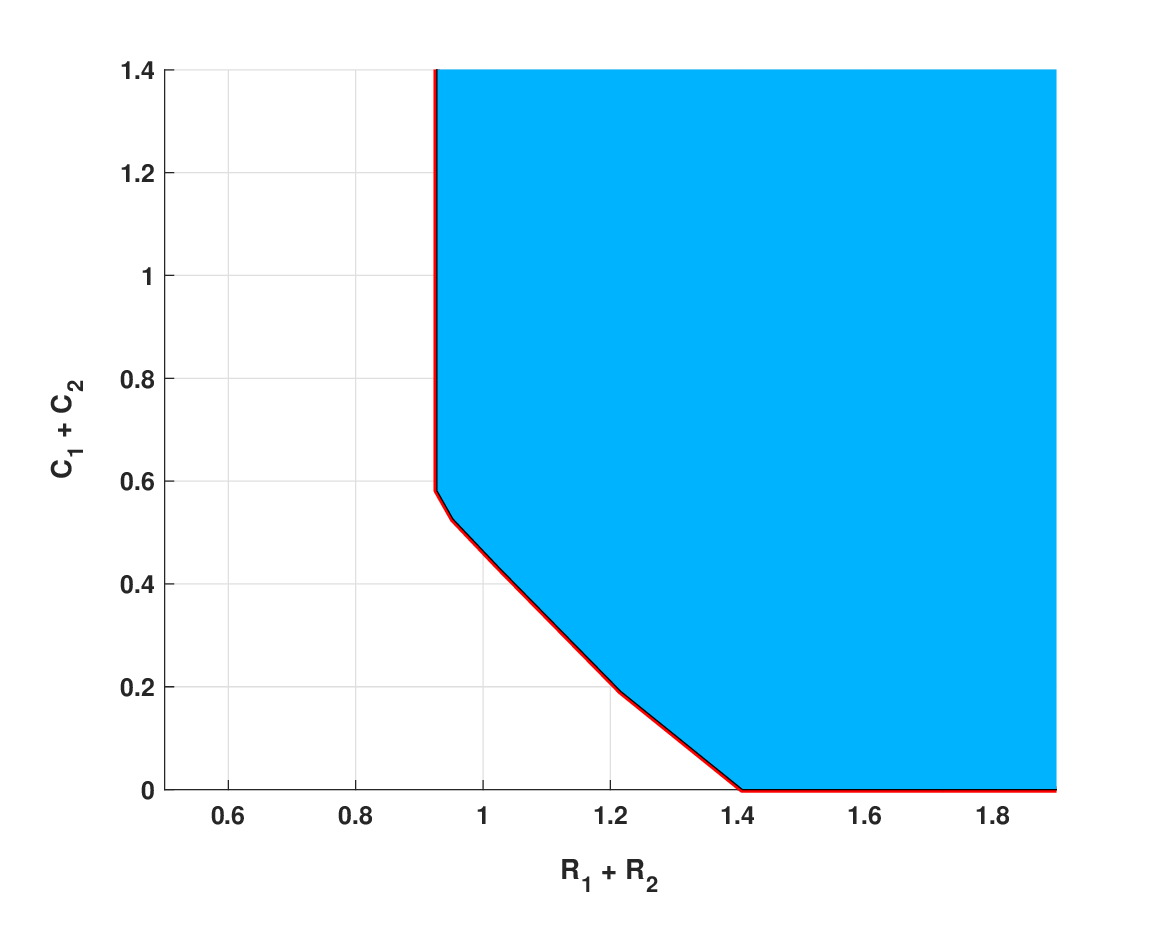}
    \vspace{-0.1in}
\caption{Figure depicting the trade-off between the sum rate and the sum common randomness.}
\vspace{-0.2in}
\label{Fig:rateregion}
\end{figure}
\end{example}
\noindent Our next main result for the distributed setting is the outer bound to the achievable rate region.  
\begin{theorem}
\label{Thm:DistributedOuterBound} For all $\epsilon>0$, let $\mathcal{P}_F(\epsilon)$ denote the collection of conditional PMFs $\tilde{P}_{JQUVY|X_1X_2}$ defined
on $\CalJ \times \CalQ \times \CalU \times \CalV  \times {\CalY}$ such that the following conditions are satisfied: (a) $(Q,J)$ is independent of $(X_1,X_2)$,  (b) $U -(X_1, Q, J)-(X_2, Q, J)-V$, and (c) $(X_1,X_2,Q) - (J,U,V) - Y$, and 
(d) $\|P_{X_1X_2Y}-P_{X_1X_2}\tilde{P}_{Y|X_1X_2}\|_1 \leq \epsilon$, 
where $\CalJ, \CalQ, \CalU$ and $\CalV$ are finite sets. Let $\mathcal{P}_R$ denote the collection of conditional PMFs $P_{W|X_1X_2}$ defined on $\CalW \times \CalX_1 \times \CalX_2$
such that the condition (e)  $X_1 - W - X_2$ is satisfied where $\CalW$ is a finite set. For a $\tilde{P}_{JQUVY|X_1X_2} \in \mathcal{P}_F$ and a $ P_{W|X_1X_2} \in \mathcal{P}_R$, let $\lambda_{\epsilon}(\tilde{P}_{JQUVY|X_1X_2}, P_{W|X_1X_2}) $ denote the set of rates and common randomness quadruple $(R_1,R_2,C_1,C_2)\in[0,\infty)^4$ that satisfy
% \begin{align}\label{eq:dist_rate_region_theorem}
%     {R}_1 & \geq  I(W;U|V) + I(X_{1},U|W, Q) \nonumber \\
%     {R}_2 & \geq  I(W;V|U) + I(X_{2},V|W, Q)   \nonumber \\
%     {R}_1 + R_2 & \geq 
%   I(W;U,V) +  I(X_{1};U|W,Q) + I(X_{2};V|W,Q )  \nonumber \\
%     % R_1 + C_1 & \geq \sum_{i=1}^n \bigg[ I(W_i;U_i|V_i) + I(X_{1i}, Y_i; U_i|V_i, Q_i, W_i)  - I(X_{1i}, Y_i; V_i|Q_i,W_i, X_1^{i-1}, Y^{i-1})\bigg].\nonumber\\
%     % {R}_2 + C_2 & \geq \sum_{i=1}^n \bigg[ I(W_i;U_i|V_i) + I(X_{2i}, Y_i; V_i|U_i, Q_i, W_i)  - I(X_{2i}, Y_i; U_i|Q_i,W_i, X_2^{i-1}, Y^{i-1})\bigg].\nonumber \\
%     R_1 + R_2 + C_1 + C_2& \geq  I(W; U,V) + I(X_{1},X_{2}, Y; U, V|Q,W),
% \end{align} 
\begin{align}\label{eq:convdist_rate_region_theorem}
    {R}_1 & \geq  I(W;U|V, J) + I(X_{1},U|W, Q, J)-\epsilon \nonumber \\
    {R}_2 & \geq  I(W;V|U, J) + I(X_{2},V|W, Q, J) -\epsilon  \nonumber \\
    {R}_1 + R_2 & \geq 
  I(W;U,V|J) +  I(X_{1};U|W,Q,J) + I(X_{2};V|W,Q,J ) -2\epsilon \nonumber \\
    % R_1 + C_1 & \geq \sum_{i=1}^n \bigg[ I(W_i;U_i|V_i) + I(X_{1i}, Y_i; U_i|V_i, Q_i, W_i)  - I(X_{1i}, Y_i; V_i|Q_i,W_i, X_1^{i-1}, Y^{i-1})\bigg].\nonumber\\
    % {R}_2 + C_2 & \geq \sum_{i=1}^n \bigg[ I(W_i;U_i|V_i) + I(X_{2i}, Y_i; V_i|U_i, Q_i, W_i)  - I(X_{2i}, Y_i; U_i|Q_i,W_i, X_2^{i-1}, Y^{i-1})\bigg].\nonumber \\
    R_1 + R_2 + C_1 + C_2& \geq  I(W; U,V|J) + I(X_{1},X_{2}, Y; U, V|Q,W,J) - g_c(\epsilon),
\end{align} 
under the Markov coupling between $\tilde{P}_{JQUVY|X_1X_2}$ and $ P_{W|X_1X_2}$, i.e., condition (f) $W - (X_1,X_2) -
(J,Q,U,V,Y) $ is satisfied, where $g_c(\epsilon) \deq 4\epsilon\left(\log(|\CalX_1||\CalX_2||\CalY|) 
-\log (\epsilon) \right)$. In other words, the joint distribution of the concerned random variables is given by $P_WP_{X_1|W}P_{X_2|W}\tilde{P}_Q \tilde{P}_J \tilde{P}_{U|X_1QJ}\tilde{P}_{V|X_2QJ}\tilde{P}_{Y|UVJ}$,
and with which  the mutual information terms are evaluated.
We have $\mathcal{R}_d(P_{\ulineX Y}) \subseteq \bigcap_{\epsilon>0}\mathcal{R}_O(P_{\ulineX Y},\epsilon),$ where
\begin{align}
    \mathcal{R}_O(P_{\ulineX Y}, \epsilon) \deq 
    \bigcap_{P_{W|X_1X_2} \in \mathcal{P}_R} \bigcup_{P_{JQUVY|X_1X_2} \in \mathcal{P}_F(\epsilon)}\lambda_{\epsilon}(P_{JQUVY|X_1X_2}, P_{W|X_1X_2}) . 
\end{align}
In other words, if $(R_1,R_2,C) \in \bigcap_{\epsilon>0}\mathcal{R}_O(P_{\ulineX Y},\epsilon)$, then $(R_1,R_2,C)$ is achievable.
\end{theorem}
\begin{proof}
The proof of the above theorem is provided in Appendix \ref{appx:converseDistributed}.
\end{proof}

\begin{remark}
Note that for every 
$P_{W|X_1X_2} \in \mathcal{P}_R$, we have an outer bound, obtained by taking the intersection over $\epsilon$ and the union over $P_{JQUVY|X_1X_2} \in 
\mathcal{P}_F(\epsilon)$, on $\mathcal{R}_d(P_{\underline{X}Y})$. Hence we have a family of outer bounds. 
\end{remark}
\begin{remark} 
One may question the computability of the outer bound provided in Theorem \eqref{Thm:DistributedOuterBound}. The computability of this bound depends on the cardinality of the auxiliary random variables defined in the theorem. Currently, we are unable to bound the cardinality of the auxiliary random variables, but aim to provide one in our future work. As a matter
of fact, the current outer bounds for the equivalent distributed rate distortion problem still suffers from the computability issue. The first outer bound to this problem was provided in \cite{berger} and a recent substantial improvement was made by authors in \cite{aaron2008ImprovedOuterBounds,kang2010new}. All these bounds suffer from the absence of cardinality bounds on
at least one of the variables used and hence cannot be claimed to be computable using finite resources. This problem still remains open. 
\end{remark}
\begin{remark}
Due to the lack of cardinality bounds, the space of probability distributions is not compact, and hence the mutual information may not be a continuous function of $\epsilon$. Therefore, the continuity of $\mathcal{R}_O(P_{\ulineX Y},\epsilon)$ at $\epsilon = 0$ still remains an open question. When the cardinality bounds become available, we will have continuity at $\epsilon=0$, and thus $\mathcal{R}_O(P_{\ulineX Y},0) = \bigcap_{\epsilon>0}\mathcal{R}_O(P_{\ulineX Y},\epsilon)$. 
\end{remark}

% In the interest of brevity, we only indicate the challenges and the new elements 
% in our proof. We refer the reader to \cite{arxiv_CurrentPaper} for a complete proof.

\section{Soft Covering with Side Information}
\label{Sec:SoftCoverWithSideInfo}
 Although our paper is mainly geared toward the distributed case (addressed in Section \ref{Sec:DistribuetdSoftCovering}), we provide  a proof of the side information scenario for pedagogical reasons. 
We provide a new proof of achievability of $\mathcal{R}_s(P_{XYZ})$. 
The proof develops a new construction of random encoders by relaxing the PMF requirement, and using refined Chernoff-Hoeffding bound, which could find applications in other problems of information theory. This relaxation and the refined bound play a central role in generalizing the results to the distributed case \ref{Fig:NetworkSoftCovering}.
As mentioned earlier, the side-information problem was addressed in  \cite{yassaee2015channel} using a different proof methodology.
% , however our proof is significantly distinct in its approach and proof techniques.
% (iii)  Furthermore, our proof leverages a 

\begin{theorem}
\label{Thm:SideInformation}
$(R,C) \in \mathcal{R}_{s}(P_{XYZ})$ if and only if there exists a PMF $P_{WXYZ}$ such that
(i) $P_{XYZ}(x,y,z) = \sum_{w \in \mathcal{W}}P_{WXYZ}(w,x,y,z)$ for all $(x,y,z)$ where $\mathcal{W}$ is the alphabet of $W$, (ii) $Z-X-W$ and $X-(Z,W)-Y$ are Markov chains, (iii) $|\mathcal{W}| \leq ( |\mathcal{X}||\mathcal{Y}||\mathcal{Z}|)^2$, and
\begin{align}\label{eq:SI_rateRegion}
    R \geq I(X;W) - I(W;Z),\quad  R+C \geq I(XYZ;W)-I(W;Z).
\end{align}
\end{theorem}
\begin{proof}
We begin the proof by describing the encoder.
% We provide the main elements (achievability in Sec.~\ref{SubSec:SideInfoAchievability} and converse in Sec.~\ref{SubSec:SideInfoConverse}) of our proof here with particular emphasis on the new elements. 
% The reader is referred to \cite{arxiv_CurrentPaper} for more technical details.
%\vspace{-0.15cm}
\subsection{Encoder Description}
\label{SubSec:SideInfoAchievability}
Fix a PMF $P_{WXYZ}$ satisfying the constraints stated in the theorem.
Throughout, $\mu \in [2^{nC}]$ denotes the $C$ bits of common randomness shared 
between the encoder and decoder. For each $\mu \in [2^{nC}]$, we shall design 
a randomized encoder $E^{(\mu)}: \mathcal{X}^{n} \rightarrow [\Theta]$ and a 
randomized decoder $D^{(\mu)}:\mathcal{Z}^{n} \times [\Theta] \rightarrow 
\mathcal{Y}^{n}$ that induce PMFs $P_{M|X^n}^{(\mu)}$ and 
$P_{Y^n|Z^nM}^{(\mu)}$ respectively, for which
\begin{align} \label{eq:main_lemma}
 \mathscr{Q}\define \frac{1}{2}\sum_{x^n,y^n,z^n}&\Bigg|P^n_{XYZ}(x^n,y^n,z^n) - 
 \!\!\!\!\sum_{\substack{ \mu \in
[2^{nC}]  }}\sum_{m \in[ \Theta]} \!\!\!
\frac{P^{n}_{XZ}(x^{n},z^{n})}{2^{nC}}
P^{(\mu)}_{M|X^{n}}(m|x^{n}
)P^{(\mu) } _ { Y^ { n } 
|Z^{n},M}(y^{n}|z^{n},m)\Bigg| \leq \varepsilon.
\end{align}
From now on we denote $\Theta = 2^{nR}.$ The design of these randomized encoders and decoders involves building a collection of codebooks
$\mathcal{C} \deq (\mathcal{C}^{(\mu)} : \mu \in [2^{nC}])$ where  $\mathcal{C}^{(\mu)} \deq (\mathtt{w}^{n}(l,\mu) \in \mathcal{W}^{n} : l \in 
[2^{n\tilde{R}}])$ for $\mu \in [2^{nC}]$,
where $\mathcal{W}$ is the alphabet of $W$ in 
the theorem statement, and $\tilde{R}$ will be specified shortly. On observing $x^{n}$ and $\mu$,  the randomized encoder chooses an index $L$ in $[2^{n\tilde{R}}]$ according to 
a PMF $E^{(\mu)}_{L|X^n}(\cdot|\cdot)$. The chosen index is then mapped to 
an index in $[2^{nR}]$ which is communicated to the decoder. Before we specify 
the PMF $E^{(\mu)}_{L|X^n}(\cdot|\cdot)$, let us describe how the chosen 
index is mapped to an index in $[2^{nR}]$. 
% In doing this, our first task is to identify and index the distinct codewords in $\mathcal{C}$. Firstly, for $\mathcal{C}^{(\mu)}$, we let $\Theta^{(\mu)}$ denote the number of distinct codewords in $\mathcal{C}^{(\mu)}$. 
% Secondly, we let $\mathcal{I}^{(\mu)}_{\mathcal{C}}: [2^{n\tilde{R}}] \rightarrow [\Theta^{(\mu)}]$ be defined such that $\mathcal{I}^{(\mu)}_{\mathcal{C}}(l)=\mathcal{I}^{(\mu)}_{\mathcal{C}}(\tilde{l})$ if and only if $\mathtt{w}(l,\mu)=\mathtt{w}(\tilde{l},\mu)$. Lastly,
We define a binning map $b^{(\mu)}:[2^{n\tilde{R}}]\rightarrow [2^{nR}]$. On 
observing $x^n$, the encoder chooses $L \in [2^{n\tilde{R}}]$ with 
respect to PMF $E^{(\mu)}_{L|X^n}(\cdot|x^{n})$, and communicates  $b^{(\mu)}(L)$ to the 
decoder.

Let us relate to the above three elements that make up the encoder. The PMF $E^{(\mu)}_{L|X^{n}}$ is analogous to the likelihood encoder $\Gamma_{J|X^{n},K}$ of Cuff \cite{201311TIT_Cuf} but with important 
changes to incorporate typicality-based encoding that permits the use of 
side-information at the decoder.
% The map $\mathcal{I}_{\mathcal{C}}^{(\mu)}$ 
% eliminates duplication of indices with identical codewords and is employed for 
% simplifying the analysis. 
The map $b^{(\mu)}$ performs standard 
information-theoretic binning \cite{197601TIT_WynZiv} to utilize 
side-information. We now specify $ E^{(\mu)}_{L|X^{n}}(\cdot|\cdot)$. Fix $\epsilon>0,\delta>0,\eta>0$, and for $x^n \in T_{\delta}(X)$ and $l \in [2^{n\tilde{R}}]$, let 
\begin{align*}
    E^{(\mu)}_{L|X^n}(l|x^n) &\deq \frac{1}{2^{n\tilde{R}}}\frac{1-\epsilon}{1+\eta}\sum_{w^n \in T_{\delta}(W|x^n)}\mathbbm{1}_{\{\mathtt{w}^n(l,\mu) = w^n\}}\frac{P^n_{X|W}(x^n|w^n)}{P^n_{X}(x^n)}.\nonumber 
\end{align*}
% \begin{align}
% \label{Eqn:EMuLGivenXn}E^{(\mu)}_{L|X^n}(l|x^n) &\deq \displaystyle
%      \frac{ (1-\epsilon)
% P^n_{X|W}(x^n|\mathtt{w}^n(l,\mu))}{(1+\eta)2^{n\tilde{R}}P^n_X(x^n)} 
% \text{if } l\neq 0, \mathtt{w}^n(l,\mu) \in 
% T_{\delta}(W|x^n),\mbox{ and}  \\
%  E^{(\mu)}_{L|X^n}(l|x^n) &= 1- 
% \displaystyle\sum_{l=1}^{2^{n\tilde{R}}}E^{(\mu)}_{L|X^n}(l|x^n) 
% \text{ if } l=0, \nonumber
% \end{align}
% and $E^{(\mu)}_{L|X^n}(l|x^n) = 1_{\{l=0\}}$ for $x^n \notin T_{\delta}(X)$. 
In specifying $E^{(\mu)}_{L|X^{n}}$, we have relaxed the requirement that $E^{(\mu)}_{L|X^n}(\cdot|x^n)$ be a PMF. This relaxation - a novelty of our work - yields analytical tractability of a random coding ensemble to be described in the sequel.
However, note that 
these maps depend on the choice of the codebook $\mathcal{C}$. We prove in Appendix \ref{AppSec:ProofOfLemma_PMF} 
that with high probability, $E^{(\mu)}_{L|X^{n}}(\cdot|x^{n}):[2^{nC}] \rightarrow \mathbb{R}$ is a PMF for 
every $x^{n} \in \notin T_{\delta}(X)$. This will form a part of our random codebook 
analysis and in fact, as we see in Lemma \ref{lem:Lemma_E_PMF}, one of the rate constraints is a consequence of the 
conditions necessary for the above definition of
$E^{(\mu)}_{L|X^{n}}(\cdot|\cdot)$ to be a PMF. We also note that $E^{(\mu)}_{L|X^n}$ being a PMF guarantees $P_{M|X^n}$ is a PMF.

\begin{comment}{ In the sequel, we refer to 
$E^{(\mu)}_{L|X^{n}}(\cdot|x^{n})$ (and the induced $P_{M|X^{n}}$)  as a PMF.}\end{comment}

Having specified $E^{(\mu)}_{L|X^{n}}(\cdot|\cdot)$, we now characterize 
$P_{M|X^n}$ for $m \in [2^{nR}]\bigcup \{0\}$ as \begin{eqnarray}\label{Eqn:PMFInducedByEncoder}
P_{M|X^{n}}^{(\mu)}(m|x^{n}) \deq 
\begin{cases}
% 1& \mbox{if } m_{1}=0 \mbox{ and }s_1^{(\mu_1)} > 1,\\
% 0& \mbox{if } m_{1}\neq 0 \mbox{ and }s_{1}^{(\mu_1)}(x_{1}^{n}) > 1,\\
\mathbbm{1}_{\{m=0\}}& \mbox{if } s^{(\mu)}(x^{n}) > 1,\\
1-s^{(\mu)}(x^{n})&\mbox{if }m=0 \mbox{ and 
}s^{(\mu)}(x^{n}) \in [0, 1],\\
\sum_{l=1}^{2^{n\tilde{R}}}E^{(\mu)}_{L|X^n}(l|x^n)\mathbbm{1}_{\{b^{(\mu)}(l) = m\}} &\mbox{if }m\neq0 \mbox{ and 
}s^{(\mu)}(x^{n}) \in [0, 1]
\end{cases}
\end{eqnarray}
for all $x^n \in T_{\delta}(X)$, and $s^{(\mu)}(x^{n})$ defined as $s^{(\mu)}(x^{n}) \deq \sum_{l=1}^{2^{n\tilde{R}}}E^{(\mu)}_{L|X^n}(l|x^n)$. For $x^n \notin T_{\delta}(X)$, we let $ P_{M|X^{n}}^{(\mu)}(m|x^{n}) = \mathbbm{1}_{\{m=0\}}.$
% %\vspace{-0.1cm}
% \begin{align}
% P^{(\mu)}_{M|X^{n}}(m|x^{n}) &= \sum_{w^n} \sum_{l=1 }^{2^{n\tilde{R}}}
% E^{(\mu)}_{L|X^n}
% (l|x^{n})\mathds{1}_{\left\{ \begin{array}{c}w^n = \mathtt{w}^n(l,\mu) , b^{(\mu)}( \mathcal{I}^{(\mu)}_{\mathcal{C}}(l))=m \end{array}\!\!\!\!\right\}}\!\!\!\!\!\!\!\!\!\!\!\nonumber\\
% &= \sum_{\substack{w^n \in\\ T_{\delta}(W|x^n)}} \sum_{l=1 }^{2^{n\tilde{R}}} \frac{(1-\epsilon) P^n_{X|W}(x^n|w^n)}{2^{n\tilde{R}}(1+\eta)P^n_X(x^n)} \mathds{1}_{\left\{\!\!\!\! \begin{array}{c}w^n = \mathtt{w}^n(l,\mu) ,  b^{(\mu)}( \mathcal{I}^{(\mu)}_{\mathcal{C}}(l))=m \end{array}\!\!\!\!\right\}}\!\!\!\!\!,
% \end{align}
% for $m\neq 0$ and $P^{(\mu)}_{M|X^{n}}(0|x^{n})  =\displaystyle 1-\sum_{m=1}^{2^{nR}}P^{(\mu)}_{M|X^{n}}(m|x^{n})$. 
It can be verified that $P_{M|X^n}$ is a valid PMF. We have thus described the encoder and $P_{M|X^n}$.
\vspace{-0.2in}
\subsection{Decoder Description}
We now describe the decoder. On observing $z^n \in \mathcal{Z}^n, \mu$ and the 
index $m \in [2^{nR}]\bigcup \{0\}$ communicated by the encoder, for $m\neq 0$, the decoder populates 
\begin{align*}
    \mathcal{D}^{(\mu)}(z^n,m)\deq\{l \in [2^{n\tilde{R}}] : 
 b^{(\mu)}(l)=m, (\mathtt{w}^{n}(l,\mu),z^n) \in 
T_{\delta}(W,Z) \}
\end{align*}
 Let%\vspace{-0.1cm}
\begin{eqnarray}
f^{(\mu)}(m,z^n) \deq \begin{cases} \mathtt{w}^{n}(l,\mu)&\mbox{if } \mathcal{D}^{(\mu)}(z^n,m)=\{l\}\\w_0&\mbox{otherwise, i.e., }|\mathcal{D}^{(\mu)}(z^n,m)| \neq 1 \text{ or } m =0.\end{cases}
\nonumber
\end{eqnarray}
The decoder chooses $y^n$ according to PMF $P^n_{Y|WZ}(y^n|f^{(\mu)}(m,z^n),z^n)$. This 
implies the PMF $P^{(\mu)}_{Y^n|Z^n M}(\cdot|\cdot)$ is given by%\vspace{-0.1cm}
\begin{eqnarray}
\label{Eqn:PMFInducedByDecoder}
P^{(\mu)}_{Y^n|Z^n M}(\cdot|z^n,m) = P^n_{Y|WZ}(\cdot|f^{(\mu)}(m_,z^n),z^n).\end{eqnarray}
\subsection{Distribution of Codebook}
To prove of existence of a codebook for which the above terms are arbitrarily small, we employ random coding. Specifically, we let the codewords of $\mathcal{C}$ to be IID with distribution
\begin{align}\label{eq:prunedDistribution}
    \widetilde{P}_{W^n}(w^n) =& \begin{cases}
     \frac{P^n_W(w^n)}{1-\epsilon} & \quad \text{if} \quad w^n \in  T_{\bar{\delta}}(W) \\
0 & \quad \text{otherwise},
    \end{cases} 
\end{align}
where $\bar{\delta} \deq \delta|\mathcal{X}+\mathcal{W}|$, and $\epsilon(\delta,n) \deq \sum_{w^n\notin 
T_{\bar{\delta}}(W)}P^n_W(w^n).$ 
Note that $\epsilon(\delta,n) \searrow 0$ as $n \rightarrow \infty$ for every $\delta>0$ sufficiently small.
The binning of the codewords is performed independently, where each $b^{(\mu)}(\cdot)$ is chosen randomly, uniformly and independently from $[2^{nR}].$
\subsection{Analysis of Total Variation}
We begin by splitting $\mathscr{Q}$ into two terms using an indicator function $\mathds{1}_{\{ \mathtt{PMF}(\mathcal{C})\}}$ as
\begin{eqnarray}
 \label{eq:SI_indicator}
    {\mathscr{Q}} = {\mathscr{Q}\cdot \mathds{1}_{\{ \mathtt{PMF}(\mathcal{C})\}}} + {\mathscr{Q}\cdot \mathds{1}^c_{\{ \mathtt{PMF}(\mathcal{C})\}}} 
    % \leq \EE{\big[\mathscr{Q}\mathds{1}_{\{ \mathtt{PMF}(\mathcal{C})\}}\big]} +2\cdot\PP\left\{\mathds{1}_{\{ \mathtt{PMF}(\mathcal{C})\}} = 0 \right\} \
\end{eqnarray}
where $\mathds{1}_{\{ \mathtt{PMF}(\mathcal{C})\}}$ is defined as
\begin{eqnarray}
\mathds{1}_{\{ \mathtt{PMF}(\mathcal{C})\}} = \begin{cases} 1 &\mbox{ if }  \displaystyle  
s^{(\mu)}(x^n) \in 
[0,1]  \text{ for all } x^n \in T_{\delta}(X), \mu \in [2^{nC}],\\ 0 &\mbox{ otherwise},
\end{cases}\nonumber\end{eqnarray} 
and recalling $s^{(\mu)}(x^n)= \sum_{l =1}^{2^{n\tilde{R}}}E^{(\mu)}_{L|X^n}(l|x^n).$
% and (\ref{eq:SI_indicator}) follows from the upper bound of 1 over the total variation.
Taking expectation over the codebooks and bounding $\mathscr{Q}$ in the 
% second term of the 
right hand side of \eqref{eq:SI_indicator} by 1\footnote{Total Variation is bounded from above by 1}  gives
\begin{align}\label{eq:SI_indicProb}
    \EE{[\mathscr{Q}]} \leq \EE{[\mathscr{Q}\mathds{1}_{\{ \mathtt{PMF}(\mathcal{C})\}}]} +\PP\left\{\mathds{1}_{\{ \mathtt{PMF}(\mathcal{C})\}} = 0 \right\}. 
\end{align}
% and the event $(C) \nsim \mathtt{PMF} $ is defined as the complement of the event $\left\{s^{(\mu)}(x^n) \in [0,1]  \text{ for all } x_1^n \in T_{\delta}(X_1), x_2^n \in T_{\delta}(X_2), \mu \in [2^{nC}]\right\}$, for a random collection $(c_{1} ,c_{2})$. 
We now show using the lemma below, that by appropriately constraining $\tilde{R}$,  $\PP\left\{\mathds{1}_{\{ \mathtt{PMF}(\mathcal{C})\}} = 0 \right\}$ can be made arbitrarily small. In other words, with high probability, we will have  $E_{L|X^n}^{(\mu)}$ such that $\displaystyle 0\leq \sum_{l_1}^{2^{n\tilde{R}}}E_{L|X^n}^{(\mu)} \leq 1 $ for all $\mu \in [2^{nC}]$ and $x^n \in T_{\delta}(X)$.
\begin{lemma}\label{lem:Lemma_E_PMF}
For any $\delta,\eta \in (0,1/2),$ if $\tilde{R} > I(X:W)+4\delta_1$   then
\begin{align}
\PP&\left(\bigcap_{\mu=1}^{2^{nC}}\bigcap_{x^n\in T_{\delta}(X)}\left(\sum_{l =1}^{2^{n\tilde{R}}}E^{(\mu)}_{L|X^n}(l|x^n) \leq 1\right)\right) \rightarrow 1 \text{ as } n \rightarrow \infty,
\end{align}
 where $\delta_1(\delta),\delta_2(\delta) \searrow 0$ as $\delta \searrow 0,$
\end{lemma}
\begin{proof}
The proof is provided in Appendix \ref{AppSec:ProofOfLemma_PMF}.
\end{proof}
Since, we have
\begin{align}
    \PP\left\{\mathds{1}_{\mathtt{PMF}(\mathcal{C})}=0\right\}  & =  1-  \PP\Bigg(\bigcap_{\mu_1=1}^{2^{nC_1}}\bigcap_{\substack{x^n\in\\ T_{\delta}(X_1)}}\left(E^{(\mu_1)}_{L_1|X_1^n}(l_1|x_1^n) \leq 1\right)\Bigg), \nonumber
\end{align}
from Lemma \ref{lem:Lemma_E_PMF}, for any $\delta \in (0,1)$, we have $\PP\left\{\mathds{1}_{\mathtt{PMF}(\mathcal{C})}=0\right\} \leq \epsilon_p $ for all sufficiently large $n$,
 where $\epsilon_p(\delta) \searrow 0 $ as $\delta \searrow 0.$

\noindent We now look at the first term in (\ref{eq:SI_indicator}), i.e., $\mathscr{Q}\cdot \mathds{1}_{\{ \mathtt{PMF}(\mathcal{C})\}}$. 
This can be expanded as
\begin{align}
    \mathscr{Q}\cdot \mathds{1}_{\{ \mathtt{PMF}(\mathcal{C})\}}  = \!\left[
     \sum_{x^n \in T_{\delta}(X))}P^n_{{X}}({x}^n)\mathscr{Q}_{{x}^n}
     + 
     \sum_{{x}^n \notin T_{\delta}(X)}P^n_{{X}}({x}^n)\mathscr{Q}_{{x}^n}\right]\cdot \mathds{1}_{\{ \mathtt{PMF}(\mathcal{C})\}},\nonumber
\end{align}
where  $\mathscr{Q}_{{x}^n}$ is defined as 
\begin{align}
    \mathscr{Q}_{{x}^n} \deq \frac{1}{2}\sum_{y^n,z^n}&\Bigg|P^n_{YZ|X}(y^n,z^n|x^n) - \sum_{\substack{ \mu \in
[2^{nC}]}}\sum_{m \in[ 2^{nR} ]\bigcup \{0\}} \frac{P^{n}_{Z|X}(z^{n}|x^n)}{2^{nC}}
P^{(\mu)}_{M|X^{n}}(m|x^{n}
)P^{(\mu) } _ { Y^ { n } 
|Z^{n},M}(y^{n}|z^{n},m)\Bigg|.  \nonumber
\end{align}
Using the standard typicality arguments\footnote{Note that $\mathscr{Q}_{x^n}$ is a total variational distance between two conditional PMFs, conditioned on $X$, for each ${x^n}$, and hence it is bounded from above by one.}, we obtain, for all sufficiently large $n$,
\begin{align}\label{eq:SI_expandQtypX}
    \mathscr{Q}\cdot \mathds{1}_{\{ \mathtt{PMF}(\mathcal{C})\}}  = \!
     \sum_{{x}^n \in T_{\delta}(X)}P^n_{{X}}({x}^n)\mathscr{Q}_{{x}^n}
  \mathds{1}_{\{ \mathtt{PMF}(\mathcal{C})\}} + \;\epsilon_t(\delta),
\end{align}
where $\epsilon_t(\delta) \searrow 0$ as $\delta \searrow 0$. Now, what remains is the first term in (\ref{eq:SI_expandQtypX}).
A major portion of our analysis from here on deals with arguing that this term can be made arbitrarily small. Further, since this term contains the indicator $ \mathds{1}_{\{ \mathtt{PMF}(\mathcal{C})\}}$, we can restrict our analysis to only the set of random collection of codebook $C$ that satisfy $0\leq \sum_{l=1}^{2^{n\tilde{R}}}E^{(\mu)}_{L|X^n}(l|x^n) \leq 1$ for all ${x}^n \in T_{\delta}(X)$ and $\mu \in [2^{nC}]$.

\noindent \textbf{Step 1: Isolating the error induced by not covering} \\We begin our analysis by isolating the error induced by not covering the product distribution $P_{XYZ}^n$. Note that under the condition that $\mathds{1}_{\{\mathtt{PMF}(\mathcal{C})\}} = 1,$ we have $  P^{(\mu)}_{M|X^n}(m|x^n) = \sum_{l =1}^{2^{n\tilde{R}}}E^{(\mu)}_{L|X^n}(l|x^n)$ when $m \neq 0$, and $ P^{(\mu)}_{M|X^n}(0|x^n) = 1-\sum_{l =1}^{2^{n\tilde{R}}}E^{(\mu)}_{L|X^n}(l|x^n)$. Using this, we substitute the definition of  randomized encoder
(\ref{Eqn:PMFInducedByEncoder}) and the decoder (\ref{Eqn:PMFInducedByDecoder}) in the second term within the modulus of $\mathscr{Q}_{x^n}$. This gives
\begin{eqnarray}
\frac{1}{2^{nC}}\sum_{\substack{ \mu 
\in [2^{nC}] }} \sum_{m \in[ 2^{nR}]\cup \{0\} }
{P^{n}_{Z|X}(z^{n}|x^n)P^{(\mu)}_{M|X^{n}}\!(m|x^{n}
)P^{(\mu) } _ { Y^ { n } |Z^{n}M}(y^{n}|z^{n},m)} =\!T_{1}\!+\!T_{2},\!
 \nonumber
\end{eqnarray}
where\footnote{For the ease of notation, we do not show the dependency of $T_1,$ and $T_2$ on $x^n, y^n$ and $z^n$.},
\begin{align} 
T_{1} &\deq \sum_{\substack{ \mu \in [2^{nC}] }}\sum_{m \in[ 2^{nR}] }\sum_{l=1 }^{2^{n\tilde{R}}} \sum_{\substack{w^n \in T_{\delta}(W|x^n)}} \frac{(1-\epsilon)}{(1+\eta)}\frac{1}{2^{n(\tilde{R}+C)}}
\frac{ P^{n}_{Z|X}(z^{n}|x^n) 
P^n_{X|W}(x^n|w^n)}{ P^n_X(x^n)}
\nonumber \\ & \hspace{2.1in} \mathds{1}_{\left\{ \substack{ w^n = \mathtt{w}^n(l,\mu) 
,b^{(\mu)}( l)=m }\right\}}
 P^n_{Y|WZ}(y^{n}|f^{(\mu)}(b^{(\mu)}(l),z^n 
),z^n)\nonumber\\&
%%%%%%%% new line %%%%%%%%%%%%%%%%%%%
= \sum_{\substack{ \mu \in [2^{nC}]  }}\sum_{l=1
}^{2^{n\tilde{R}}} \sum_{\substack{w^n \in T_{\delta}(W|x^n)}} \frac{(1-\epsilon) }{(1+\eta)}\frac{1}{2^{n(\tilde{R}+C)}}
\frac{P^{n}_{Z|X}(z^{n}|x^n)
P^n_{X|W}(x^n|w^n)}{P^n_X(x^n)}\mathds{1}_{\left\{  w^n = \mathtt{w}^n(l,\mu) 
\right\}} \nonumber \\ & \hspace{3.5in}
P^n_{Y|WZ}(y^{n}|f^{(\mu)}(b^{(\mu)}(l),z^n 
),z^n),\nonumber \\
%%%%%%%% new line %%%%%%%%%%%%%%%%%%%
T_{2}&\deq \frac{1}{2^{nC}}\sum_{\substack{ \mu \in [2^{nC}]  
}}{P^{n}_{Z|X}(z^{n}|x^{n})\left[1-\sum_{l =1}^{2^{n\tilde{R}}}E^{(\mu)}_{L|X^n}(l|x^n)\right]}{}P^n_{Y|WZ}(y^{n}|w_0,z^n).\nonumber
\end{align}

Substituting $T_{1},T_{2}$ for the second term within the modulus of $\mathscr{Q}_{x^n}$, and applying triangle inequality, we obtain 
% one can verify\footnote{This is a standard application of triangular inequality after adding an appropriate term.} that 
$\mathscr{Q}_{x^n}\11_{\mathtt{PMF}(\mathcal{C})} \leq \left[\frac{1}{2}
\sum_{y^{n},z^{n}} (S + \widetilde{S})\right] \11_{\mathtt{PMF}(\mathcal{C})} \leq \frac{1}{2}
\sum_{y^{n},z^{n}} (S + \widetilde{S}\11_{\mathtt{PMF}(\mathcal{C})})$, where
\begin{align}
    S &  \deq \Bigg|P^n_{YZ|X}(y^n,z^n|x^n) -  \nonumber \\& \hspace{0.1in}
\frac{(1-\epsilon)}{(1+\eta)}\frac{1}{2^{n(\tilde{R}+C)}} \sum_{\substack{ \mu ,l}} \!\sum_{\substack{w^n \in T_{\delta}(W|x^n)}}\!\!\!\!\!\!\!\!\!
\left.\frac{P^{n}_{Z|X}(z^{n}|x^{n})
P^n_{X|W}(x^n|w^n)}{P^n_X(x^n)}  P^n_{Y|WZ}(y^{n}|f^{(\mu)}(b^{(\mu)}(l),z^n ),z^n)\mathds{1}_{\left\{ 
\substack{w^n = \mathtt{w}^n(l,\mu)}\right\}} \right|,
 \nonumber\\
  %%%%%%%%%%%%%% new line %%%%%%%%%%%%%%%%%%%
 \widetilde{S} & \deq \left|\frac{1}{2^{nC}}\sum_{\substack{ \mu \in [2^{nC}]  
}}P^{n}_{Z|X}(z^{n}|x^n) \left(1-\sum_{l =1}^{2^{n\tilde{R}}}E^{(\mu)}_{L|X^n}(l|x^n)\right)P^n_{Y|WZ}(y^{n}|w_0,z^n)\right|. \nonumber
\end{align}

 Note that the term corresponding to $\tilde{S}$ captures the error induced by not covering the product distribution $P_{XYZ}^n(\cdot)$ and we bound this term employing the following proposition.
\begin{prop}\label{propSI:S_tilde}
There exist functions  $\epsilon_{\widetilde{S}}(\delta), $ and $\delta_{\widetilde{S}}(\delta)$, 
such that for  all sufficiently small $\delta$ and sufficiently large $n$, we have $\EE[\sum_{x^n \in T_\delta(X)}P^n_X(x^n)\sum_{y^{n},z^n}\widetilde{S}\11_{\mathtt{PMF}(\mathcal{C})}] \leq\epsilon_{\widetilde{S}}(\delta) $, if  $\tilde{R} >  I(X;W) + \delta_{\widetilde{S}}, $ where $\epsilon_{\widetilde{S}}, \delta_{\widetilde{S}} \searrow 0$ as $\delta \searrow 0$. 
\end{prop}
\begin{proof}
The proof is provided in Appendix \ref{appx:propSI:S_tilde}
\end{proof}
Now we move on to isolating the error component of $S$ caused by binning the randomized encoders.
\noindent \textbf{Step 2: Error caused by binning}\\
We now consider the term corresponding to $S$. By adding and subtracting an appropriate term within the modulus of $S$ and using triangle inequality, $S$ can be bounded as $S \leq S_1 + S_2$, where
\begin{align}
 &S_{1} \deq \Bigg|P^n_{YZ|X}(y^n,z^n|x^n) - \nonumber 
 \sum_{\substack{ \mu 
,l}} \!\!\!\sum_{\substack{w^n \in\\ T_{\delta}(W|x^n)}}\!\!\!\!\!\!
\left.\frac{(1-\epsilon) 
P^n_{X|W}(x^n|w^n)P^{n}_{Z|X}(z^{n}|x^n) P^n_{Y|WZ}(y^{n}|w^n,z^n 
)}{2^{n(\tilde{R}+C)}(1+\eta)P^n_X(x^n)}\mathds{1}_{\left\{ 
\substack{w^n = \mathtt{w}^n(l,\mu)}\right\}} \right|,
 \nonumber\\
 %%%%%%%%%%%%%% new line %%%%%%%%%%%%%%%%%%%
 &S_{2} \deq \Bigg| \frac{(1-\epsilon)}{(1+\eta)}\frac{1}{2^{n(\tilde{R}+C)}}\sum_{\substack{ \mu ,l}}  \sum_{\substack{w^n \in 
\\T_{\delta}(W|x^n)}} \!\!\!
\frac{  P^{n}_{Z|X}(z^{n}|x^n) 
P^n_{X|W}(x^n|w^n)}{ P^n_X(x^n)}\mathds{1}_{\left\{ 
\substack{w^n = \mathtt{w}^n(l,\mu)}\right\}} \nonumber\\
&\hspace{2.2in}\left(\!P^n_{Y|WZ}(y^{n}|w^n,z^n ) - 
P^n_{Y|WZ}(y^{n}|f^{(\mu)}(b^{(\mu)}(l),z^n 
),z^n)\!\right)\!\!\Bigg|\nonumber.
%%%%%%%%%%%%%% new line %%%%%%%%%%%%%%%%%%%%%%%%
% &S_{3} = \left|\frac{1}{2^{nC}}\sum_{\substack{ \mu \in [2^{nC}]  
% }}P^{n}_{XZ}(x^{n},z^{n}) \left({1-\sum_{m=1}^{2^{nR}}P^{(\mu)}_{M|X^{n}}(m|x^{n}
% )}\right)P^n_{Y|WZ}(y^{n}|w_0,z^n)\right|.\nonumber
\end{align}

Note that the term $S_2$ captures the error introduced due to the binning operation. To bound this term, we provide the following proposition.
\begin{prop}[Mutual Packing]\label{propSI:Lemma for S_2}
There exist  $\epsilon_{S_{2}}(\delta),$ such that for  all sufficiently small $\delta$ and sufficiently large $n$, we have $\EE\left[\sum_{x^n \in T_{\delta}(X)}P_X^n(x^n)\sum_{y^n,z^n}{S}_2\right] \leq \epsilon_{{S_2}}(\delta) $, if  $(\tilde{R}_1- R_1) \leq  I(W;Z)+\delta_{S_2}$,  where  $\epsilon_{{S}_2}, \delta_{S_2}(\delta) \searrow 0$ as $\delta \searrow 0$.
\end{prop}
\begin{proof}
The proof is provided in Appendix  \ref{appx:propSI:Lemma for S_2}.
\end{proof}
Now we are left with the analysis of the term $S_1$. 
% For this, we segregate the effect of two encoders within the term $S_1$,  and separately analyze each of them, starting with the Alice's encoder.

\noindent\textbf{Step 3: Bounding the approximation/covering error} \\
In this last step, we analyze the term $S_1$ which captures the action of the encoder in approximating the product distribution $P_{{X}YZ}^n(\cdot)$. For that, we split $S_{1}$ as $ S_{1} \leq S_{11} + S_{12} $, where
\begin{align}
    \!S_{11} =  \!\Bigg|&P^n_{YZ|X}(y^n,z^n|x^n) -  2^{-n(\tilde{R}+C)}\!\sum_{\substack{ \mu ,l}} 
P^{n}_{Z|X}(z^{n}|x^{n}) \!\!\!\!\!\!\sum_{w^n \in T_{\bar{\delta}}(W)}\!\!\!\!\!\!\!\! \frac{ P^n_{X|W}(x^n|w^n)}{P^n_X(x^n)}\mathds{1}_{\left\{ \substack{w^n = \\\mathtt{w}^n(l,\mu)}\right\}} P^n_{Y|WZ}(y^{n}|w^n,z^n )\Bigg| \nonumber 
\end{align}
\begin{align}
 S_{12} & =  2^{-n(\tilde{R}+C)}\Bigg| \sum_{\substack{ \mu ,l}} 
P^{n}_{Z|X}(z^{n}|x^{n}) \sum_{w^n \in T_{\bar{\delta}}(W)}
 \frac{ P^n_{X|W}(x^n|w^n)}{P^n_X(x^n)}\mathds{1}_{\left\{ \substack{w^n = \\\mathtt{w}^n(l,\mu)}\right\}} P^n_{Y|WZ}(y^{n}|w^n,z^n) - \nonumber \\ 
 &\hspace{1in} \left(\frac{1-\epsilon}{1+\eta}\right)  \sum_{\substack{ \mu ,l}} 
P^{n}_{Z|X}(z^{n}|x^n) \sum_{w^n \in T_{\delta}(W|x^n)}
\frac{ P^n_{X|W}(x^n|w^n)}{P^n_X(x^n)}\mathds{1}_{\left\{ \substack{w^n = \\\mathtt{w}^n(l,\mu)}\right\}} P^n_{Y|WZ}(y^{n}|w^n,z^n)\Bigg|. \label{eq:S12_sideInf}
\end{align}
% Now we have $\mathscr{Q} \leq \displaystyle \!\!\!\!\sum_{x^{n},y^{n},z^{n}}\!\!\!\! S_{11}+S_{12}+S_{2}+S_{3}$. 
Using the Markov chains $Z-X-W$ and $X-(Z,W)-Y$ which $P_{WXYZ}$ satisfies, and the fact that $\sum_{w^n \in T_{\bar{\delta}}(W)}\mathds{1}_{\left\{ {w^n = \\\mathtt{w}^n(l,\mu)}\right\}} = 1, $ we can simplify the second term in $ S_{11}$ as
\begin{align}
    & 2^{-n(\tilde{R}+C)} \!\sum_{\substack{ \mu ,l}} 
P^{n}_{Z|X}(z^{n}|x^{n}) \!\!\!\!\!\!\sum_{w^n \in T_{\bar{\delta}}(W)}\!\!\!\!\!\!\!\! \frac{ P^n_{X|W}(x^n|w^n)}{P^n_X(x^n)}\mathds{1}_{\left\{ \substack{w^n = \mathtt{w}^n(l,\mu)}\right\}} P^n_{Y|WZ}(y^{n}|w^n,z^n) \nonumber \\ 
%%%%%%%%%%%%%% new line %%%%%%%%%%%%%%
& = \frac{1}{2^{n(\tilde{R}+C)}} \sum_{\substack{ \mu ,l}}\sum_{w^n \in T_{\bar{\delta}}(W)}\!\!\!\!\!\frac{P_{X|W}^n(x^n|\mathtt{w}^n(l,\mu))}{P^n_X(x^n)} P_{Z|XW}^n(z^n|x^n, \mathtt{w}^n(l,\mu)) P_{Y|WXZ}^n(y^n|\mathtt{w}^n(l,\mu),x^n,z^n)\mathds{1}_{\left\{ \substack{w^n = \\\mathtt{w}^n(l,\mu)}\right\}} \nonumber\\
%%%%%%%%%%%%%% new line %%%%%%%%%%%%%%
& = \frac{1}{2^{n(\tilde{R}+C)}} \sum_{\substack{ \mu ,l}}\frac{P_{XYZ|W}^n(x^n,y^n,z^n|\mathtt{w}^n(l,\mu))}{P_X^n(x^n)}. \nonumber
\end{align}
% where in the last equality above, we use  $\left( \sum_{w^n \in T_{\bar{\delta}}(W)}\mathds{1}_{\left\{ \substack{w^n = \mathtt{w}^n(l,\mu)}\right\}} \right)  =1$. 
 Substituting the above simplification into the expression for $S_{11}$ gives
\begin{align}\label{eq:softcoveringterm}
    S_{11} = \Bigg|P^n_{YZ|X}(y^n,z^n|x^n) -  \frac{1}{2^{n(\tilde{R}+C)}}\sum_{\substack{ \mu ,l}} 
 \frac{P_{XYZ|W}^n(x^n,y^n,z^n|\mathtt{w}^n(l,\mu))}{P_X^n(x^n)}\Bigg|.
 \end{align}
 Substituting this simplification in $\EE[\sum_{x^n\in T_{\delta}(X)}P^n_X(x^n) (\sum_{y^{n},z^{n}} S_{11})]$, we obtain
 \begin{align}
    \EE&[\sum_{x^n\in T_{\delta}(X)}P^n_X(x^n) (\sum_{y^{n},z^{n}} S_{11})] \nonumber \\
    &=\EE\left[ \sum_{x^n\in T_{\delta}(X)}\sum_{y^{n},z^{n}}\Bigg|P^n_{XYZ}(x^n,y^n,z^n) -  \frac{1}{2^{n(\tilde{R}+C)}}\sum_{\substack{ \mu ,l}} 
 {P^n_{XYZ|W}(x^n,y^n,z^n|\mathtt{w}^n(l,\mu))}\Bigg|\right] \nonumber \\
%  & \leq \EE\left[ \sum_{x^n, y^{n},z^{n}}\Bigg|P^n_{XYZ}(x^n,y^n,z^n) -  \frac{1}{2^{n(\tilde{R}+C)}}\sum_{\substack{ \mu ,l}} {P^n_{XYZ|W}(x^n,y^n,z^n|\mathtt{w}^n(l,\mu))}\Bigg|\right] \\
  & \leq \EE\left[ \sum_{x^n, y^{n},z^{n}}\Bigg|\widetilde{P}^n_{XYZ}(x^n,y^n,z^n) -  \frac{1}{2^{n(\tilde{R}+C)}}\sum_{\substack{ \mu ,l}} 
 {P^n_{XYZ|W}(x^n,y^n,z^n|\mathtt{w}^n(l,\mu))}\Bigg|\right] + {2\epsilon} \label{eq:S11_expanded},
\end{align}
where the last inequality follows by defining $\widetilde{P}^n_{XYZ}(\cdot)$ as $\widetilde{P}^n_{XYZ}(x^n,y^n,z^n) \deq \sum_{w^n \in T_{\bar{\delta}}(W)}P^n_{XYZ|W}(x^n,y^n,z^n|w^n)\widetilde{P}_{W^n}(w^n)$.
%  Our analysis so far has been for a specific codebook $\mathcal{C}$. To prove of existence of a codebook for which the above terms are arbitrarily small, we employ random coding. Specifically, we let the codewords of $\mathcal{C}$ to be IID with distribution
% \begin{align}\label{eq:prunedDistribution}
%     \tilde{p}_{W^n}(w^n) =& \begin{cases}
%      \frac{P^n_W(w^n)}{\sum_{w^n\in 
% T_{\bar{\delta}}(W)}P^n_W(w^n)} & \quad \text{if} \quad w^n \in  T_{\bar{\delta}}(W) \\
% 0 & \quad \text{otherwise}.
%     \end{cases} 
% \end{align}
%  Henceforth, we consider the expected value of the above terms. 

\begin{lemma}[One-shot Soft Covering]\label{lem:Change Measure Soft Covering Variance Based}
	Let $P_{AB}$ be a joint PMF defined on $\mathcal{A}\times\mathcal{B}$  with	$\mathcal{A}$ and $\mathcal{B}$ being  finite sets. Further, suppose we are given a subset $\mathcal{T} \subset \mathcal{A}$ and a collection of subsets $\mathcal{T}_b \subset \mathcal{A}$ for all $b \in \mathcal{B}$ which satisfy the following hypotheses for all $b \in \mathcal{B}$:
	\begin{subequations}\label{Soft_Covering-constraints}
	\begin{align}
		 P_{A}(\mathcal{T}) & \geq 1-\epsilon, \label{Soft_Covering-constraints1} \\
		P_{A|B}(\mathcal{T}_b|b) & \geq 1-\epsilon,  \label{Soft_Covering-constraints2} \\
		 \left(\sum_{a\in\CalT}\sqrt{P_A(a)}\right)^2 & \leq D, \label{Soft_Covering-constraints3} \quad \text{and}\\
		 P_{A|B}(a|b) & \leq \frac{1}{d}, \forall a \in \mathcal{T}_b, \label{Soft_Covering-constraints4}
	\end{align}
  	\end{subequations}
	for some $\epsilon \in (0,1)$ and $d < D$.
	Let $M $ be a finite non-negative integer and let a random covering code $\mathbbm{C}\deq \{C_m\}_{m \in [1,M]}$ be defined as a collection of codewords $C_m$ that are chosen pairwise independently according to the distribution $P_B$ from $\mathcal{B}$.
	Then we have
	\begin{align}
        \EE_{\mathbbm{C}}\left[\sum_{a\in{\mathcal{A}}}\Big|P_{A}(a) - \frac{1}{M}\sum_{m=1}^M P_{A|B}(a|C_m)\Big|\right] \leq \sqrt{\frac{D}{Md}} +4\delta(\epsilon).
	\end{align}
%Futhermore, for $\Tilde{\sigma}_x$
%  defined as $\Tilde{\sigma}_x \deq \Pi\Pi_x{\sigma}_x\Pi_x\Pi$, we have
%  \begin{align}
%         \EE_{\mathbbm{C}}\left[\Big\|\sum_{x\in\bar{\mathcal{X}}}\lambda_x\tilde{\sigma}_x - \frac{1}{M}\sum_{m=1}^M \frac{\lambda_{C_m}}{\mu_{C_m}}\tilde{\sigma}_{C_m}\Big\|_1\right] \leq \sqrt{\frac{\kappa D}{Md}}. \label{lem:ChangeMeasure:ineq2}
% 	\end{align}
 \end{lemma}
\begin{proof}
The proof is provided in Appendix \ref{appx:proofOneShotLemma}
\end{proof}

Now we prove the above term is small by using the following identification. Identify $\mathcal{A} $ by $ (\mathcal{X}^n \times \mathcal{Y}^n \times \mathcal{Z}^n), $ $\mathcal{B}$ by $T_{\bar{\delta}}(W)$, $\mathcal{T}$ by $T_{\delta}(XYZ)$,  $\mathcal{T}_{b}$ by $T_{\delta}(XYZ|w^n)$ for all $w^n \in T_{\bar{\delta}}(W)$, and $P_{AB}$ by ${P}_{XYZ|W}^n\widetilde{P}_{W^n}$ (with $\widetilde{P}_{W^n}(\cdot)$ as defined in \eqref{eq:prunedDistribution}).  Using this identification we first compute $D$ and $d$ that satisfy the hypothesis of the lemma.   This gives $D = 2^{n(H(X,Y,Z) + \delta_{XYZ})} $ and $d = 2^{n(H(XYZ|W) - \delta'_{XYZ})}.$
To satisfy \eqref{Soft_Covering-constraints1}, we use the fact that if $\| P_A - Q_A\| \leq \varepsilon_A,$ for $P_A$ and $Q_A$ defined as two distributions on $\CalA$ then for any subset $\bar{\CalA} \subset \CalA$, we have $ P_A(\bar{\CalA}) \geq Q_A(\bar{\CalA}) -\varepsilon_A. $ Since $ \|P^n_{XYZ} - \widetilde{P}_{X^nY^nZ^n}\| \leq 2\epsilon $, we have $\widetilde{P}_{X^nY^nZ^n}(T_{\delta(XYZ)}) \geq 1-3\epsilon(\delta)$ which can be made arbitrarily close to $1$ for a sufficiently large $n$. The hypotheses $\eqref{Soft_Covering-constraints2}$ and $\eqref{Soft_Covering-constraints4}$ can be shown to be true using the basic typicality arguments. For the hypothesis $\eqref{Soft_Covering-constraints3}$ we use the bound $\widetilde{P}_{X^nY^nZ^n}(\cdot) \leq \frac{1}{(1-\epsilon)}P^n_{XYZ}(\cdot)$, which gives the $D$ mentioned above.

Using this identification and applying Lemma \eqref{lem:Change Measure Soft Covering Variance Based} on \eqref{eq:S11_expanded} we obtain 
$\EE[\sum_{x^n\in T_{\delta}(X)}P^n_X(x^n) (\sum_{y^{n},z^{n}} S_{11})] \leq \epsilon_{S_{11}},$ if $\tilde{R}+C \geq I(XYZ;W) + \delta_{S_{11}}$ for sufficiently $n$, where $\delta_{S_{11}}(\delta), \epsilon_{S_{11}}(\delta) \searrow 0$ as $\delta\searrow0$.

% Using this simplification, $\EE[\sum_{x^n\in T_{\delta}(X)}P^n_X(x^n) (\sum_{y^{n},z^{n}} S_{11})]$ can be shown to be small if $\tilde{R} + C \geq I(XYZ;W)$, for a sufficiently large $n$, using a standard application of \cite[Lem.~19]{CuffPhDThesis}.

Now consider $S_{12}$. This term can be split into the $S'_{12}$ and $S''_{12}$ such that $S_{12} = S'_{12} + S''_{12}$, where
\begin{align}
    S'_{12}  &\deq  2^{-n(\tilde{R}+C)}\Bigg|\left(1-\frac{1-\epsilon}{1+\eta}\right)  \sum_{\substack{ \mu ,l}} 
P^{n}_{Z|X}(z^{n}|x^n) \hspace{-10pt}\sum_{\substack{w^n \in\\  T_{\delta}(W|x^n)\cap T_{\bar{\delta}}(W)}}
 \hspace{-10pt}\frac{ P^n_{X|W}(x^n|w^n)}{P^n_X(x^n)}\mathds{1}_{\left\{ \substack{w^n = \\\mathtt{w}^n(l,\mu)}\right\}} P^n_{Y|WZ}(y^{n}|w^n,z^n) \Bigg|, \nonumber \\
 S''_{12}  &\deq  2^{-n(\tilde{R}+C)} \Bigg|\sum_{\substack{ \mu ,l}} 
P^{n}_{Z|X}(z^{n}|x^{n}) \sum_{\substack{w^n \notin T_{\delta}(W|x^n)\\w^n \in T_{\bar{\delta}}(W)}}
\frac{ P^n_{X|W}(x^n|w^n)}{P^n_X(x^n)}\mathds{1}_{\left\{ \substack{w^n = \\\mathtt{w}^n(l,\mu)}\right\}} P^n_{Y|WZ}(y^{n}|w^n,z^n)\Bigg|. \nonumber
\end{align}
Now, we apply expectation over each of the following to obtain,
\begin{align}
    \Expectation\left[\sum_{y^n,z^n}S'_{12}\right] = & 2^{-n(\tilde{R}+C)}\frac{\eta+\epsilon}{1+\eta} \!\sum_{y^n,z^n}\!\sum_{\substack{ \mu ,l}}  
P^{n}_{Z|X}(z^{n}|x^{n}) \hspace{-15pt}\sum_{\substack{w^n \in\\  T_{\delta}(W|x^n)\cap T_{\bar{\delta}}(W)}}\hspace{-15pt}
 \frac{ P^n_{X|W}(x^n|w^n)}{P^n_X(x^n)}\frac{P_{W}^n(w^n)}{1-\epsilon} P^n_{Y|WZ}(y^{n}|w^n,z^n) \nonumber \\
 %%%%%%%%%%%%%%%%% new line %%%%%%%%%%%%%%%%%%%%%%%%
 \leq &\left(\frac{\eta+\epsilon}{1+\eta}\right) \frac{1}{(1-\epsilon)} \sum_{y^n,z^n} 
 \sum_{w^n \in T_{\delta}(W|x^n)}
 P^n_{WYZ|X}(w^n,y^n,z^n|x^n) 
 \leq \;\;\; \frac{(\eta + \epsilon)}{(1+\eta)(1-\epsilon)}.
\end{align}
And similarly, we have
\begin{align}
    \Expectation\left[\sum_{y^n,z^n}S''_{12}\right] = & \frac{1}{(1-\epsilon)}\sum_{y^n,z^n} \sum_{\substack{w^n \notin T_{\delta}(W|x^n)\\w^n \in T_{\bar{\delta}}(W)}}
P^n_{WYZ|X}(w^n,y^n,z^n|x^n) \nonumber \\
 \leq & \frac{1}{(1-\epsilon)} \sum_{\substack{w^n \notin T_{\delta}(W|x^n)}}
P^n_{W|X}(w^n|x^n) \leq \frac{\epsilon''}{1-\epsilon},
\end{align}
% In the interest of brevity, we consider a simple version of this term. Observe that
% \begin{align}
%     \EE&\bigg[\sum_{y^n}|\frac{1}{2^{nR}}\sum_{x^n \in T_{\delta}(X)}\sum_{l=1}^{2^{nR}}
%     \mathbbm{1}_{\{X^n(l) = x^n\}} 
%      P^n_{Y|X}(y^n|x^n) - \frac{1}{2^{nR}}\sum_{x^n\in T_{\delta}(X|y^n)}\sum_{l=1}^{2^{nR}}
%     \mathbbm{1}_{\{X^n(l) = x^n\}} 
%      P^n_{Y|X}(y^n|x^n)|\bigg] \nonumber\\
%           &= \EE\bigg[\sum_{y^n}| \frac{1}{2^{nR}}\sum_{x^n\notin T_{\delta}(X|y^n)}\sum_{l=1}^{2^{nR}}
%     \mathbbm{1}_{\{X^n(l) = x^n\}} 
%      P^n_{Y|X}(y^n|x^n)|\bigg]
%      \nonumber \\
%      &= \sum_{y^n}\sum_{x^n\notin T_{\delta}(X|y^n)}P^n_X(x^n) 
%      P^n_{Y|X}(y^n|x^n) \leq \epsilon'. \nonumber
% \end{align}
% An analogous sequence of steps can be used to show $\Expectation[\sum_{x^{n},y^{n},z^{n}} S_{12}]$ is arbitrarily small. Refer \cite{arxiv_CurrentPaper} for details. 
% Lemma \ref{lem:conditionSoftCovering} proves a basic version of this term, and can be easily extended to prove that $\sum_{x^{n},y^{n},z^{n}} S_{12}$ is also small in expected sense. A more complete proof is provided in  \cite{arxiv_CurrentPaper}.
where $\epsilon''(\delta) \searrow 0$ as $\delta \searrow 0$.
We have argued that terms $S_1, S_2$ and $\widetilde{S}$ are small in expectation for sufficiently large $n$, which implies $\EE[\mathscr{Q}_{x^n}\11_{\{\mathtt{PMF}(\mathcal{C})\}}] \leq \epsilon_{Q}$ for sufficiently large $n$, where $\epsilon_Q(\delta) \searrow 0$ as $\delta \searrow 0.$ Using this in \eqref{eq:SI_expandQtypX}, and subsequently in \eqref{eq:SI_indicProb}, and eventually in    \eqref{eq:SI_indicator} gives $\EE[\mathscr{Q}] \leq \epsilon$, for sufficiently large $n$ if 
% and that the collection of $p^{(\mu)}_{M|X^n}(\cdot|\cdot)$ is a valid PMF with high probability.
% To connect the high probability and  
% We provide an explanation, using these bounds, on the existence of a collection of encoders satisfying definition (\ref{Defn:AlternateSideInfo}) in \cite{arxiv_CurrentPaper}. 
% By using Markov Lemma we guarantee the existence of a collection of encoders satisfying (\ref{Defn:AlternateSideInfo}) if   
\begin{align}\label{eq:SI_rateRegion1}
    \tilde{R} \geq R \geq 0,\quad  C\geq 0, \quad  \tilde{R} \geq I(X;W), \quad \tilde{R} - R \leq I(W;Z) \quad \text{and} \quad  \tilde{R} + C \geq I(XYZ;W).
\end{align} 
% where $\epsilon(\delta) \searrow 0$ as $\delta \searrow 0$.
Lastly, the proof is completed using the Fourier-Motzkin Elimination \cite{fourier-motzkin}.
\end{proof}
\section{Proof of Theorem \ref{Thm:Distributed}} \label{sec:ProofOfDistTheoremInnerBound}
\label{Sec:ProofOfDistTheorem}

Having designed a randomized encoding scheme based on typicality for the side-information case, we are  in a position to employ the same encoder for the distributed scenario. 
In contrast to the side-information scenario, both randomized encoders choose 
codewords resulting in the need to prove that the the individual codewords 
chosen by the distributed randomized encoders are with high probability jointly 
typical with the observed source sequences. This involves new elements in the 
 context of soft covering. 
% Specifically, we leverage Suen's 
% \cite{janson2011random} inequality to upper bound the probability of sum of 
% random variables that are dependent.
Fix a PMF $P_{QW_1W_2X_1X_2Y}$ satisfying the constraints stated in the theorem. Since $Q$, the time sharing random variable, is employed in the standard way, for ease of exposition, we provide the proof of the special case of $Q = 0.$ Its generalization can be obtained in a straightforward way.
 Let $\mu \in [2^{nC}]$  denote the common randomness shared amidst 
all terminals. The first encoder uses a part of the entire common randomness available to it, say $C_1$ bits out of the $C$ bits, which is denoted by $\mu_1 \in [2^{nC_1}]$. Similarly, let $\mu_2 \in [2^{nC_2}]$ denote the common randomness used by the second encoder.  Note that $C = C_1 \times C_2$ and $\mu \deq (\mu_1,\mu_2)$. Our goal is to prove the existence of PMFs 
$P_{M_1|X_1^n}^{(\mu_1)}(m_1|x_1^n) : x_1^n \in \mathcal{X}_1^n, m_1 \in 
[\Theta_1], \mu_1 \in [2^{nC_1}], P^{\mu_2}_{M_2|X_2^n}(m_2|x_2^n) : x_2^n \in 
\mathcal{X}_2^n, m_2 \in [\Theta_2], \mu_2 \in [2^{nC_2}]$, $P^{(\mu)}_{Y^n|M_1,M_2}(y^n| m_1,m_2) : y^n \in 
\mathcal{Y}^n, (m_1,m_2) \in 
[\Theta_1] \times [\Theta_2]$ such that 
\begin{align}
 \mathscr{Q}\define \!\frac{1}{2}&\!\!\sum_{x_1^n,x_2^n,y^n}\Bigg|P^n_{X_1X_2Y}(x_1^n,x_2^n,y^n)  - \sum_{\substack{ \mu \in
[2^{nC}]  }}\sum_{m_1 \in[ \Theta_1]}\sum_{m_2 \in[ \Theta_2]} \!\!\!
\frac{P^{n}_{X_1X_2}(x_1^{n},x_2^n)}{2^{nC}} P^{(\mu_1)}_{M_1|X_1^{n}}(m_1|x_1^{n})\nonumber \\ & \hspace{2.7in}P^{(\mu_2)}_{M_2|X_2^{n}}(m_2|x_2^{n}
)P^{(\mu) } _ { Y^ { n } 
|M_1,M_2}(y^{n}|m_1,m_2)\Bigg| \leq \varepsilon, 
\label{eq:dist_main_lemma}
\end{align}
for $\frac{\log \Theta_j}{n} =R_j : j \in [2]$ and for all sufficiently large $n$. Consider the collections $\CalC_{1} \deq 
(\mathcal{C}_{1}^{(\mu_1)}:1 \leq \mu_1 \leq 2^{nC_1})$ where $\mathcal{C}_{1}^{(\mu_1)} \deq 
(\mathtt{w}_{1}(l_{1},\mu_1) : 1\leq l_{1} \leq 2^{n\tilde{R}_{1}})$ and $\mathcal{C}_2 \deq 
(\mathcal{C}_{2}^{(\mu_1)}:1 \leq \mu_2 \leq 2^{nC_2})$
where $\mathcal{C}_2^{(\mu_2)}
\deq (\mathtt{w}_{2}(l_{2},\mu_2): 1 \leq l_{2} \leq 
2^{n\tilde{R}_{2}})$. For this 
collection, we let
% \begin{eqnarray}
%  s_{1}^{(\mu)}(x_{1}^{n}) =  \frac{1}{2^{n\tilde{R_1}}} 
% \frac{1-\epsilon}{1+\eta} 
% \sum_{l_{1}=1}^{2^{n\tilde{R_1}}}\frac{P^{n}_{X_1|W_1}(x_1^n|\mathtt{w}_1^n(l_1
% ,\mu))}{P^n_{X_1}
% (x_1^n)}\mathbbm{1}_{\{ \mathtt{w}_1^n(l_1
% ,\mu)
% \in T_{\delta}(W_1|x_1^n)\}}\nonumber\\
% s_{2}(x_{2}^{n}) = \frac{1}{2^{n\tilde{R_2}}} \frac{1-\epsilon}{1+\eta} 
% \sum_{l_{2}=1}^{2^{n\tilde{R_2}}}\frac{P^{n}_{X_2|W_2}(x_2^n|\mathtt{w}
% _2^n(l_2,\mu_2))}{P^n_{X_2}
% (x_2^n)}\mathbbm{1}_{\{ \mathtt{w}_2^n(l_2,\mu_2) 
% \in T_{\delta}(W_2|x_2^n)\}}\nonumber
% \end{eqnarray}
\begin{align}
    E^{(\mu_1)}_{L_1|X_1^n}(l_1|x_1^n) &\deq \frac{1}{2^{n\tilde{R}_1}}\frac{1-\epsilon_1}{1+\eta}\sum_{w_1^n \in T_{\delta}(W_1|x_1^n)}\mathbbm{1}_{\{\mathtt{w}^n(l_1,\mu_1) = w_1^n\}}\frac{P^n_{X_1|W_1}(x_1^n|w_1^n)}{P^n_{X_1}(x_1^n)}\nonumber \\
    %%%%%%%%%%%%%%% new line %%%%%%%%%%%%%%%%
    E_{L_2|X_2^n}^{(\mu_2)}(l_2|x_2^n) &\deq \frac{1}{2^{n\tilde{R}_2}}\frac{1-\epsilon_2}{1+\eta}\sum_{w_2^n \in T_{\delta}(W_2|x^n_2)}\mathbbm{1}_{\{\mathtt{w}^n(l_2,\mu_2) = w_2^n\}}\frac{P^n_{X_2|W_2}(x_2^n|w_2^n)}{P^n_{X_2}(x_2^n)}\nonumber
\end{align}
where  $\bar{\delta}_i \deq \delta|\mathcal{X}_i+\mathcal{W}_i|$ and $\epsilon_i = 1-P_W^n(T_{\bar{\delta}_i}(W_i)); i=1,2.$ The definition of $E^{(\mu_1)}_{L_1|X_1^n}$ and $E^{(\mu_2)}_{L_2|X_2^n}$ can be thought of as encoding rules that do not exploit the additional rebate obtained by using binning techniques.
\subsection{Binning of Random Encoders}
Further, we define maps $b_1^{(\mu_1)}:[2^{n\tilde{R}_1}]\rightarrow [2^{nR_1}]$ and $b_2^{(\mu_2)}:[2^{n\tilde{R}_2}]\rightarrow [2^{nR_2}]$ performing standard 
information-theoretic binning, with $0 < R_2 \leq \tilde{R_2}$ and $0 < R_2 \leq \tilde{R_2}$. Using these maps, we induce the PMF $P_{M_{1}|X_{1}^{n}}^{(\mu_1)}$ on the message to be transmitted by the first encoder as
% $P_{M_{1}|X_{1}^{n}}^{(\mu_1)}$ and $P_{M_{2}|X_{2}^{n}}^{(\mu_2)}$ as
\begin{eqnarray}\label{Eqn:dist_PMFInducedByEncoder1}
P_{M_{1}|X_{1}^{n}}^{(\mu_1)}(m_{1}|x_{1}^{n}) = 
\begin{cases}
% 1& \mbox{if } m_{1}=0 \mbox{ and }s_1^{(\mu_1)} > 1,\\
% 0& \mbox{if } m_{1}\neq 0 \mbox{ and }s_{1}^{(\mu_1)}(x_{1}^{n}) > 1,\\
\mathbbm{1}_{\{m_1=0\}}& \mbox{if } s_{1}^{(\mu_1)}(x_{1}^{n}) > 1,\\
1-s_{1}^{(\mu_1)}(x_{1}^{n})&\mbox{if }m_{1}=0 \mbox{ and 
}s_{1}^{(\mu_1)}(x_{1}^{n}) \in [0, 1],\\
\sum_{l_1=1}^{2^{n\tilde{R}_1}}E^{(\mu_1)}_{L_1|X_1^n}(l_1|x_1^n)\mathbbm{1}_{\{b_1^{(\mu_1)}(l_1) = m_1\}} &\mbox{if }m_{1}\neq0 \mbox{ and 
}s_{1}^{(\mu_1)}(x_{1}^{n}) \in [0, 1]
\end{cases}
\end{eqnarray}
for all $x_1^n \in T_{\delta}(X_1)$ and $s_1^{(\mu_1)}(x_{1}^{n})$ defined as $s_1^{(\mu_1)}(x_{1}^{n}) = \sum_{l_1=1}^{2^{n\tilde{R}_1}}E^{(\mu_1)}_{L_1|X_1^n}(l_1|x_1^n)$. For $x_1^n \notin T_{\delta}(X_1)$, we let $ P_{M_{1}|X_{1}^{n}}^{(\mu_1)}(m_{1}|x_{1}^{n}) = \mathbbm{1}_{\{m_1=0\}}.$

We similarly define the PMF $P_{M_{2}|X_{2}^{n}}^{(\mu_2)}$ for the second encoder as
\begin{eqnarray}\label{Eqn:dist_PMFInducedByEncoder2}
P_{M_{2}|X_{2}^{n}}^{(\mu_2)}(m_{2}|x_{2}^{n}) = 
\begin{cases}
% 1& \mbox{if } m_{1}=0 \mbox{ and }s_1^{(\mu_1)} > 1,\\
% 0& \mbox{if } m_{1}\neq 0 \mbox{ and }s_{1}^{(\mu_1)}(x_{1}^{n}) > 1,\\
\mathbbm{1}_{\{m_2=0\}}& \mbox{if } s_{2}^{(\mu_2)}(x_{2}^{n}) > 1,\\
1-s_{2}^{(\mu_2)}(x_{2}^{n})&\mbox{if }m_{2}=0 \mbox{ and 
}s_{2}^{(\mu_2)}(x_{2}^{n}) \in [0, 1],\\
\sum_{l_2=1}^{2^{n\tilde{R}_2}}E_{L_2|X_2^n}^{(\mu_2)}(l_2|x_2^n)\mathbbm{1}_{\{b_2(l_2) = m_2\}} &\mbox{if }m_{2}\neq0 \mbox{ and 
}s_{2}^{(\mu_2)}(x_{2}^{n}) \in [0, 1]
\end{cases}
\end{eqnarray}
for all $x_2^n \in T_{\delta}(X_2)$ and $s^{(\mu_2)}_2(x_{2}^{n})$ defined as $s^{(\mu_2)}_2(x_{2}^{n}) = \sum_{l_2=1}^{2^{n\tilde{R}_2}}E^{(\mu_1)}_{L_2|X_2^n}(l_2|x_2^n)$. For $x_2^n \notin T_{\delta}(X_2)$, we let $ P_{M_{2}|X_{2}^{n}}^{(\mu_2)}(m_{2}|x_{2}^{n}) = \mathbbm{1}_{\{m_2=0\}}$.

With this definition note that, $\displaystyle \sum_{m_1=0}^{2^{nR_1}}P_{M_1|X_1^n}^{(\mu_1)}(m_1|x_1^n) = 1$ for all $\mu_1 \in [2^{nC_1}]$ and $x^n_1 \in \mathcal{X}_1^n$ and similarly, $\displaystyle \sum_{m_2=0}^{2^{nR_2}}P_{M_2|X_2^n}^{(\mu_2)}(m_2|x_2^n) = 1$ for all $\mu_2 \in [2^{nC_2}]$ and $ x^n_2 \in \mathcal{X}_2^n$.

\subsection{Decoder Mapping}
We now describe the decoder. On observing $\mu$ and the 
indices $m_{1},m_{2}\in [2^{nR_1}]\times [2^{nR_2}]$ communicated by the encoders, the decoder first deduces $(\mu_1,\mu_2)$ from $\mu$ and then populates 
\begin{align}
    \mathcal{D}^{(\mu_1,\mu_2)}(m_1,m_2)=\left\{
    \begin{array}{cc}
        (l_1,l_2) \in [2^{n\tilde{R}_1}]\times[2^{n\tilde{R}_2}]:  b_1^{(\mu_1)}(l_1)=m_1, b_2^{(\mu_2)}(l_2)=m_2, &  \\
        (\mathtt{w}_1^{n}(l_1,\mu_1),\mathtt{w}_2^{n}(l_2,\mu_2)) \in 
T_{\delta}(W_1,W_2) & 
    \end{array}   \right\}.
\end{align}
% where $\delta_2$ will be specified later in Proposition \ref{prop:Lemma for S_2}.
Let
\begin{eqnarray}
f^{(\mu)}(m_1,m_2) = \begin{cases} (\mathtt{w}_1^{n}(l_1,\mu_1),\mathtt{w}_2^{n}(l_2,\mu_2))&\mbox{if } \mathcal{D}^{(\mu_1,\mu_2)}(m_1,m_2)=\{(l_1,l_2)\}\\(\tilde{w}_1^n,\tilde{w}_2^n)&\mbox{otherwise, i.e., }|\mathcal{D}^{(\mu_1,\mu_2)}(m_1,m_2)| \neq 1\end{cases}.
\nonumber
\end{eqnarray}
The decoder chooses $y^n$ according to PMF $P^n_{Y|W_1 W_2}(y^n|f^{(\mu)}(m_1,m_2))$. This 
implies the PMF $P^{(\mu_1)}_{Y^n|M_1 M_2}(\cdot|\cdot)$ is given by
\begin{eqnarray}
\label{Eqn:dist_PMFInducedByDecoder}
P^{(\mu)}_{Y^n|M_1M_2}(\cdot|m_1,m_2) = P^n_{Y|W_1W_2}(y^{n}|f^{(\mu)}(m_1,m_2)).
\end{eqnarray}

 \subsection{Distribution of Codebooks}
 The PMF defined on the ensemble of codebooks is as specified below. The codewords of the random codebook  $\CalC_{1}^{(\mu_1)} = (\mathtt{W}_{1}(l_1,\mu_1 ) : 1\leq l_1 \leq 2^{n\tilde{R}_{1}})$ for each $\mu_1 \in 2^{nC_1}$ are mutually independent and distributed with PMF 
\begin{eqnarray}
 \mathbb{P}(\mathtt{W}_{1}(l_1,\mu_1 ) = w_{1}^{n}) = \frac{P_{W_{1}}^n(w_{1}^{n})}{(1-\epsilon_1)}\mathds{1}_{\{ w_{1}^{n} \in T_{\delta}^{n}({W_{1}}) \}} \nonumber
\end{eqnarray} 
% where $\epsilon_1 = \sum_{\w_1^n \notin T_{\delta}(W_1)}P^n_{W_1}^{w_1^n}$
 Similarly, $\CalC_{2}^{(\mu_2)} = (\mathtt{W}_{2}(l_2,\mu_2 ) : 1\leq l_2 \leq 2^{n\tilde{R}_{2}})$ for each $\mu_2 \in [2^{nC_2}]$ are mutually independent and distributed with PMF 
\begin{eqnarray}
 \mathbb{P}(\mathtt{W}_{2}(l_2,\mu_2 ) = w_{2}^{n}) = \frac{P_{W_{2}}^n(w_{2}^{n})}{(1-\epsilon_2)}\mathds{1}_{\{ w_{2}^{n} \in T_{\delta}^{n}({W_{2}}) \}} \nonumber
\end{eqnarray}
where, recall  $\epsilon_i = 1-P_W^n(T_{\delta}(W_i)); i=1,2.$
Finally, the binning functions $b_1^{(\mu_1)}(\cdot) $ and $b_2^{(\mu_2)}(\cdot)$ are chosen random, uniformly and independently from the sets $[2^{nR_1}]$ and $[2^{nR_2}], $ respectively.

We now begin our analysis of (\ref{eq:dist_main_lemma}). Our goal is to prove the existence of a collections $\mathcal{C}_{1},\mathcal{C}_{2}$ for which (\ref{eq:dist_main_lemma}) holds. We do this via random coding. 
 Specifically, we prove that $\Expectation{\mathscr{Q}} \leq \epsilon$ where the expectation is over the ensemble of codebooks.
\subsection{Analysis of Total Variation}
We begin by splitting $\mathscr{Q}$ into two terms using an indicator function $\mathds{1}_{\{ \mathtt{PMF}(\CalC_{1},\CalC_{2})\}}$ as
\begin{eqnarray}
 \label{eq:indicator}
    \Expectation{\mathscr{Q}} = \Expectation{\big[\mathscr{Q}\cdot \mathds{1}_{\{ \mathtt{PMF}(\CalC_{1},\CalC_{2})\}}\big]} + \Expectation{\big[\mathscr{Q}\cdot \mathds{1}^c_{\{ \mathtt{PMF}(\CalC_{1},\CalC_{2})\}}\big]} \leq 
    \EE{\big[\mathscr{Q}\mathds{1}_{\{ \mathtt{PMF}(\CalC_{1},\CalC_{2})\}}\big]} +\PP\left\{\mathds{1}_{\{ \mathtt{PMF}(\CalC_{1},\CalC_{2})\}} = 0 \right\} \
\end{eqnarray}
where $\mathds{1}_{\{ \mathtt{PMF}(\CalC_{1},\CalC_{2})\}}$ is defined as
\begin{eqnarray}
\mathds{1}_{\{ \mathtt{PMF}(\CalC_{1},\CalC_{2})\}} = \begin{cases} 1 &\mbox{ if }  \displaystyle  
s_1^{(\mu_1)}(x_1^n) \in 
[0,1] \mbox{ and } \displaystyle  
s_2^{(\mu_2)}(x_2^n)\in 
[0,1] \\ & \text{ for all } x_1^n \in T_{\delta}(X_1), x_2^n \in T_{\delta}(X_2), \mu_1 \in [2^{nC_1}], \mu_2 \in [2^{nC_2}]\\ 0 &\mbox{ otherwise},
\end{cases}\nonumber\end{eqnarray} 
and (\ref{eq:indicator}) follows from the upper bound of 1 over the total variation.
% Taking expectation over the codebooks and bounding the $\mathscr{Q}$ in the second term of the right hand side\footnote{Total Variation is bounded from above by 2} by 2  gives
% \begin{align}
%     \EE{[\mathscr{Q}]} \leq \EE{[\mathscr{Q}\mathds{1}_{\{ \mathtt{PMF}(\CalC_{1},\CalC_{2})\}}]} +2\cdot\PP\left\{\mathds{1}_{\{ \mathtt{PMF}(\CalC_{1},\CalC_{2})\}} = 0 \right\} \nonumber
% \end{align}
% where the event $(C_{1},C_{2}) \nsim \mathtt{PMF} $ is defined as the complement of the event $\left\{s_1^{(\mu_1)}(x_1^n) \in 
% [0,1] \mbox{ and } s_2(x_2^n)\in 
% [0,1] \text{ for all } x_1^n \in T_{\delta}(X_1), x_2^n \in T_{\delta}(X_2), \mu \in [2^{nC}]\right\}$, for a random collection $(c_{1} ,c_{2})$. 
We now show using the lemma below, that by appropriately constraining $\tilde{R}_1$ and $\tilde{R}_2$,  $\PP\left\{\mathds{1}_{\{ \mathtt{PMF}(\CalC_{1},\CalC_{2})\}} = 0 \right\}$ can be made arbitrarily small. In other words, with high probability, we will have  $E_{L_1|X_1^n}^{(\mu_1)}$ and $E_{L_2|X_2^n}^{\mu_2}$ such that $\displaystyle 0\leq \sum_{l_1=1}^{2^{n\tilde{R_1}}}E_{L_1|X_1^n}^{(\mu_1)} \leq 1 $ for all $\mu_1 \in [2^{nC_1}]$ and $x_1^n \in T_{\delta}(X_1)$ , and $\displaystyle 0\leq \sum_{l_2=1}^{2^{n\tilde{R}_2}}E^{(\mu_2)}_{L_2|X_2^n} \leq 1 $ for all $\mu_2 \in [2^{nC_2}]$ and $x_2^n \in T_{\delta}(X_2)$.

\begin{prop}\label{lem:2PMFWHP}
For any $\delta,\eta \in (0,1/2),$ if $\tilde{R}_1 > I(X_1:W_1)+4\delta_1$ and $\tilde{R}_2 > I(X_2:W_2)+4\delta_2$, where $\delta_1(\delta),\delta_2(\delta) \searrow 0$ as $\delta \searrow 0,$   then
\begin{align}
\PP&\left[\left(\bigcap_{\mu=1}^{2^{nC_1}}\bigcap_{x^n\in T_{\delta}(X_1)}\left(\sum_{l_1 =1}^{2^{n\tilde{R}_1}}E^{(\mu_1)}_{L_1|X_1^n}(l_1|x_1^n) \leq 1\right)\right)\bigcap\left(\bigcap_{\mu_2=1}^{2^{nC_2}}\bigcap_{x_2^n\in T_{\delta}(X_2)}\left(\sum_{l_2 =1}^{2^{n\tilde{R}_2}} E_{L_2|X_2^n}^{(\mu_2)}(l_2|x_2^n)  \leq 1\right)\right)\right]  \nonumber \\
& \hspace{1cm}\rightarrow 1 \text{ as } n \rightarrow \infty.
\end{align}
\end{prop}
\begin{proof}
The proof follows from Lemma \ref{lem:Lemma_E_PMF}.
% Using the lemma () from Appendix $(\ref{AppSec:ProofOfLemma})$ twice, we get the following
% \begin{align}
%     \PP\left[\bigcap_{\mu_1=1}^{2^{nC_1}}\bigcap_{x_1^n\in T_{\delta(X_1)}}\left(\sum_{l_1 =1}^{2^{n\tilde{R}_1}}  E^{(\mu_1)}_{L_1|X_1^n}(l_1|x_1^n)\right) \leq 1\right] &\geq 1 - 2\cdot2^{nC_1}|T_{\delta}(X_1)|\exp{\bigg(-\frac{\eta^2 2^{n(\tilde{R}_1 - I(X_1,W_1)-4\delta_1)} }{4 \ln{2}}\bigg)} \nonumber \\
%     \PP\left[\bigcap_{\mu=2}^{2^{nC_2}}\bigcap_{x_2^n\in T_{\delta(X_2)}}\left(\sum_{l_2 =1}^{2^{n\tilde{R}_2}}  E^{(\mu_1)}_{L_2|X_2^n}(l_2|x_2^n)\right) \leq 1\right] &\geq 1 - 2\cdot2^{nC_2}|T_{\delta}(X_2)|\exp{\bigg(-\frac{\eta^2 2^{n(\tilde{R}_2 - I(X_2,W_2)-4\delta_2)} } {4 \ln{2}}\bigg)} \nonumber
% \end{align}
% Applying union bound to the above inequalities gives
% \begin{align}
%     &\PP\left[\left(\bigcap_{\mu=1}^{2^{nC_1}}\bigcap_{x^n\in T_{\delta}(X_1)}\left(\sum_{l_1 =1}^{2^{n\tilde{R}_1}}E^{(\mu_1)}_{L_1|X_1^n}(l_1|x_1^n) \leq 1\right)\right)\bigcap\left(\bigcap_{\mu_2=1}^{2^{nC_2}}\bigcap_{x_2^n\in T_{\delta}(X_2)}\left(\sum_{l_2 =1}^{2^{n\tilde{R}_2}} E_{L_2|X_2^n}^{(\mu_2)}(l_2|x_2^n) \leq 1\right)\right)\right] \nonumber \\
%     & \geq 1- 2\!\cdot\!2^{nC_1}|T_{\delta}(X_1)|\!\exp{\!\bigg(\!\!\!-\frac{\eta^2 2^{n(\tilde{R}_1 - I(X_1,W_1)-4\delta_1)} }{4 \ln{2}}\bigg)} - 2\!\cdot\!2^{nC_2}|T_{\delta}(X_2)|\!\exp{\!\bigg(\!\!\!-\frac{\eta^2 2^{n(\tilde{R}_2 - I(X_2,W_2)-4\delta_2)} } {4 \ln{2}}\bigg)} \nonumber
% \end{align}
% and hence the result follows.
\end{proof}
% Since, we have
% \begin{align}
%     \PP\left\{\mathds{1}_{\mathtt{PMF}(\CalC_{1},\CalC_{2})}=1\right\}  & =  1-  \PP\left[\!\left(\bigcap_{\mu_1=1}^{2^{nC_1}}\bigcap_{\substack{x^n\in\\ T_{\delta}(X_1)}}\!\!\!\!\left(E^{(\mu_1)}_{L_1|X_1^n}(l_1|x_1^n) \leq 1\right)\!\right)\!\bigcap\!\left(\bigcap_{\mu_2=1}^{2^{nC_2}}\bigcap_{\substack{x_2^n\in\\ T_{\delta}(X_2)}}\!\!\!\!\left(E_{L_2|X_2^n}^{(\mu_2)}(l_2|x_2^n) \leq 1\right)\!\right)\!\right] \nonumber 
% \end{align}
% From the above Lemma \ref{lem:2PMFWHP}, for any $\delta \in (0,1)$, we have $\PP\left\{(C_{1},C_{2})\nsim \mathtt{PMF}\right\} \leq \epsilon_p $
%  where $\epsilon_p(\delta) \searrow 0 $ as $\delta \searrow 0.$

\noindent We now look at the first term in (\ref{eq:indicator}), i.e., $\mathscr{Q}\cdot \mathds{1}_{\{ \mathtt{PMF}(\CalC_{1},\CalC_{2})\}}$. This can be expanded as
\begin{align}\label{eq:expandQtypX}
    \mathscr{Q}\cdot \mathds{1}_{\{ \mathtt{PMF}(\CalC_{1},\CalC_{2})\}}  = \!\left[
     \sum_{\underline{x}^n \in T_{\delta}(\underline{X})}P^n_{\underline{X}}(\underline{x}^n)\mathscr{Q}_{\underline{x}^n}
     + \!\!
     \sum_{\underline{x}^n \notin T_{\delta}(\underline{X})}P^n_{\underline{X}}(\underline{x}^n)\mathscr{Q}_{\underline{x}^n}\right]\cdot \mathds{1}_{\{ \mathtt{PMF}(\CalC_{1},\CalC_{2})\}},
\end{align}
where $T_{\delta}(\underline{X})$ is defined as $T_{\delta}(\underline{X}) \define \{\underline{x}^n: (x_1^n, x_2^n) \in T_{\delta}(X_1,X_2) \}$ and $\mathscr{Q}_{\underline{x}^n}$ is defined as 
\begin{align}
    \mathscr{Q}_{\underline{x}^n} \deq \frac{1}{2}&\sum_{y^n}\left|P^n_{Y|\underline{X}}(y^n|\underline{x}^n) - \!\!\!\!\sum_{\substack{ \mu_1,\mu_2 }}\frac{1}{2^{n{(C_1+C_2)}}}\hspace{-17pt}
\sum_{\substack{m_1 \in[ 2^{nR_1}]\cup \{0\}\\ m_2 \in[2^{nR_2}]\cup \{0\}}} \hspace{-17pt}
P^{(\mu_1)}_{M_1|X_1^{n}}(m_1|x_1^{n})P^{(\mu_2)}_{M_2|X_2^{n}}(m_2|x_2^{n}
)P^{(\mu) } _ { Y^ { n } 
|M_1,M_2}(y^{n}|m_1,m_2)\right|. \nonumber
\end{align}
Since, using the standard typicality arguments one can argue $\sum_{\underline{x}^n \notin T_{\delta}(\underline{X})}P^n_{\underline{X}}(\underline{x}^n) \leq \epsilon_t$, where $\epsilon_t(\delta) \searrow 0$ as $\delta \searrow 0$. We bound $\mathscr{Q}_{\underline{x}^n}$ within the second summation in the right hand side of the above equation\footnote{Note that $\mathscr{Q}_{x_1^n,x_2^n}$ is a total variational distance between two conditional PMFs, conditioned on $(X_1,X_2)$, for each ${(x_1^n,x_2^n)}$ and hence it is bounded from above by one.} to obtain,
\begin{align}
    \mathscr{Q}\cdot \mathds{1}_{\{ \mathtt{PMF}(\CalC_{1},\CalC_{2})\}}  = \!
     \sum_{\underline{x}^n \in T_{\delta}(\underline{X})}P^n_{\underline{X}}(\underline{x}^n)\mathscr{Q}_{\underline{x}^n}
   \mathds{1}_{\{ \mathtt{PMF}(\CalC_{1},\CalC_{2})\}} + \;\epsilon_t(\delta)\label{eq:Dist_Q_typ}.
\end{align}
Now, what remains is the first term in (\ref{eq:Dist_Q_typ}). A major portion of our analysis from here on deals with arguing that this term can be made arbitrarily small. Further, since this term contains the indicator $ \mathds{1}_{\{ \mathtt{PMF}(\CalC_{1},\CalC_{2})\}}$, we can restrict our analysis to only the set of random codebooks $(\CalC_{1},\CalC_{2})$ that satisfy $0\leq \sum_{l_1=1}^{2^{n\tilde{R}_1}}E^{(\mu_1)}_{L_1|X_1^n}(l_1|x_1^n) \leq 1$ and  $0\leq \sum_{l_2=1}^{2^{n\tilde{R}_2}}E_{L_2|X_2^n}^{(\mu_2)}(l_2|x_2^n) \leq 1$ for all $\underline{x}^n \in T_{\delta}(\underline{X})$ and $\mu_1 \in [2^{nC_1}], \mu_2 \in [2^{nC_2}]$.

\noindent \textbf{Step 1: Isolating the error induced by not covering}\\As a first step, we separate the error induced by not covering the product distribution $P_{X_1X_2Y}^n(\cdot)$ through the randomized encoders and provide a bound to it. Note that under the condition that $\mathds{1}_{\{\mathtt{PMF}(\mathcal{C}_1,\mathcal{C}_2)\}} = 1,$ we have $  P^{(\mu_i)}_{M_i|X_i^n}(m_i|x_i^n) = \sum_{l_i =1}^{2^{n\tilde{R}_i}}E^{(\mu_i)}_{L_i|X^n_i}(l_i|x_i^n)$ when $m_i \neq 0$, and $ P^{(\mu_i)}_{M_i|X_i^n}(0|x_i^n) = 1-\sum_{l_i =1}^{2^{n\tilde{R}_i}}E^{(\mu_i)}_{L_i|X_i^n}(l_i|x_i^n)$, for $i \in \{1,2\}$. Using this, we substitute the definition of the randomized encoders (\ref{Eqn:dist_PMFInducedByEncoder1}), (\ref{Eqn:dist_PMFInducedByEncoder2}) and the decoder (\ref{Eqn:dist_PMFInducedByDecoder}) in the second 
term within the modulus of $\mathscr{Q}_{\underline{x}^n}$ for $\underline{x}^n \in T_{\delta}(\underline{X}),$ which gives,
\begin{eqnarray}
\sum_{\substack{ \mu_1 
\in [2^{nC_1}],  \mu_2 \in [2^{nC_2}] \\m_1 \in[ 2^{nR_1}]\cup \{0\}\\m_2 \in [2^{nR_2}]\cup\{0\} }} 
\frac{P^{(\mu_1)}_{M_1|X_1^{n}}\!(m_1|x^{n}_1)P^{(\mu_2)}_{M_2|X_2^{n}}\!(m_2|x^{n}_2 )P^{(\mu) } _ { Y^ { n } |M_1 M_2}(y^{n}|m_1,m_2)}{2^{n(C_1+C_2)}} =\!T_{1}\!+\!T_{2}+T_3+T_4,\!
 \nonumber
\end{eqnarray}
where\footnote{For the ease of notation, we do not show the dependency of $T_1, T_2, T_3$ and $T_4$ on $\underline{x}^n$, however, in principle they depend on $\underline{x}^n$ and in fact, are only defined for $\underline{x}^n \in T_{\delta}(\underline{X})$  },
%\vspace{-10pt}
\begin{align} 
T_{1} &\deq \!\!\sum_{\substack{ \mu_1,\mu_2 \\m_1 \in[ 2^{nR_1}]\\ m_2 \in[ 2^{nR_2}]  }}
\!\sum_{l_1,l_2 } \sum_{\substack{w_1^n \in \\T_{\delta}(W_1|x_1^n)}}\sum_{\substack{w_2^n \in \\T_{\delta}(W_2|x_2^n)}}  
\!\!\!\!\!\frac{(1-\epsilon_1)(1-\epsilon_2)
P^n_{X_1|W_1}(x_1^n|w_1^n) P^n_{X_2|W_2}(x_2^n|w_2^n) }{2^{n(\tilde{R}_1+\tilde{R}_2+C)}(1+\eta)^2 P^n_{X_1}(x_1^n) P^n_{X_2}(x_2^n) } \nonumber\\ 
%%%%%%%%%%%%%%%%%% new sub line %%%%%%%%%%%%%%%%%%
&\hspace{20pt}\mathds{1}_{\left\{ \substack{ w_1^n = \mathtt{w}_1^n(l_1,\mu_1), b_1^{(\mu_1)}(l_1)=m_1}\right\}}\mathds{1}_{\left\{ \substack{ w_2^n = \mathtt{w}_2^n(l_2,\mu_2), b_2(l_2)=m_2 }\right\}} 
P^n_{Y|W_1 W_2}(y^{n}| f^{(\mu)}(b_1^{(\mu_1)}(l_1), b_2^{(\mu_2)}(l_2) ) )\nonumber \\&
%%%%%%%%%%%%%%%%%% new line %%%%%%%%%%%%%%%%%%
= \sum_{\substack{ \mu_1,\mu_2  }}\sum_{l_1,l_2 
} \sum_{\substack{w_1^n \in \\T_{\delta}(W_1|x_1^n)}}\sum_{\substack{w_2^n \in \\T_{\delta}(W_2|x_2^n)}} 
\frac{(1-\epsilon_1)(1-\epsilon_2)
P^n_{X_1|W_1}(x_1^n|w_1^n) P^n_{X_2|W_2}(x_2^n|w_2^n) }{2^{n(\tilde{R}_1+\tilde{R}_2+C_1+C_2)}(1+\eta)^2 P^n_{X_1}(x_1^n) P^n_{X_2}(x_2^n) } \nonumber \\ &
%%%%%%%%%%%%%%%%%% new sub line %%%%%%%%%%%%%%%%%%
\hspace{60pt}\mathds{1}_{\left\{w_1^n = \mathtt{w}_1^n(l_1,\mu_1)\right\}}\mathds{1}_{\left\{w_2^n = \mathtt{w}_2^n(l_2,\mu_2)\right\}}P^n_{Y|W_1W_2}(y^{n}|f^{(\mu)}(b_1^{(\mu_1)}(l_1),b_2^{(\mu_2)}(l_2) 
)), \nonumber
\end{align}
\begin{align}
T_{2}&\deq \!\sum_{\substack{ \mu_1,\mu_2,l_2 }}\!\sum_{\substack{w_2^n \in \\T_{\delta}(W_2|x_2^n)}}\!\!\!\!\!\frac{ \left[1-\sum_{l_1=1}^{2^{n\tilde{R}_1}}E^{(\mu_1)}_{L_1|X_1^{n}}(l_1|x_1^{n}
)\right]}{2^{n(C_1+C_2)}}\frac{(1-\epsilon_1) 
P^n_{X_2|W_2}(x_2^n|w_2^n)  }{2^{n\tilde{R}_2}(1+\eta)P^n_{X_2}(x_2^n)}\mathds{1}_{\left\{  w_2^n = \mathtt{w}_2^n(l_2,\mu_2) 
\right\}}P^n_{Y|W_1W_2}(y^{n}|\tilde{w}^n_1,\tilde{w}^n_2), \nonumber
\end{align}
\begin{align}
T_{3}&\deq \!\sum_{\substack{ \mu_1,\mu_2,l_1 }}\!\sum_{\substack{w_1^n \in \\T_{\delta}(W_1|x_1^n)}}\!\!\!\!\!\frac{ \left[1-\sum_{l_2=1}^{2^{n\tilde{R}_2}}E^{(\mu_2)}_{L_2|X_2^{n}}(l_2|x_2^{n}
)\right]}{2^{n(C_1+C_2)}}\frac{(1-\epsilon_2) 
P^n_{X_1|W_1}(x_1^n|w_1^n)  }{2^{n\tilde{R}_1}(1+\eta)P^n_{X_1}(x_1^n)}\mathds{1}_{\left\{  w_1^n = \mathtt{w}_1^n(l_1,\mu_1)\right\}}P^n_{Y|W_1W_2}(y^{n}|\tilde{w}^n_1,\tilde{w}^n_2),\nonumber
\end{align}
\begin{align}
T_{4}&\deq \sum_{\substack{ \mu_1,\mu_2}}\frac{\left[1-\sum_{l_1=1}^{2^{n\tilde{R}_1}}E^{(\mu_1)}_{L_1|X_1^{n}}(l_1|x_1^{n}
)\right]\left[1-\sum_{l_2=1}^{2^{n\tilde{R}_2}}E^{(\mu_2)}_{L_2|X_2^{n}}(l_2|x_2^{n}
)\right]}{2^{n(C_1+C_2)}}P^n_{Y|W_1W_2}(y^{n}|\tilde{w}^n_1,\tilde{w}^n_2)\nonumber
\end{align}
% As pointed earlier, the indicator function $\mathbbm{1}_{\left\{\mathtt{PMF}(\CalC_{1},\CalC_{2})\right\}}$ restricts
The above simplification in the expression for $T_1$ is obtained by using $\sum_{m_1\in[2^{nR_1}]}\mathds{1}_{\left\{  w_1^n = \mathtt{w}_1^n(l_1,\mu_1)\right\}} = 1$ and $\sum_{m_2\in[2^{nR_2}]}\mathds{1}_{\left\{  w_2^n = \mathtt{w}_2^n(l_2,\mu_2)\right\}} = 1$, which follows from the definition of the maps $b_1^{(\mu_1)}$ and $b_2^{(\mu_2)}$.
A similar simplification for the expressions $T_2$ and $T_3$ is used while substituting $P_{M_1|X_1^n}^{(\mu_1)}(0|x_1^n) = 1-\sum_{l_1=1}^{2^{n\tilde{R}_1}}E^{(\mu_1)}_{L_1|X_1^n}(l_1|x_1^n)$ and $P_{M_2|X_2^n}^{(\mu_2)}(0|x_2^n) = 1-\sum_{l_2=1}^{2^{n\tilde{R}_2}}E^{(\mu_2)}_{L_2|X_2^n}(l_2|x_2^n)$, respectively. Finally, $T_4$ uses the substitution for both $P_{M_1|X_1^n}^{(\mu_1)}(0|x_1^n)$ and $P_{M_2|X_2^n}^{(\mu_2)}(0|x_2^n)$.
Substituting $T_{1},T_{2},T_3$ and $T_4$ for the second term within the modulus of 
(\ref{eq:dist_main_lemma}), we obtain  $\mathscr{Q}_{\underline{x}^n}\11_{\{\mathtt{PMF}(\CalC_1,\CalC_2)\}} \leq  
\frac{1}{2}\sum_{y^{n}}\big( S+\tilde{S}\big)\11_{\{\mathtt{PMF}(\CalC_1,\CalC_2)\}} \leq \frac{1}{2}\sum_{y^{n}}\big( S+\tilde{S}\11_{\{\mathtt{PMF}(\CalC_1,\CalC_2)\}}\big),$ where
\begin{align*}
    S & \deq \Bigg| P_{Y|X_1X_2}^n(y^n|x_1^n,x_2^n) -  \sum_{\substack{ \mu_1,\mu_2 }}\sum_{l_1,l_2} \sum_{\substack{w_1^n \in \\T_{\delta}(W_1|x_1^n)}}\sum_{\substack{w_2^n \in \\T_{\delta}(W_2|x_2^n)}} 
\!\!\!\!\!\!\frac{(1-\epsilon_1)(1-\epsilon_2)
P^n_{X_1|W_1}(x_1^n|w_1^n) P^n_{X_2|W_2}(x_2^n|w_2^n) }{2^{n(\tilde{R}_1+\tilde{R}_2+C_1+C_2)}(1+\eta)^2 P^n_{X_1}(x_1^n) P^n_{X_2}(x_2^n) }\nonumber \\ 
& \hspace{2in}
 \mathds{1}_{\left\{ w_1^n = \mathtt{w}_1^n(l_1,\mu_1) \right\}} 
\mathds{1}_{\left\{ w_2^n = \mathtt{w}_2^n(l_2,\mu_2) \right\}} P^n_{Y|W_1W_2}(y^{n}|f^{(\mu)}(b^{(\mu_1)}_1(l_1),b_2^{(\mu_2)}(l_2))) \Bigg|,
\end{align*}
and $\tilde{S} \deq |T_2| + |T_3| + |T_4|.$ Note that the term corresponding to $\tilde{S}$ captures the error induced by not covering the product distribution $P_{X_1X_2Y}^n(\cdot)$ and we bound this term employing the following proposition.
\begin{prop}\label{prop:S_tilde}
    There exist functions  $\epsilon_{\widetilde{S}}(\delta), $ and $\delta_{\widetilde{S}}(\delta)$, 
    such that for  all sufficiently small $\delta$ and sufficiently large $n$, we have $\EE\left[\frac{1}{2}\sum_{\underline{x}^n \in T_{\delta}(\underline{X})}\sum_{y^{n}}P_{\ulineX}^n(\ulinex^n)\widetilde{S}\11_{\{\mathtt{PMF}(\CalC_1,\CalC_2)\}}\right] \leq\epsilon_{\widetilde{S}}(\delta) $, if  $\tilde{R}_1 >  I(X_1;W_1) + \delta_{\widetilde{S}} $ and $\tilde{R}_2 > I(X_2;W_2) + \delta_{\widetilde{S}},$ where $\epsilon_{\widetilde{S}}, \delta_{\widetilde{S}} \searrow 0$ as $\delta \searrow 0$. 
\end{prop}
\begin{proof}
    The proof is provided in Appendix \ref{appx:prop:S_tilde}.
\end{proof}
Now we move on to isolating the error component of $S$ caused by binning the randomized encoders.
\noindent \textbf{Step 2: Error caused by binning}\\
By adding and subtracting
an appropriate term within the modulus of $S$ and using triangle inequality, $S$ can be bounded as $S \leq S_1 + S_2$, where
% \begin{align}
%     \sum_{\substack{ \mu_1,\mu_2  }}&\sum_{l_1,l_2 
% } \sum_{\substack{w_1^n \in \\T_{\delta}(W_1|x_1^n)}}\sum_{\substack{w_2^n \in \\T_{\delta}(W_2|x_2^n)}} 
% \!\!\!\!\frac{(1-\epsilon_1)(1-\epsilon_2)
% P^n_{X_1|W_1}(x_1^n|w_1^n) P^n_{X_2|W_2}(x_2^n|w_2^n) }{2^{n(\tilde{R}_1+\tilde{R}_2+C_1+C_2)}(1+\eta)^2 P^n_{X_1}(x_1^n) P^n_{X_2}(x_2^n) }\mathds{1}_{\left\{ \!\!\!\!\begin{array}{c} w_1^n = \mathtt{w}_1^n(l_1,\mu_1) 
% \end{array}\!\!\!\!\right\}} \nonumber \\ &
% \hspace{100pt}\mathds{1}_{\left\{ \!\!\!\!\begin{array}{c} w_2^n = \mathtt{w}_2^n(l_2,\mu_2) 
% \end{array}\!\!\!\!\right\}}P^n_{Y|W_1W_2}(y^{n}|w_1^n,w^n_2),\nonumber 
% \end{align}
% and using triangle inequality, we obtain $\mathscr{Q}_{\underline{x}^n} \leq \displaystyle 
% \frac{1}{2}\sum_{y^{n}}P_{\ulineX}^n(\ulinex^n)\left[ S_{11}+S_{12}+S_{2}+S_3+S_4\right]$, where
\begin{align}
    S_{1} \deq \Bigg| &P_{Y|X_1X_2}^n(y^n|x_1^n,x_2^n) - \!
    \sum_{\substack{ \mu_1,\mu_2 }}\sum_{l_1,l_2 
}\!\sum_{\substack{w_1^n \in \\T_{\delta}(W_1|x_1^n)}}\sum_{\substack{w_2^n \in \\T_{\delta}(W_2|x_2^n)}} 
\!\!\!\!\!\!\!\frac{(1-\epsilon_1)(1-\epsilon_2)
P^n_{X_1|W_1}(x_1^n|w_1^n) P^n_{X_2|W_2}(x_2^n|w_2^n) }{2^{n(\tilde{R}_1+\tilde{R}_2+C_1+C_2)}(1+\eta)^2 P^n_{X_1}(x_1^n) P^n_{X_2}(x_2^n) } \nonumber \\ &
\hspace{2.6in}\mathds{1}_{\left\{ w_1^n = \mathtt{w}_1^n(l_1,\mu_1)
\right\}}\mathds{1}_{\left\{  w_2^n = \mathtt{w}_2^n(l_2,\mu_2) 
\right\}}P^n_{Y|W_1W_2}(y^{n}|w_1^n,w^n_2)
))  \Bigg| \nonumber
\end{align}
\begin{align}
    S_{2} \deq &\sum_{\substack{ \mu_1,\mu_2 }}\sum_{l_1,l_2} \sum_{\substack{w_1^n \in \\T_{\delta}(W_1|x_1^n)}}\sum_{\substack{w_2^n \in \\T_{\delta}(W_2|x_2^n)}} 
\!\!\!\!\!\!\frac{(1-\epsilon_1)(1-\epsilon_2)
P^n_{X_1|W_1}(x_1^n|w_1^n) P^n_{X_2|W_2}(x_2^n|w_2^n) }{2^{n(\tilde{R}_1+\tilde{R}_2+C_1+C_2)}(1+\eta)^2 P^n_{X_1}(x_1^n) P^n_{X_2}(x_2^n) }\nonumber \\
& \hspace{0.2in} \mathds{1}_{\left\{ w_1^n = \mathtt{w}_1^n(l_1,\mu_1) \right\}} 
\mathds{1}_{\left\{ w_2^n = \mathtt{w}_2^n(l_2,\mu_2) \right\}}
\Bigg|P^n_{Y|W_1W_2}(y^{n}|w_1^n,w^n_2) - P^n_{Y|W_1W_2}(y^{n}|f^{(\mu)}(b^{(\mu_1)}_1(l_1),b_2^{(\mu_2)}(l_2))) \Bigg|.
\nonumber 
\end{align}
Note that the term $S_2$ captures the error introduced due to the binning operation. To bound this term, we provide the following proposition.
\begin{prop}[Mutual Packing]\label{prop:Lemma for S_2}
There exist  $\epsilon_{S_{2}}(\delta),$ such that for  all sufficiently small $\delta$ and sufficiently large $n$, we have $\EE\left[\sum_{\ulinex^n \in T_\delta(\ulineX)}P^n_{\ulineX}(\ulinex^n){S}_2\right] \leq \epsilon_{{S_2}}(\delta) $, if  $(\tilde{R}_1- R_1) + (\tilde{R}_2  -R_2) \leq  I(W_1;W_2)+\delta_{S_2}$,  where  $\epsilon_{{S}_2}, \delta_{S_2}(\delta) \searrow 0$ as $\delta \searrow 0$.
\end{prop}
\begin{proof}
The proof is provided in Appendix  \ref{appx:prop:Lemma for S_2}.
\end{proof}
Now, we are left with the analysis of the term $S_1$. For this, we segregate the effect of two encoders within the term $S_1$,  and separately analyze each of them, starting with the Alice's encoder.
\begin{figure}
 \centering
\includegraphics[width=4in]{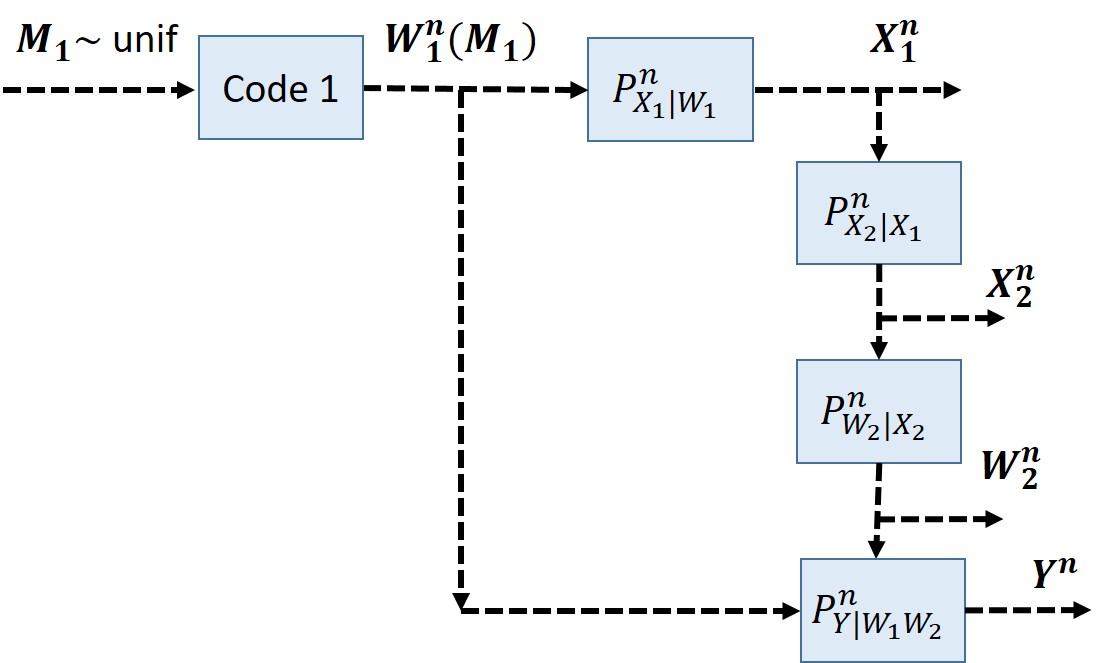}%\vspace{-0.22in}
\caption{Depiction of the approximation performed by Alice (encoder 1) while assuming a product distribution on Bob's (encoder 2) end.}%\vspace{-0.25in}
\label{Fig:AliceEncoding}
\end{figure} 

\noindent\textbf{Step 3: Term concerning Alice's encoding} \\
For notational convenience, we first define $E^{(\mu_1)}_{W_1^n|X_1^n} $ and $E^{(\mu_2)}_{W_2^n|X_2^n}, $ as 
\begin{align}
    E^{(\mu_1)}_{W_1^n|X_1^n}(w_1^n|x_1^n) & \deq \frac{1}{2^{n\tilde{R}_1}}\frac{(1-\epsilon_1)}{(1+\eta)}\sum_{l_1=1}^{2^{n\tilde{R}_1}}\frac{P_{X_1|W_1}^n(x_1^n|w_1^n)}{P_{X_1^n}^n(x_1^n)}   \mathds{1}_{\left\{ w_1^n = \mathtt{w}_1^n(l_1,\mu_1),w_1^n \in T_{\delta}(W_1|x_1^n)\right\}},
    \nonumber \\
    E^{(\mu_2)}_{W_2^n|X_2^n}(w^n_2|x_2^n) & \deq \frac{1}{2^{n\tilde{R}_2}}\frac{(1-\epsilon_2)}{(1+\eta)}\sum_{l_2=1}^{2^{n\tilde{R}_2}}\frac{P_{X_2|W_2}^n(x_2^n|w_2^n)}{P_{X_2^n}^n(x_2^n)}   \mathds{1}_{\left\{ w_2^n = \mathtt{w}_2^n(l_2,\mu_2),w_2^n \in T_{\delta}(W_2|x_2^n)\right\}}.  \nonumber 
\end{align}
Note that when $\PMFIndicator=1$, we also have $0 \leq \sum_{w_1^n \in \mathcal{W}_1^n}E^{(\mu_1)}_{W_1^n|X_1^n}(w_1^n|x_1^n) \leq 1 $ and $0\leq \sum_{w_2^n \in \mathcal{W}_2^n}E^{(\mu_2)}_{W_2^n|X_2^n}(w^n_2|x_2^n) \leq 1$. 
This simplifies $S_{1}$ as
\begin{align}
    S_{1} = \Bigg| P_{Y|X_1X_2}^n&(y^n|x_1^n,x_2^n) - \frac{1}{2^{n(C_1+C_2)}}\sum_{\substack{ \mu_1,\mu_2  }} \sum_{\substack{w_1^n,w_2^n }} 
\!\!
E^{(\mu_1)}_{W_1^n|X_1^n}(w_1^n|x^n_1)E^{(\mu_1)}_{W_2^n|X_2^n}(w_2^n|x^n_2)P^n_{Y|W_1W_2}(y^{n}|w_1^n,w^n_2)
))  \Bigg|. \nonumber
\end{align}
Now we add and subtract a term that separates the action of first encoder from that of second encoder allowing us to separately bound the error introduced by each of these encoders. This term essentially assumes that the second encoder is simply a conditional product PMF $P^n_{W_2|X_2}$ as opposed to the n-letter PMF, while keeping the first encoder the same. Figure \ref{Fig:AliceEncoding} illustrates the dynamics of this term.  The term is  given as
\begin{align*}
    \frac{1}{2^{nC_1}}\sum_{\substack{ \mu_1 \in [2^{nC_1}]  }} \sum_{\substack{w_1^n,w_2^n }} 
\!\! E^{(\mu_1)}_{W_1^n|X_1^n}(w_1^n|x^n_1)P^n_{W_2|X_2}(w_2^n|x^n_2)P^n_{Y|W_1W_2}(y^{n}|w_1^n,w^n_2).
\end{align*}
By adding and subtracting this term and using triangle inequality we obtain $S_{1} \leq Q_1 + Q_2,$ where
\begin{align}
Q_1 \deq     &\Bigg| P_{Y|X_1X_2}^n(y^n|x_1^n,x_2^n) -  \frac{1}{2^{nC_1}}\!\sum_{\substack{ \mu_1  }} \sum_{\substack{w_1^n,w_2^n }} 
\!\! E^{(\mu_1)}_{W_1^n|X_1^n}(w_1^n|x^n_1)P^n_{W_2|X_2}(w_2^n|x^n_2)P^n_{Y|W_1W_2}(y^{n}|w_1^n,w^n_2)
 \Bigg|, \nonumber \\
Q_2 \deq     &\Bigg|  \frac{1}{2^{nC_1}}\sum_{\substack{ \mu_1 
]  }} \sum_{\substack{w_1^n,w_2^n }} 
\!\!
E^{(\mu_1)}_{W_1^n|X_1^n}(w_1^n|x^n_1)P^n_{W_2|X_2}(w_2^n|x^n_2)P^n_{Y|W_1W_2}(y^{n}|w_1^n,w^n_2)
  \nonumber \\
 & \hspace{0.8in}- \frac{1}{2^{n(C_1+C_2)}}\sum_{\substack{ \mu_1,\mu_2  }} \sum_{\substack{w_1^n,w_2^n }} 
\!\!
E^{(\mu_1)}_{W_1^n|X_1^n}(w_1^n|x^n_1)E^{(\mu_2)}_{W_2^n|X_2^n}(w_2^n|x^n_2)P^n_{Y|W_1W_2}(y^{n}|w_1^n,w^n_2) \Bigg|. \nonumber
\end{align}
Our objective here is to show $\displaystyle 1/2\hspace{-15pt}\sum_{\underline{x}^n\in T_{\delta}(\underline{X}),y^n}\hspace{-15pt}P_{\underline{X}^n}(\underline{x}^n)  S_{1} \cdot \PMFIndicator$ $\leq $ $\displaystyle 1/2\hspace{-15pt}\sum_{\underline{x}^n\in T_{\delta}(\underline{X}),y^n}\hspace{-15pt}P_{\underline{X}^n}(\underline{x}^n)  [Q_{1} +Q_2] \cdot \PMFIndicator$ is small, which eventually leads to (while also showing other terms corresponding to $S_{2}$ and $\tilde{S}$, are small), establishing 
$\sum_{\underline{x}^n \in T_{\delta}(\underline{X})}P^n_{\underline{X}}(\underline{x}^n)\mathscr{Q}_{\underline{x}^n}\cdot \PMFIndicator$ vanishes in expectation. With this partition, the terms within the modulus of $Q_1$ differ only in the action of Alice's encoding/approximation, and similarly, the terms within $Q_2$ differ only in the action of Bob's encoding/approximation. Showing that these two terms are small forms a major portion of the achievability proof. To begin with, let us consider $Q_1$.

\noindent\textit{Analysis of $Q_{1}$}:
To prove $ \left[\frac{1}{2}\sum_{\underline{x}^n\in T_{\delta}(\underline{X}),y^n}P_{\underline{X}^n}(\underline{x}^n)  Q_1\cdot \PMFIndicator\right]$ is small, we characterize the rate constraints which ensure that an upper bound to $ Q_1 $ can be made to vanish in an expected sense. In addition, this upper bound becomes useful in obtaining a single-letter characterization for the rate needed to make the term corresponding to $ Q_2 $ vanish. For this, we define $ J $ for each $\underline{x}^n \in T_{\delta}(\underline{X})$ as,
% To show $\displaystyle 1/2\hspace{-15pt}\sum_{\underline{x}^n\in T_{\delta}(\underline{X}),y^n}\hspace{-15pt}P_{\underline{X}^n}(\underline{x}^n)  Q_1\cdot \PMFIndicator$ is small, we prove a stronger result in an expected sense. 
% (Applying monotonicity of total variation to the latter result proves the former).
% Using this stronger result and the monotonicity of total variation, we indeed prove that the expectation of the term corresponding $Q_1$ can be   made arbitrarily small. 
% Further, this stronger result becomes useful in obtaining a single letter characterization for the rate needed to make the term corresponding to $Q_2$ to vanish. 
% and show that it is small in the expected sense.
% For this, let us define $J$ for each $\underline{x}^n \in T_{\delta}(\underline{X})$ as,
\begin{align}
J=  \Bigg| P_{YW_2|\underline{X}}^n&(y^n,w_2^n|\underline{x}^n) -  \frac{1}{2^{nC_1}}\sum_{\substack{ \mu_1 \in [2^{nC_1}]  }} \sum_{\substack{w_1^n}}\!
E^{(\mu_1)}_{W_1^n|X_1^n}(w_1^n|x^n_1)P^n_{W_2|X_2}(w_2^n|x^n_2)P^n_{Y|W_1W_2}(y^{n}|w_1^n,w^n_2),
  \Bigg| \label{eq:strongerResult}
\end{align}
where $E^{(\mu_1)}_{W_1^n|X_1^n}(w_1^n|x^n_1) \in [0,1] $ for all $x_1^n \in T_{\delta}(X_1)$. 
By defining $J$ we have added the random variable $W_2$ into the collecion of random variables which first encoder is trying to approximate. Hence, this encoder now approximates the joint product PMF $P_{\ulineX W_2 Y}$. To make $J$ small, we expect the sum of the encoding rate of first encoder and common randomness i.e.,  $\tilde{R}_1 + C_1$ to be larger then $I(W_1;X_1X_2W_2Y)$.  We prove this by bounding $  \sum_{\underline{x}^n \in T_{\delta}(\underline{X})} \sum_{y^n,w_2^n}P_{\underline{X}}^n(\underline{x}^n)J$ using the following proposition.
\begin{prop}
\label{prop:Lemma for J}
There exist  $\epsilon_{J}(\delta), \delta_{J}(\delta)$ such that for  all sufficiently small $\delta$ and sufficiently large $n$, we have $\EE\left[ \sum_{\underline{x}^n \in T_{\delta}(\underline{X})} \sum_{y^n,w_2^n}P_{\underline{X}}^n(\underline{x}^n)J\cdot \right] \leq \epsilon_{J}$ if $S_1 + C_1 \geq I(W_1;X_1X_2YW_2) + \delta_{J}$, where  $\epsilon_{{J}}, \delta_{{J}} \searrow 0$ as $\delta \searrow 0$.
\end{prop}
\begin{proof}
    The proof is provided in Appendix \ref{appx:prop:Lemma for J}.
\end{proof}
Now in regards to $Q_1$, applying triangle inequality on the summation over $w_2$ gives
\begin{align}
    \sum_{\underline{x}^n \in T_{\delta}(\underline{X}),y^n}P_{\underline{X}}^n{(\underline{x}^n)}Q_1  \leq \sum_{\underline{x}^n \in T_{\delta}(\underline{X})}\sum_{y^n,w_2^n}P_{\underline{X}}^n{(\underline{x}^n)}J  
\end{align}
Using the above proposition concludes the proof for the term corresponding to $Q_1$. Now, we move on to bounding the term $Q_2$.\\
\begin{figure}
 \centering
\includegraphics[width=4in]{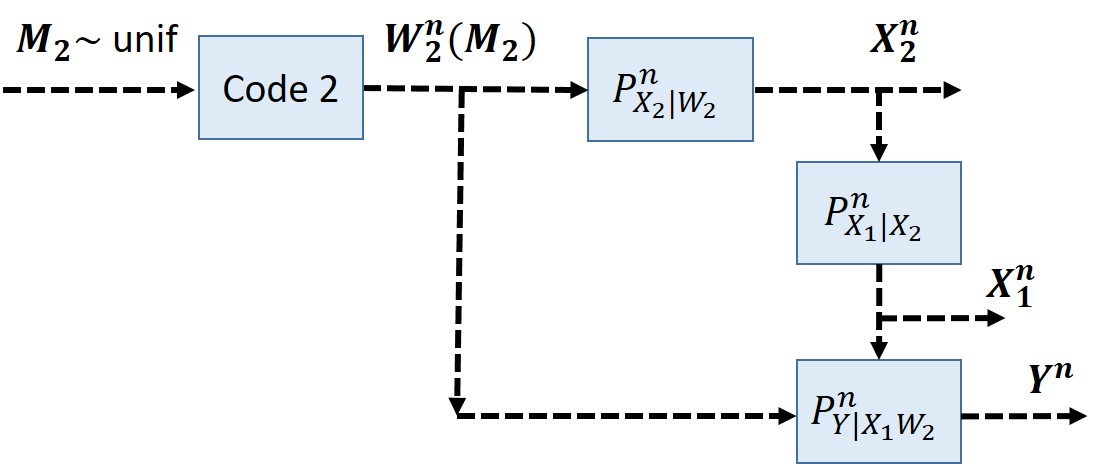}%\vspace{-0.22in}
\caption{Depiction of the approximation performed by Bob (encoder 2).}%\vspace{-0.25in}
\label{Fig:BobEncoding}
\end{figure} 

\noindent \textbf{Step 4: Analysis of Bob's encoding} Using the term $J$ in Step 3, we ensured that the random variables $X_1X_2YW_2$ are close to a product PMF in total variation. In this step, we approximate the PMF of random variables $X_1X_2Y$ using the Bob's encoding rule and bound the term corresponding to $Q_2$ (as illustrated in Figure \ref{Fig:BobEncoding}).  We proceed with the following proposition. 

\begin{prop} \label{prop:Lemma for Q2}
There exist functions  $\epsilon_{Q_{2}}(\delta) $ and $\delta_{{Q}_{2}}(\delta)$, such that for  all sufficiently small $\delta$ and sufficiently large $n$, we have $\EE[{Q}_2] \leq\epsilon_{{Q_2}}$, if $ S_1+C_1\geq  I(W_1;X_1X_2YW_2)  + \delta_{Q_2}$ and $ S_2+C_2 \geq  I(W_2;X_1X_2Y)  + \delta_{Q_2}$, where $\epsilon_{Q_2},\delta_{{Q}_2}  \searrow 0$ as $\delta \searrow 0$.
\end{prop}
\begin{proof}
 The proof is provided in Appendix \ref{appx:prop:Lemma for Q2}.
\end{proof}

% Invoking \cite[Corollary VII.4]{201311TIT_Cuf}, we note that the RHS above is at most \epsilon$ if $\tilde{R}_2 \geq \frac{1}{n}I_{\overline{s}}(X_1^n X_2^n Y^n;W_2^n)$ and $n$ is chosen sufficiently large. Observe that we proved $\sum_{\underline{x}^n,y^{n}}J \leq \epsilon$ for sufficiently large $n$ in (\ref{eq:Q1}). Hence $|\overline{s}_{\ulineX^{n} W_{2}^{n} Y^{n}}-P^{n}_{\ulineX W_{2} Y}|_{TV} \leq \epsilon$. From \cite[Thm.~17.3.3]{2006EIT_CovTho}, we have $ \lim_{n\rightarrow \infty }\frac{1}{n}I_{\overline{s}}(X_1^n X_2^n Y^n;W_2^n) = I(\ulineX,Y; W_{2})$. We therefore have the RHS of (\ref{Eqn:Q2Again}) to be dominated by $\epsilon$ for sufficiently large $n$ if $\tilde{R}_{2} \geq I(\ulineX,Y; W_{2})$.

% Now, we are left to prove the terms $S_2, S_3$ and $S_4$ are small. For this, we 
% provide additional constraints on the rates $\tilde{R}_1$ and $\tilde{R}_2$, and 
% as a result prove that these terms are small with high probability.

Hence, in bounding the terms corresponding to $Q_1$ and $Q_2$, we have obtained the following constraints: 
\begin{align}
    \tilde{R}_1 + C_1 \geq I(W_1;X_1X_2YW_2), \quad
    \tilde{R}_2+C_2 \geq  I(W_2;X_1X_2Y). \label{SCrate1}
\end{align}
% By doing an exact symmetric analysis, but by replacing the first encoder by a product distribution instead of the second encoder in $S_1$, we obtain the following constraints
% \begin{align}
%     \tilde{R}_1 + C_1 &\geq I(W_1;X_1X_2Y) , \nonumber \\
%     \tilde{R}_2+C_2 &\geq  I(W_2;X_1X_2YW_1) . \label{SCrate2}
% \end{align}

% \begin{lemma}

% \end{lemma}

% By time sharing between the above rates \eqref{SCrate1} and \eqref{SCrate2}, one can obtain the following rate constraints
% \begin{align}
%     \tilde{R}_1+ C_1 & \geq I(W_1;X_1X_2Y), \nonumber\\
%     \tilde{R}_2+C_2 &\geq  I(W_2;X_1X_2Y), \nonumber\\
%     \tilde{R}_1 +\tilde{R}_2 +C_1 +C_2 &\geq I(W_1W_2;X_1X_2Y) + I(W_1;W_2).  
% \end{align}

\subsection{Rate Constraints}
To sum-up, we showed that  \eqref{eq:dist_main_lemma} holds for sufficiently large $n$ and with probability sufficiently close to 1, if the following bounds holds while incorporating the time sharing random variable $Q$ taking values over the finite set $\mathcal{Q}$:
\begin{align}\label{eq:rate-region1}
    \tilde{R}_1 & \geq I(X_1;W_1|Q), \quad %+4\delta
    \tilde{R}_2  \geq I(X_2;W_2|Q), \nonumber \\ %+4\delta
    \tilde{R}_1+ C_1 & \geq I(X_1X_2YW_2;W_1|Q), \quad  %+\delta'
    \tilde{R}_2+ C_2  \geq I(X_1X_2Y;W_2|Q), \nonumber \\ %+\delta'
    % \tilde{R}_1 +\tilde{R}_2 +C_1 +C_2 &\geq I(X_1X_2Y;W_1W_2|Q) + I(W_1;W_2|Q), \nonumber \\
    \tilde{R}_1+\tilde{R}_2 - (R_1+R_2)  &\leq I(W_1;W_2|Q), \quad
    0 \leq R_1 \leq \tilde{R}_1,\quad
    0 \leq R_2 \leq \tilde{R}_2,\quad
    % 0\leq & \;C_1+C_2\leq C
    C_1  \geq 0, C_2 \geq 0.
\end{align}
Let us denote the above achievable rate-region by $\CalR_1$. By doing an exact symmetric analysis, but by replacing the first encoder by a product distribution instead of the second encoder in $S_1$, all the constraints remain the same, except that the constraints on $\tilde{R}_1+C_1$ and $\tilde{R}_2 + C_2$ change as follows
\begin{align}
    \tilde{R}_1 + C_1 \geq I(W_1;X_1X_2Y|Q) , \quad
    \tilde{R}_2+C_2 \geq  I(W_2;X_1X_2YW_1|Q) . \label{SCrate2}
\end{align}
Let us denote the above region by $\CalR_2$.
By time sharing between the any two points of $\CalR_1$ and $\CalR_2$ one can achieve any point in the convex hull of $(\CalR_1\bigcup\CalR_2).$ The following lemma gives a symmetric  characterization of the convex hull of the union of the above achievable rate-regions.
\begin{lemma}
For the above defined rate regions $\CalR_1$ and $\CalR_2$, we have $\CalR_3 = \text{Convex Hull}(\CalR_1 \bigcup \CalR_2)$, where $\CalR_3$ is given by the set of all the sextuples $(\tilde{R}_1, \tilde{R}_2, R_1, R_2, C_1, C_2)$ satisfying the following constraints: 
\begin{align}\label{eq:rate-regionR3}
    \tilde{R}_1 & \geq I(X_1;W_1|Q), \quad
    \tilde{R}_2  \geq I(X_2;W_2|Q), \nonumber \\ %+4\delta
    \tilde{R}_1+ C_1 & \geq I(X_1X_2Y;W_1|Q), \quad %+\delta'
    \tilde{R}_2+ C_2  \geq I(X_1X_2Y;W_2|Q), \nonumber \\ %+\delta'
    \tilde{R}_1 +\tilde{R}_2& +C_1 +C_2 \geq I(X_1X_2Y;W_1W_2|Q) + I(W_1;W_2|Q), \nonumber \\ 
    \tilde{R}_1+\tilde{R}_2 - (R_1+R_2)  &\leq I(W_1;W_2|Q) \quad
    0 \leq R_1 \leq \tilde{R}_1\quad
    0 \leq R_2 \leq \tilde{R}_2\quad
    % 0\leq & \;C_1+C_2\leq C
    C_1  \geq 0, C_2 \geq 0 
\end{align}
\end{lemma} 

\begin{proof}
The proof follows from elementary set-theoretic analysis, and hence is omitted. 
\end{proof}

% By time sharing between the above rates \eqref{SCrate1} and \eqref{SCrate2}, one can obtain the following rate constraints
% \begin{align}
%     \tilde{R}_1+ C_1 & \geq I(W_1;X_1X_2Y), \nonumber\\
%     \tilde{R}_2+C_2 &\geq  I(W_2;X_1X_2Y), \nonumber\\
%     \tilde{R}_1 +\tilde{R}_2 +C_1 +C_2 &\geq I(W_1W_2;X_1X_2Y) + I(W_1;W_2).  
% \end{align}

\begin{lemma}
Let $\bar{\mathcal{R}}_3$ denote the set of all quadruples $(R_1,R_2,C_1,C_2)$ for which there exists $(\tilde{R}_1,\tilde{R}_2)$ such that the sextuple $(R_1,R_2, C_1, C_2, \tilde{R}_1, \tilde{R}_1)$ satisfies the inequalities in (\ref{eq:rate-regionR3}). Let $\mathcal{R}_F$ denote the set of all quadruples $(R_1, R_2, C_1, C_2)$ that satisfy  the inequalities in \eqref{eq:dist_rate_region_theorem} given in the statement of the theorem. Then, $
\bar{\mathcal{R}}_3=\mathcal{R}_F$.
\end{lemma}
\begin{proof}
This follows by Fourier-Motzkin elimination \cite{fourier-motzkin}.
\end{proof}
The cardinality bounds on the auxilliary random variables follows from an argument using supporting hyperplanes of convex sets \cite{supportingHyperplanes}, and Caratheodary theorem \cite{201710TIT_LiElg}.

% For that, we eliminate $(\tilde{R}_1, \tilde{R}_2)$ from the system of inequalities given by (\ref{eq:rate-region1}). This gives us an equivalent rate-region described by all $(R_1,R_2,C)$ that satisfies the following set of inequalities:
% \begin{align}
%     {R}_1 & \geq I(X_1;W_1|Q) - I(W_1;W_2|Q), \nonumber \\
%     {R}_2 & \geq I(X_2;W_2|Q) - I(W_1;W_2|Q), \nonumber \\
%     {R}_1 + R_2 & \geq I(X_1;W_1|Q) + I(X_2;W_2|Q)-  I(W_1;W_2|Q), \nonumber \\
%     {R}_1 + C & \geq I(X_1X_2Y;W_1|Q) -  I(W_1;W_2|Q), \nonumber \\
%     {R}_2 + C & \geq I(X_1X_2Y;W_2|Q) -  I(W_1;W_2|Q), \nonumber \\
%     R_1+R_2+C & \geq I(X_1X_2Y;W_1W_2|Q).
% \end{align}
% Taking a convex closure of the above rate-region completes the proof.

% \textbf{Binned Random Encoders} To prove the achievablility of the above theorem, we introduce a binned version of the above encoders defined as follows. Each of the source first generates a collection of random encoders as defined above ($\ref{eq:randomEncoder}$). Let $E_{W^n|X^n}^{(\mu_1)}$ be one such encoder. We know that $E_{W^n|X^n}^{(\mu_1)} \neq 0$ only on the set $w^n \in \{T_{\delta}(W|x^n)\cap w_0\}$. Let N denote the size of the set and let [1,N] be some indexing of this set. Given K for which N is divisible, partition [1,N] into K equal bins and for each $i \in [1,K]$, let $B(i)$ denote the $i^{th}$ bin. The binned encoder $E_{B|X^n}^{(\mu_1)} $  is given by 
% \begin{align*}
%     E_{B|X^n}^{(\mu_1)}(i|x^n) = \sum_{w^n\in B(i)} E_{W^n|X^n}^{(\mu_1)}(w^n|x^n)
% \end{align*}
% \section{Conclusion}
\label{Sec:Conclusion}
\appendices
\section{Proof of Lemmas}
\subsection{Proof of Lemma \ref{lem:Lemma_E_PMF}: \texorpdfstring{$E_{L|X^n}^{(\mu)}(\cdot|\cdot)$} is a PMF with high probability}
\label{AppSec:ProofOfLemma_PMF}
% \begin{lemma}
% \label{Lem:PMFWHP}
% % \begin{align}
% %   \sum_{l =1}^{2^{n\tilde{R}}} E^{(\mu)}_{L|X^n}(l|x^n) \leq 1  \text{ with high probability. } \nonumber
% % \end{align}
% % $\sum_{l =1}^{2^{n\tilde{R}}} E^{(\mu)}_{L|X^n}(l|x^n) \leq 1 $ with high probability if $\tilde{R} \geq I(X;W) + 4\delta$, for all $\delta,\eta \in (0,1)$ and $x^n \in T_{\delta}(X)$.
% For any $\delta,\eta \in (0,1),$ if $\tilde{R} > I(X:W)+4\delta$, then
% \begin{align}
% \PP\left[\bigcap_{\mu=1}^{2^{nC}}\bigcap_{x^n\in T_{\delta}(X)}\left\{E^{(\mu)}_{L|X^n}(l|x^n) \leq 1\right\}\right]  \rightarrow 1 \text{ as } n \rightarrow \infty
% \end{align}
% \end{lemma}

% \begin{proof}
From the definition of $E^{(\mu)}_{L|X^n}(l|x^n)$, we have \text{ for } $x^n \in T_{\delta}(X), $
\begin{align*}
%   \sum_{l =1}^{2^{n\tilde{R}}}& E^{(\mu)}_{L|X^n}(l|x^n) \nonumber \\ 
   \sum_{l =1}^{2^{n\tilde{R}}} E^{(\mu)}_{L|X^n}(l|x^n) =  &\frac{1}{2^{n\tilde{R}}} \left(\frac{1-\epsilon}{1+\eta}\right) \sum_{\substack{w^n \in \\T_{\delta}(W|x^n)}} \sum_{l=1}^{2^{n\tilde{R}}} \mathbbm{1}_{\{\mathtt{w}^n(l,\mu) = w^n\}} \frac{P^n_{X|W}(x^n|w^n)}{P^n_X(x^n)} .
\end{align*}
% Since  $x^n \in T_{\delta}(X)$, and the summation is over $w^n \in T_{\delta}(W|x^n)$, we have $Q^n_{X|W}(x^n|w^n) \leq $

Let us define $Z^{(\mu)}_l(x^n)$, for $x^n \in T_{\delta}(X)$ as 
\begin{align}\label{def:ZtZ}
    Z_l^{(\mu)}(x^n) & = \sum_{w^n \in 
T_{\delta}(W|x^n)}\mathbbm{1}_{\{\mathtt{w}^n(l,\mu) = w^n\}} P^n_{X|W}(x^n|w^n) 
(1-\epsilon)
\end{align}
and let $D = 2^{n(H(X|W)-\delta_1)}$, where $\delta_1(\delta) \searrow 0$ as $\delta \searrow 0$. This gives us the following bound on the expectation of the empirical average of $\{Z^{(\mu)}_l(x^n)\}_{l \in [2^{n\tilde{R}}]}$ as 
\begin{align}
\label{eq:con1}
    \EE\bigg[\frac{1}{N}\sum_{l=1}^N D Z^{(\mu)}_l(x^n) \bigg]  
    & = 2^{n(H(X|W)-\delta_1)} \sum_{w^n \in T_{\delta}(W|x^n)\cap T_{\bar{\delta}}(W)}P_W^n(w^n) 
P^n_{X|W}(x^n|w^n)  \nonumber \\
& = 2^{n(H(X|W)-\delta_1)} \sum_{w^n \in T_{\delta}(W|x^n)}P_W^n(w^n) 
P^n_{X|W}(x^n|w^n) \nonumber \\
    & \geq 2^{n(H(X|W)-\delta_1)} 2^{-n(H(X,W)+2\delta_1)} 2^{n(H(W|X)-\delta_1)}  \geq 2^{-n(I(X;W)+4\delta_1)},
\end{align}
for all sufficiently large $n$, where in the above equations we use the fact that $\EE[\mathbbm{1}_{\{W^n(l,\mu) = w^n\}}] = 
\frac{P_W^n(w^n)}{1-\epsilon}$ for $w^n \in T_{\bar{\delta}}(W)$, and the fact that if $x^n \in T_{\delta}(X)$ and 
$w^n \in T_{\delta}(W|x^n)$, then 
$(x^n,w^n) \in T_{\bar{\delta}}(X,W)$,
and consequently $w^n \in T_{\bar{\delta}}(W)$. 
Furthermore, for sufficiently large $n$, we also have
\begin{align} \label{eq:con2}
    D Z^{(\mu)}_l(x^n) & \leq  2^{n(H(X|W)-\delta_1)} 2^{-n(H(X|W)-\delta_1)}(1-\epsilon)\!\!\!\sum_{w^n \in T_{\delta}(W|x^n)}\mathbbm{1}_{\{W^n(l,\mu) = w^n\}}   \leq   1 
\end{align}
where we have bounded $\displaystyle\sum_{w^n \in T_{\delta}(W|x^n)}\mathbbm{1}_{\{W^n(l,\mu) = w^n\}}$ by $1$.
% if $x^n \in T_{\delta}(X)$ and $w^n \in T_{\delta}(W|x^n)$ then $(x^n,w^n) \in T_{2\delta}(X,W)$ and $w^n \in T_{\delta_1}(W)$, where $\delta_1= 2|\mathcal{X}|\delta$.

Since $\{Z^{(\mu)}_l(x^n)\}_{l}$ is a sequence of IID Random variables, we can approximate its empirical average, for $x^n \in T_{\delta}(X)$, using a refined Chernoff-Hoeffding bound given by
% \footnote{The Statement of the bound used and its proof is provided in \cite{arxiv_CurrentPaper}}
% to obtain

% \begin{lemma} \label{lem:chernoff}
% Let $\{Z_n\}_{n=1}^N$ be a sequence of N IID random variables bounded between $[0,1]$, such that $\EE\bigg[\frac{1}{N}\sum_{n=1}^{N}Z_n\bigg] = \mu  \geq \theta$ where $\theta \in (0,1)$ then
% \begin{align}
%     \PP\bigg(\frac{1}{N}\sum_{n=1}^{N}Z_n \in & [(1-\eta)\mu,(1+\eta)\mu] \bigg)  \geq 1 - 2\exp{\bigg(-\frac{N\eta^2 \theta}{4 \ln{2}}\bigg)}
% \end{align}
% \end{lemma}
\begin{lemma} \label{lem:chernoff}
Let $\{Z_n\}_{n=1}^N$ be a sequence of N IID random variables bounded between zero and one, i.e., $ Z_n \in [0,1] \quad \forall n \in [N]$, and suppose $\EE\bigg[\frac{1}{N}\sum_{n=1}^{N}Z_n\bigg] = \mu$ be bounded below by a positive constant $\theta$ as $ \mu \geq \theta$ where $\theta \in (0,1)$, then for every $\eta \in (0,1/2)$ and $(1+\eta)\theta < 1$, we can bound the probability that the ensemble average of the sequence $\{Z_n\}_{n=1}^{N}$ lies in $(1\pm \eta)\mu$ as
\begin{align}
    \PP\bigg(\frac{1}{N}\sum_{n=1}^{N}Z_n \in & [(1-\eta)\mu,(1+\eta)\mu] \bigg)  \geq 1 - 2\exp{\bigg(-\frac{N\eta^2 \theta}{4 \ln{2}}\bigg)}
\end{align}
\end{lemma}
\begin{proof}
Follows from Operator Chernoff Bound \cite{ahlswede2002strong}.
\end{proof}

Note that $\{DZ^{(\mu)}_l(x^n)\}_{l}$ satisfies the  constraints of the above lemma from Eqns.  (\ref{eq:con1}) and (\ref{eq:con2}).
Thus applying Lemma (\ref{lem:chernoff}) to $\{DZ^{(\mu)}_l(x^n)\}_{l}$ for every $x^n \in T_{\delta}(X)$ gives
% \begin{lem}{Chernoff Bound} \label{lem:chernoff}
% Let $\{Z_n\}_{n=1}^N$ be a sequence of N IID random variables bounded between $[0,1]$, such that $\EE\bigg[\frac{1}{N}\sum_{n=1}^{N}Z_n\bigg] = \mu  \geq \theta$ where $\theta \in (0,1)$ then
% \begin{align}
%     \PP\bigg(\frac{1}{N}\sum_{n=1}^{N}Z_n \in [(1-\eta)\mu,(1+\eta)\mu] \bigg) \geq 1 - 2\exp{\bigg(-\frac{N\eta^2 a}{4 \ln{2}}\bigg)}
% \end{align}
% \end{lem}
% \begin{proof}
% Follows from Operator Chernoff Bound
% \end{proof}
% Note that $\{DZ^{(\mu)}_l(x^n)\}_{l}$ satisfies the Applying the constraints from the above lemma from Eqns.  (\ref{eq:con1} and \ref{eq:con2}).
% Thus applying Lemma (\ref{lem:chernoff}) to $\{DZ^{(\mu)}_l(x^n)\}_{l}$ for some $x^n \in T_{\delta}(X)$ gives
\begin{align}\label{eq:chernoff1_Z}
    \!\!\PP\left(\!\! \frac{1}{2^{n\tilde{R}}}\sum_{l=1}^{2^{n\tilde{R}}} Z^{(\mu)}_l(x^n)\in [(1-\eta)\EE[ Z^{(\mu)}(x^n)],(1+\eta)\EE[ Z^{(\mu)}(x^n)]]\!\!\right) \geq 1 - 2\exp{\bigg(\!\!-\frac{\eta^2 2^{n(\tilde{R} - I(X;W)-4\delta_1)} }{4 \ln{2}}\bigg)}, 
\end{align}
% where we have used the additivity of expectation to simplify $\EE\bigg[\frac{1}{N}\sum_{l=1}^N Z^{(\mu)}_l(x^n) \bigg]$ and $Z(x^n)$ is used to denote a generic random variable from the IID sequence.
where $Z^{(\mu)}(x^n) \deq \frac{1}{2^{n\tilde{R}}}\sum_{l=1}^{2^{n\tilde{R}}} Z^{(\mu)}_l(x^n) $, the ensemble mean of the IID sequence $\{Z^{(\mu)}_l(x^n)\}_{l}$.
Substituting the following simplification
\begin{align}
 \frac{1}{2^{n\tilde{R}}}\sum_{l =1}^{2^{n\tilde{R}}}Z_l^{(\mu)}(x^n) = (1+\eta)P_{X}^n(x^n) \sum_{l =1}^{2^{n\tilde{R}}}  E^{(\mu)}_{L|X^n}(l|x^n), 
\end{align} 
which follows from the definition of $Z_l^n(x^n)$  in (\ref{eq:chernoff1_Z}) gives
\begin{align}
    \PP\left((1+\eta)P_{X}^n(x^n) \sum_{l =1}^{2^{n\tilde{R}}}  E^{(\mu)}_{L|X^n}(l|x^n) \leq (1+\eta)\EE{[Z^{(\mu)}(x^n)]}\right)  \geq 1 - 2\exp{\bigg(-\frac{\eta^2 2^{n(\tilde{R} - I(X,W)-4\delta_1)} }{4 \ln{2}}\bigg)}. 
\end{align}
Further we can bound $\EE[Z(x^n)]$ as
\begin{align}
    \frac{\EE[Z^{(\mu)}(x^n)]}{P^n_X(x^n)}  \leq  \frac{(1-\epsilon)}{P_X^n(x^n)} \sum_{w^n \in T_{\delta}(W|x^n)}\tilde{P}_W^n(w^n) P^n_{X|W}(x^n|w^n) \leq  \frac{1}{P_X^n(x^n)} \sum_{w^n} P^n_{X,W}(x^n,w^n) =1. \nonumber
\end{align}
% where the last inequality above is obtained by adding more terms in the summation and using the definition of $\tilde{p}_W^n(w^n)$ from (\ref{eq:prunedDistribution}).
This simplifies the above probability term as
\begin{align}
    \PP\left(\sum_{l =1}^{2^{n\tilde{R}}}  E^{(\mu)}_{L|X^n}(l|x^n) \leq 1\right)  \geq 1 - 2\exp{\bigg(-\frac{\eta^2 2^{n(\tilde{R} - I(X,W)-4\delta_1)} }{4 \ln{2}}\bigg)}. \nonumber
\end{align}
Using the union bound, we extend the above probability to the intersection of all $\mu \in [2^{nC}]$ and $x^n \in T_{\delta}(X)$ as
\begin{align}
\PP\left[\bigcap_{\mu=1}^{2^{nC}}\bigcap_{x^n\in T_{\delta}(X)}\left(\sum_{l =1}^{2^{n\tilde{R}}}  E^{(\mu)}_{L|X^n}(l|x^n)\right) \leq 1\right] &\geq 1-\sum_{\mu=1}^{2^{nC}}\sum_{x^n \in T_{\delta}(X)} \PP\left(\sum_{l =1}^{2^{n\tilde{R}}}  E^{(\mu)}_{L|X^n}(l|x^n) \leq 1\right)\nonumber \end{align}
\begin{align}\hspace{20pt}\geq 1 - 2^{nC}|T_{\delta}(X)|2\exp{\bigg(-\frac{\eta^2 2^{n(\tilde{R} - I(X,W)-4\delta_1)} }{4 \ln{2}}\bigg)} \label{eq:unionProb}.
\end{align}
Therefore, if $\tilde{R} > I(X;W) + 4\delta_1$, the second term in the right hand side of (\ref{eq:unionProb}) decays exponentially to zero and as a result the probability of the above intersections goes to 1.
This completes the proof of the lemma.

\subsection{Proof of Lemma \ref{lem:Change Measure Soft Covering Variance Based}}
\label{appx:proofOneShotLemma}
Be begin by defining $K$ as 
\begin{align*}
    K &\deq \sum_{a\in{\mathcal{A}}}\Big|P_{A}(a) - \frac{1}{M}\sum_{m=1}^M P_{A|B}(a|C_m)\Big| \\
    & = \sum_{a\in{\mathcal{A}}}\Big|\sum_{b \in \CalB}P_{AB}(a,b) - \frac{1}{M}\sum_{m=1}^M\sum_{b \in \CalB} P_{A|B}(a|b)\11_{\{C_m = b\}}\Big|,
\end{align*}
where in the above equality we have used $\sum_{b \in \CalB}\11_{\{C_m = b\}} = 1.$
Using triangle inequality, we obtain $K \leq K_1 + K_2 + K_3+K_4$, where
\begin{align}
    K_1 &\deq \sum_{a\in{\mathcal{T}}}\Big|\sum_{b \in \CalB}P_{AB}(a,b)\11_{\{a\in \CalT_b\}} - \frac{1}{M}\sum_{m=1}^M\sum_{b \in \CalB} P_{A|B}(a|b)\11_{\{C_m = b\}}\11_{\{a\in \CalT_b\}}\Big|, \nonumber \\
    K_2 \deq \sum_{a\in{\mathcal{T}}}&\sum_{b \in \CalB}P_{AB}(a,b)\11_{\{a\notin \CalT_b\}}, \quad 
    K_3 \deq \sum_{a\in{\mathcal{T}}}\frac{1}{M}\sum_{m=1}^M\sum_{b \in \CalB} P_{A|B}(a|b)\11_{\{C_m = b\}}\11_{\{a\notin \CalT_b\}},\quad \text{and}\nonumber\\
    & \hspace{1in} K_4 \deq \sum_{a\in\CalA \backslash \CalT}\Big|P_{A}(a) - \frac{1}{M}\sum_{m=1}^M P_{A|B}(a|C_m)\Big|. \nonumber
\end{align}
We begin by first bounding the terms corresponding to $K_2, K_3$ and $K_4$, finally and delve into bounding the  main term corresponding to $K_1$.
Note that $K_2$ can be written as 
\begin{align*}
    K_2 & \leq \sum_{b \in \CalB}P_B(b)\sum_{a\notin \CalT_b}P_{A|B}(a|b) = \sum_{b \in \CalB}P_B(b)(1-P_{A|B}(\CalT_b|b)) \leq \epsilon,
\end{align*}
where the last inequality uses the hypothesis \eqref{Soft_Covering-constraints2} from the statement of the lemma.
Considering the term $K_3$, applying expectation yields
\begin{align*}
    \EE[K_3] \leq \sum_{b \in \CalB}P_B(b)\sum_{a\notin \CalT_b}P_{A|B}(a|b) = \sum_{b \in \CalB}P_B(b)(1-P_{A|B}(\CalT_b|b)) \leq \epsilon,
\end{align*}
where the last inequality again uses the hypothesis \eqref{Soft_Covering-constraints2} from the statement of the lemma.
Considering the term $K_4$, we use the fact that $ \EE[\frac{1}{M}\sum_{m=1}^M P_{A|B}(a|C_m)] = P_{A}(a),$ and bound $K_4$ as 
\begin{align*}
    \EE[K_4] & \leq 2\sum_{a\in \CalA\backslash\CalT} \EE\left[\frac{1}{M}\sum_{m=1}^M P_{A|B}(a|C_m)\right] = 2\sum_{a\in \CalA\backslash\CalT}P_A(a) = 2(1- P_A(\CalT)) \leq 2\epsilon.
\end{align*}
Finally, we consider the term $K_1$.
Using the concavity of the square-root function, we have
\begin{align}
    \EE[K_1]& \leq \sum_{a\in \CalT}\sqrt{Var\left(\frac{1}{M}\sum_{m=1}^M\sum_{b \in \CalB} P_{A|B}(a|b)\11_{\{C_m = b\}}\11_{\{a\in \CalT_b\}}\right)}\label{eq:JensensUsage}.
\end{align}
Further, the term within the variance can be simplified as
\begin{align*}
    \mbox{Var}\left(\frac{1}{M}\sum_{m=1}^M\sum_{b \in \CalB} P_{A|B}(a|b)\11_{\{C_m = b\}}\11_{\{a\in \CalT_b\}}\right) 
    & \leq \frac{1}{M}\EE\left[ \sum_{b\in \CalB}P_{A|B}^2(a|b)\11_{\{a\in \CalT_b \}}\11_{\{b = C_1\}} \right]\\
    &  = \frac{1}{M}\sum_{b\in \CalB}P_{A|B}^2(a|b)P_{B}(b)\11_{\{a\in \CalT_b \}}\\
    & \leq \frac{1}{M}\frac{1}{d}\sum_{b\in \CalB}P_{A|B}(a|b)P_{B}(b) = \frac{P_A(a)}{Md},
\end{align*}
where in the first inequality we use (i) the fact that codewords are generated pairwise independently from $P_B$, and (ii) $Var(\cdot) \leq \EE[(\cdot)^2]$, in the  first equality we have used $\EE[\11_{\{C_m = b\}}] = P_B(b)$, and in the second inequality we have used the hypothesis $\eqref{Soft_Covering-constraints4}$ from the statement of the lemma
Finally, substituting the above bounds in $\eqref{eq:JensensUsage}$, and using the hypothesis \eqref{Soft_Covering-constraints3}, we obtain
\begin{align}
    \EE[K_1] & \leq \sqrt{\frac{D}{Md}}\nonumber.
\end{align}
Combining all the bounds on $K_1, K_2, K_3$ and $K_4$ completes the proof.

\section{Proof of Theorem \ref{Thm:DistributedOuterBound}}\label{appx:converseDistributed}
%  Our proof of the converse follows a similar sequence of arguments as in \cite[Sec.~VI]{201311TIT_Cuf}. 
%  Here, we derive lower bounds on rate quadruple $(R_1,R_2,C_1,C_2) \in \mathcal{S}_{\epsilon}:\epsilon >0$, where $\mathcal{S}_{\epsilon}$ is a set analogous to $\mathcal{S}_{\epsilon}$ defined as
 
%  in \cite[Sec.~VI]{201311TIT_Cuf}.
%  the side information case. proof of the converse follows a similar sequence of arguments as in \cite[Sec.~VI]{201311TIT_Cuf}. Here, we derive lower bounds on rate pairs $(R,C) \in \mathcal{S}_{\epsilon}:\epsilon >0$, where $\mathcal{S}_{\epsilon}$ is a set analogous to $\mathcal{S}_{\epsilon}$ in \cite[Eqn.~80]{201311TIT_Cuf}). 
%  In view of the similarity in arguments to \cite[Sec.~VI]{201311TIT_Cuf}, we refer to \cite{arxiv_CurrentPaper} for further details. 
Let $(R_1, R_2, C_1, C_2)$ be an achievable quadruple. Fix an arbitrary $\epsilon >0$ and a sufficiently large $n$. From Definition \ref{Defn:Distributed}
it follows that there exists $2^{nC_1}\times 2^{nC_2}$ randomized encoder pairs $(E_{1}^{(\mu_1)}, E_{2}^{(\mu_2)} )$, $  j \in [2],$ and a  corresponding collection of $2^{nC} $ randomized decoders $D^{(\mu)}$ that satisfy the following constraints:
$\frac{1}{n} \log \Theta_j \leq R_j+\epsilon$, 
and $\| P_{\underline{X}Y}^n - P_{\underline{X}^nY^n} \|_1 \leq \epsilon$.
Let $M_1$ and $M_2$ be the messages communicated by the first and second encoders, respectively, and let $K_1 \in 
[2^{nC_1}]$ and $K_2 \in 
[2^{nC_2}]$ denote the common randomness shared among the first encoder and the decoder, and the second encoder and the decoder, respectively. 
The source sequence pair $(X_1^n,X_2^n)$ and 
$K_1$ and $K_2$ are mutually independent.
Let $P_{M_j|X^n_jK_j}: j\in\{0,1\}$ denote the two distributed stochastic encoders,
respectively.  $Y^n \sim P_{Y^n|M_1 M_2, K_1, K_2}$ be the samples generated by the decoder using the messages received and the common randomness available. Lastly, let $R_1$ and $C_1$ denote communication rate and the common randomness rate of the first encoder, respectively, and similarly, let $R_2$ and $C_2$ denote communication rate and the common randomness rate of the second encoder, respectively. Additionally, let $P_{W|X_1X_2}$ be an arbitrary distribution in $\CalP_{R}$. 
Recall that $P_{X_1X_2} P_{W|X_1X_2}$ satisfies the Markov chain $X_1 -W -X_2$. We generate $n$ copies of the auxilliary random variable $W$ denoted as $W^n$,  from $(X_1^n,X_2^n)$ in a memoryless fashion  using $P_{W|X_1,X_2}$
to yield (i) $X_1^n - W^n- X_2^n$. Enforce a Markov coupling of this with the $n$-letter encoders and decoder to result in the following
$n$-letter Markov chains:
% let $W^n$ denote an auxiliary random variable, associated with joint distribution of the sources $X_1^n$ and $X_2^n$, 
(ii) $ M_1 - (X_1^n, K_1) - (W^n, X_2^n, K_2, M_2) $, and (iii) $ M_2 - (X_2^n,K_2) - (W^n,X_1^n, K_1, M_1)$.  This simplifies  the joint distribution as
\begin{align}\label{eq:dist_converse}
    &P_{K_1K_2X_1^nX_2^nW^nM_1M_2Y^n}(k_1,k_2,x_1^n,x_2^n,w^n,m_1,m_2,y^n) \nonumber  \\ 
    & \hspace{10pt}= P_{K_1}(k_1)P_{K_2}(k_2)P^n_W(w^n)P_{X_1|W}^n(x_1^n|w^n)P_{X_2|W}^n(x_2^n|w^n) \nonumber \\ & \hspace{1in} P_{M_1|X_1^nK_1}(m_1|x_1^n,k_1)P_{M_2|X_2^nK_2}(m_2|x_2^n,k_2)
    P_{Y^n|M_1K_1M_2K_2}(y^n|m_1,k_1,m_2,k_2).
\end{align}
Further, define for $i\in [n]$, $U_i$ and $V_i$ as $U_i \deq (M_1,K_1,W^{i-1}) $ and $V_i \deq (M_2,K_2,W^{i-1})$.

\noindent \textbf{Step 1: Rate Constraints:} Using this, we have
\begin{align}\label{eq:convR_1}
    n(R_1 + \epsilon) & \geq H(M_1) \nonumber\\ & \geq H(M_1|M_2,K_1,K_2) \nonumber\\
    & \geq I(X_1^n;M_1|M_2,K_1,K_2) \nonumber \\
    & \labelrel={eq:conv1} I(X_1^n;M_1|M_2,K_1,K_2) + I(W^n; M_1|M_2,K_1,K_2,X_1^n) \nonumber\\
    & \labelrel={eq:conv2} I(X_1^n,W^n;M_1|M_2,K_1,K_2) + I(X_1^n,W^n;K_1|M_2,K_2) \nonumber\\
    & = I(X_1^n,W^n;M_1,K_1|M_2,K_2)\nonumber \\
    & = I(W^n;M_1,K_1|M_2,K_2) +  I(X_1^n;M_1,K_1|M_2,K_2,W^n) \nonumber\\ 
    & \labelrel={eq:conv3} I(W^n;M_1,K_1|M_2,K_2) +  I(X_1^n;M_1,K_1|W^n) \nonumber\\
    & = \sum_{i=1}^{n}\bigg[ I(W_i;M_1,K_1|M_2,K_2,W^{i-1}) +  I(X_{1i};M_1,K_1|W^n,X_1^{i-1}) \bigg] \nonumber\\
    & \labelrel={eq:conv4} \sum_{i=1}^{n}\bigg[ I(W_i;U_i|V_i) + H(X_{1i}|W^n) -   H(X_{1i}|M_1,K_1,W^n,X_1^{i-1}) \bigg] \nonumber\\
    & \labelrel\geq{eq:conv5} \sum_{i=1}^{n}\bigg[ I(W_i;U_i|V_i) + H(X_{1i}|W_i,Q_i) -   H(X_{1i}|M_1,K_1,W_i,Q_i) \bigg] \nonumber\\
    & = \sum_{i=1}^{n}\bigg[ I(W_i;U_i|V_i) + I(X_{1i},U_i|W_i, Q_i) \bigg] \nonumber \\
    & \labelrel={eq:conv6} n\bigg(I(W;U|V,J) + I(X_{1},U|W, Q, J)\bigg),
\end{align}
where \eqref{eq:conv1} follows from the fact that 
\begin{align*}
    I(W^n; M_1|M_2,K_1,K_2,X_1^n) & = I(M_2K_2W^n;M_1|X_1^n,K_1) - I(M_2K_2;M_1|X_1^n,K_1) \\ 
    &= - I(M_2K_2;M_1|X_1^n,K_1) \leq 0,
\end{align*}
which is only true if $I(W^n; M_1|M_2,K_1,K_2,X_1^n) = 0$, \eqref{eq:conv2} follows from $I(X_1^n,W^n;K_1|M_2,K_2) = 0$ which is true given the decomposition in \eqref{eq:dist_converse},  \eqref{eq:conv3} uses the fact that for a Markov Chain $(A,B) - C - D$, we have $I(A;B|C,D) = I(A;B|C)$, \eqref{eq:conv4} is obtained using the definitions of $U_i$ and $V_i$, and the memoryless nature of the source $X_1^n$,  \eqref{eq:conv5} follows from defining for all $i\in [n]$, $Q_i \deq W^{n\backslash i}$ and using the result - conditioning reduces entropy, and finally \eqref{eq:conv6} follows by (i) defining a time-sharing random variable $J$ which is uniformly distributed in $[1,n]$ and independent of $(W^n, U^n, V^n, Q^n, X^n_1, X_2^n, Y^n)$, and (ii) defining $W, U, V, Q, X_1$ and $ X_2$ as $W_J, U_J, V_J, Q_J, X_{1J}$ and $X_{2J}$, respectively. 
Using identical steps for the bound $R_2$, we get the following bound for $R_2$
\begin{align*}
    n(R_2 + \epsilon) \geq n\big( I(W;V|U) + I(X_{2},V|W, Q) \big) .
\end{align*}

We now provide a bound on the sum rate $R_1+R_2$ as
\begin{align}\label{eq:conv_r12}
    n(R_1 + R_2 + \epsilon) &\geq H(M_1,M_2) \nonumber\\
    & \geq I(X_1^n,X_2^n,K_1,K_2;M_1,M_2) \nonumber\\
    & \labelrel={eq:conv2_1} I(W^n,X_1^n,X_2^n,K_1,K_2;M_1,M_2) \nonumber\\
    & = I(W^n;M_1,M_2) + I (K_1,K_2;M_1,M_2|W^n) + I(X_1^n,X_2^n;M_1,M_2|W^n, K_1,K_2 )\nonumber \\
    & \labelrel\geq{eq:conv2_2} I(W^n;M_1,M_2) + I (K_1,K_2;W^n|M_1,M_2) + I(X_1^n,X_2^n;M_1,M_2, K_1,K_2|W^n )\nonumber\\
    & \labelrel={eq:conv2_3} I(W^n;M_1,M_2,K_1,K_2) + I(X_1^n;M_1, K_1|W^n) + I(X_2^n;M_2, K_2|W^n ) \nonumber\\
     & = \sum_{i}^n \bigg[I(W_i;M_1,M_2,K_1,K_2|W^{i-1}) + I(X_{1i};M_1,K_1|W^n,X_1^{i-1})  \nonumber \\
     & \hspace{3.15in}+ I(X_{2i};M_2,K_2|W^n,X_2^{i-1} ) \bigg]\nonumber\\
     & = \sum_{i}^n \bigg[I(W_i;M_1,M_2,K_1,K_2|W^{i-1}) + I(X_{1i};U_i|W_i,Q_i) + I(X_{2i};V_i|W_i,Q_i ) \bigg] \nonumber\\
     & = \sum_{i}^n \bigg[I(W_i;M_1,M_2,K_1,K_2|W^{i-1}) + I(W_i;W^{i-1}) \nonumber \\
     & \hspace{2.4in}+ I(X_{1i};U_i|W_i,Q_i) + I(X_{2i};V_i|W_i,Q_i ) \bigg] \nonumber\\
     & \labelrel={eq:conv2_4} \sum_{i}^n \bigg[I(W_i;U_i,V_i) +  I(X_{1i};U_i|W_i,Q_i) + I(X_{2i};V_i|W_i,Q_i ) \bigg]  \nonumber \\
     & =  n\big(I(W;U,V|J) +  I(X_{1};U|W,Q,J) + I(X_{2};V|W,Q,J )\big),
\end{align}
where \eqref{eq:conv2_1} follows from the Markov Chain $W^n - (X_1^n,X_2^n,K_1,K_2) - (M_1,M_2)$ which makes $I(W^n;M_1,M_2|X_1^n,X_2^n,K_1,K_2) = 0$, \eqref{eq:conv2_2} follows from 
\begin{align*}
    (i) \quad I (K_1,K_2;M_1,M_2|W^n) & = H(K_1,K_2|W^n) - H(K_1,K_2|W^n,M_1,M_2) \\
    & = H(K_1,K_2) - H(K_1,K_2|W^n,M_1,M_2) \\
    & \geq I (K_1,K_2;W^n|M_1,M_2), \\
    \text{and} \quad (ii) \quad I(K_1,K_2;X_1^n,X_2^n|W^n) &= 0.
\end{align*}
\eqref{eq:conv2_3} follows from the Markov Chain $ M_1 - (X_1^n, K_1) - W^n - (X_2^n, K_2) - M_2$, and \eqref{eq:conv2_4} follows from similar arguments as in (\ref{eq:convR_1}-\ref{eq:conv4}).

We now provide the bound for  $R_1 + R_2 +C_1 +C_2$ as follows.
\begin{align}\label{eq:converse_r12+c_12}
    n(R_1 + R_2 + C_1 + C_2+\epsilon) & \geq H(M_1, M_2, K_1, K_2) \nonumber \\ 
    & \geq I(M_1, M_2,K_1, K_2; X_1^n,X_2^n, Y^n) \nonumber \\
    & \labelrel={eq:conv5_1} I(X_1^n, X_2^n, Y^n, W^n; M_1, M_2, K_1, K_2) \nonumber \\
    & =I(W^n; M_1, M_2,K_1,K_2) + I(X_1^n, X_2^n, Y^n; M_1,M_2 ,K_1, K_2| W^n), 
\end{align}
where \eqref{eq:conv5_1} follows from using the Markov chain $ W^n - (X_1^n, X_2^n) - (M_1,M_2,K_1,K_2) - Y^n$ which implies $I(W^n;M_1, M_2, K_1, K_2|X_1^n,X_2^n, Y^n ) = 0.$ Again the first term in the right hand side of \eqref{eq:converse_r12+c_12} can simplified following the approach in \eqref{eq:conv_r12} as $ I(W^n; M_1, M_2,K_1,K_2)  = \sum_{i=1}^n I(W_i; U_i,V_i)$. For the second term, we have
\begin{align*}
    I&(X_1^n, X_2^n, Y^n;  M_1,M_2 ,K_1, K_2| W^n) \\
    & = \sum_{i=1}^n \bigg[ I(X_{1i},X_{2i}, Y_i; U_i, V_i X_1^{i-1}, X_2^{i-1}, Y^{i-1}|Q_i,W_i) - I(X_{1i},X_{2i}, Y_i;X_1^{i-1}, X_2^{i-1}, Y^{i-1}|Q_i, W_i ) \bigg] \\
    & \geq \sum_{i=1}^n I(X_{1i},X_{2i},Y_i; U_i, V_i|Q_i,W_i) - ng_c(\epsilon),
\end{align*}
where in the last inequality above we use $\| P_{X_1X_2Y}^n - P_{X_1^nX_2^nY^n} \|_1 \leq \epsilon$,  implies $ I(X_{1i}, X_{2i}^n, Y_i; X_1^{i-1}, X_2^{i-1}, Y^{i-1}| Q_i, W_i) \leq ng_c(\epsilon)$, and define $g_c(\epsilon)$ as in the statement of the theorem using Lemma VI.3 from \cite{201311TIT_Cuf} obtaining $g_c(\epsilon ) \searrow 0$ as $\epsilon \searrow 0$ which follows from the memoryless nature of $P_{W^n|X^n_1X^n_2}$. Substituting the above simplification in \eqref{eq:converse_r12+c_12}, we obtain
\begin{align}
    n(R_1 + R_2 + C_1 + C_2+\epsilon) & \geq \sum_{i=1}^n \bigg[I(W_i; U_i,V_i) + I(X_{1i},X_{2i}, Y_i; U_i, V_i|Q_i,W_i) - g_c(\epsilon)\bigg] \nonumber \\
     & = n\bigg(I(W; U,V|J) + I(X_{1},X_{2}, Y; U, V|Q,W,J) - g_c(\epsilon)\bigg),
\end{align}
where the equality above follows by defining $J$ as an averaging random variable which is uniformly distributed in $[1,n]$ and $Y$ as $Y_J$.

\noindent \textbf{Step 2: Single-letter $l_1$ constraint (d):} Since the encoders and decoder satisfy the $l_1$ distance constraint
$\| P_{X_1X_2Y}^n - P_{X_1^nX_2^nY^n} \|_1 \leq \epsilon$
 (as in Definition \ref{Defn:Distributed}), using the Lemma VI.2 from \cite{201311TIT_Cuf} we have
\begin{align}
    \|P_{X_1X_2Y} - P_{X_{1J}X_{2J}Y_J}\|_1 \leq \|P^n_{X_1X_2Y} - P_{X_{1}^nX_{2}^nY^n}\|_1 \leq \epsilon.
\end{align}

\noindent \textbf{Step 3: Markov Chains:} We now argue that the Markov Chains $(a), (b), (c)$, $(e)$ and $(f)$ stated in the theorem statement hold. 
The Markov Chain $(a)$ follows from 
the standard information-theoretic arguments 
with time-sharing random variables \cite{2006EIT_CovTho}
and using the fact that (i) $J$ is independent of 
$(X_1^n,X_2^n,Q^n)$ and (ii) the stationary and memoryless nature of the sources $(X_1^n,X_2^n, W^n)$ which makes $(X_1,X_2)$ independent of $Q.$ Moving on to the next, the Markov chain $(b)$ $U - (X_1,Q,J) - (X_2,Q,J) - V$ holds true from the following arguments. 
Since $J$ is uniform and is independent of the sources $X_1^n$ and $X_2^n$, this is equivalent to showing 
$U_i - (X_{1i},Q_i) -
(X_{2i},Q_i) - V_i$ for $i \in [n]$. 
This is  equivalent to $(M_1,K_1,W^{i-1}) -
(X_{1i},W^{i-1},W^{n}_{i+1}) - (X_{2i},W^{i-1},W^{n}_{i+1}) - (M_2,K_2,W^{i-1})$. Hence we need to show 
$(M_1,K_1) - (X_{1i},W^{i-1},W^{n}_{i+1}) - (X_{2i},W^{i-1},W^{n}_{i+1}) - (M_2,K_2)$.
We show this in the following. For an arbitrary $i \in [n]$ and for $m_j \in [2^{nR_j}], k_j \in [2^{nC_j}] :j=1,2$, 
$x_{1i} \in \CalX_1$, $x_{2i} \in \CalX_2$,   $w^{[i]} \deq w^{n\backslash i} \in \CalW^{n-1}$, we have  
\begin{align}
P[M_1=&m_1, K_1 = k_1|X_{1i}=x_{1i},X_{2i}=x_{2i},W^{[i]}=w^{[i]},M_2=m_2, K_2 = k_2] \nonumber \\
&=\frac{P[M_1=m_1,K_1 = k_1, X_{1i}=x_{1i},X_{2i}=x_{2i},W^{[i]}=w^{[i]},M_2=m_2, K_2 = k_2]}{P[X_{1i}=x_{1i},X_{2i}=x_{2i},W^{[i]}=w^{[i]},M_2=m_2, K_2 = k_2]} \nonumber \\
&=\frac{P(K_1 = k_1)P(K_2 = k_2)P(W^{[i]}=w^{[i]},X_{1i}=x_{1i},X_{2i}=x_{2i})
}{P(K_2 = k_2)P(W^{[i]}=w^{[i]},X_{1i}=x_{1i},X_{2i}=x_{2i})    } \nonumber \\
& \hspace{0.75in}\times  \frac{\sum_{x_1^{[i]}}
  P(X_1^{[i]}=x_1^{[i]}|W^{[i]}=w^{[i]})
  P(M_1=m_1|X_1^n=x_1^n,K_1 = k_1)}{\sum_{x_2^{[i]}}
  P(X_2^{[i]}=x_2^{[i]}|W^{[i]}=w^{[i]})P(M_2=m_2|X_2^n=x_2^n, K_2 = k_2)} \nonumber \\
  & \hspace{1.5in}\times {\sum_{x_2^{[i]}}
  P(X_2^{[i]}=x_2^{[i]}|W^{[i]}=w^{[i]})
P(M_2=m_2|X_2^n=x_2^n, K_2 = k_2)} \nonumber \\
% &=\frac{\sum_{x_1^{[i]}}
%   P(X_1^{[i]}=x_1^{[i]}|W^{[i]}=w^{[i]})   P(M_1=m_1|X^n=x^n, K_1 = k_1) }{ \sum_{x_2^{[i]}}
% P(X_2^{[i]}=x_2^{[i]}|W^{[i]}=w^{[i]}) P(M_2=m_2|X_2^n=x_2^n, K_2 = k_2)} \nonumber \\ 
% & \hspace{2in}\times\sum_{x_2^{[i]}} P(X_2^{[i]}=x_2^{[i]}|W^{[i]}=w^{[i]})
%  P(M_2=m_2|X_2^n=x_2^n, K_2 = k_2) \nonumber \\
&=P(K_1 = k_1)\left( \sum_{x_1^{[i]}}
  P(X_1^{[i]}=x_1^{[i]}|W^{[i]}=w^{[i]})   P(M_1=m_1|X_1^n=x_1^n, K_1 = k_1) \right). \nonumber
\end{align}
Note that the right hand side in the above simplification does not depend on
$(x_{2i},m_2, k_2)$. Hence we have shown $(M_1,K_1,W^{i-1}) - (X_{1i},W^{n\backslash i}) -
(X_{2i},W^{n\backslash i},M_2, K_2)$. Similarly, using identical arguments, we can show 
$(M_2,K_2,W^{i-1}) - (X_{2i},W^{n\backslash i}) -(X_{1i},W^{n\backslash i},M_1, K_1)$. These imply that 
$U_i - (X_{1i}Q_i) - (X_{2i}Q_i) - V_i$ for all $i=1,2,\ldots,n$. 
% This is the condition (b) in Definition \ref{def:bt_outer_1}. 

To prove the next Markov Chain  $(c) $ given by  $  (X_1, X_2, Q) - (J,U,V) - Y$, consider the following arguments: Since $J$ is uniform and independent of the sources, the Markov chain $(c)$ is equivalent to  $ (X_{1i},X_{2i}, W^{n \backslash i}) - (W^{i-1},M_1, K_1,  M_2, K_2) - Y_i$ for all $ i \in [n]$. We prove this using the following. For an arbitrary $i \in [n]$ and for $m_j \in [2^{nR_j}], k_j \in [2^{nC_j}] :j=1,2$, 
$x_{1i} \in \CalX_1$, $x_{2i} \in \CalX_2$, $y_i \in \CalY,$   $w^{[i]} \deq w^{n\backslash i} \in \CalW^{n-1}$, we have  
\begin{align}
&P(X_{1i}=x_{1i},X_{2i}=x_{2i},W^{[i]}=w^{[i]}|M_1=m_1,M_2=m_2,K_1 = k_1, K_2 = k_2, W^{i-1}=w^{i-1},\! Y_i = y_i) \nonumber \\
&=\frac{P(M_1=m_1,M_2=m_2, K_1 = k_1, K_2 = k_2,W^{[i]}=w^{[i]},X_{1i}=x_{1i},X_{2i}=x_{2i},Y_i = y_i)}{P(M_1=m_1,M_2=m_2, K_1 = k_1, K_2 = k_2,W^{i-1}=w^{i-1},Y_i = y_i)} \nonumber \\
&= \frac{\bigg[\sum_{x_1^{[i]},x_2^{[i]}, w_i} P(W^n=w^n,X_1^n=x^n,X_2^n=x_2^n)}{\bigg[\sum_{x_1^{n},x_2^{n}, w_i^n} P(W^n=w^n,X_1^n=x^n,X_2^n=x_2^n)}\nonumber \\
& \hspace{1.2in} \times \frac{P(M_1=m_1|X_1^n=x^n, K_1 = k_1)P(M_2=m_2|X_2^n=x_2^n, K_2 = k_2)\bigg]} {P(M_1=m_1|X_1^n=x^n, K_1 = k_1)P(M_2=m_2|X_2^n=x_2^n, K_2 = k_2) \bigg]}\nonumber \\
& \hspace{2.4in} \times\frac{\sum_{y^{[i]}} P(Y^n = y^n| M_1 = m_1, M_2 = m_2, K_1 = k_1, K_2= k_2)}{\sum_{y^{[i]}} P(Y^n = y^n| M_1 = m_1, M_2 = m_2, K_1 = k_1, K_2= k_2)} \nonumber \\
%%%%%%%%%%%%%%%%%%% new line %%%%%%%%%%%%%%%%%
& = \frac{P(X_{1i} = x_{1i}, X_{2i} = x_{2i}, W^{[i]} = w^{[i]}, M_1 = m_1, M_2 = m_2 | K_1 = k_1, K_2 = k_2)}{P(W^{i-1} = w^{i-1}, M_1 = m_1, M_2 = m_2 | K_1 = k_1, K_2 = k_2)}\nonumber.
\end{align}
Since the right hand side of the above simplification is independent of $y_i,$ we therefore have the Markov chain $(c)$ to be satisfied. Progressing ahead, we have the Markov chain $(e)$ given by $X_{1J} -W_J - X_{2J}$ which is satisfied from the choice of $P_{W|X_1,X_2}$ similar to the arguments made in showing the Markov chain $(a)$. 
% Since $(X^n,Y^n,W^n)$ is memoryless, we have the second
% condition as well: $Q_i$ is independent of $(X_i,X_{2i})$. 
% Note that since the pair $(\hat{X}_i,\hat{Y}_i)$ is a
% function of the pair $(M_1,M_2)$, it is also a function of the pair
% $(U_i,V_i,Q_i)$. Hence condition (iii) is also satisfied. Condition (iv) is satisfied
% by the property of the assumed transmission system, i.e.,
% \[
% n(D_1+\e) \geq \sum_{i=1}^n \mathbb{E} d_1(X_i,\hat{X}_i), \ \ \ \mbox{and} 
% \ \ n(D_2+\e) \geq \sum_{i=1}^n \mathbb{E} d_2(Y_i,\hat{Y}_i).
% \]
% The condition (v) of Definition \ref{def:bt_outer_2} 
% is satisfied by choice of $P_{W|XY}$ made in the beginning.
Finally, toward showing the Markov chain $(f)$ given by $W - (X_1,X_2) - (J,Q,U,V,Y)$ consider the following set of arguments:
% Again using the fact that $J$ is uniform and independent of the sources, the Markov chain $(e)$ is equivalent to  $W_i - (X_{1i},X_{2i}) - (W^{n\backslash i},M_1, K_1,  M_2, K_2,Y_i)$ for all $ i \in [n]$. We show this using the following. 
For an arbitrary $i \in [n]$ and for $m_j \in [2^{nR_j}], k_j \in [2^{nC_j}] :j=1,2$, 
$x_{1i} \in \CalX_1$, $x_{2i} \in \CalX_2$, $y_i \in \CalY,$   $w^{[i]} \deq w^{n\backslash i} \in \CalW^{n-1}$, we have  
\begin{align}
&P(W_i = w_i|M_1=m_1,M_2=m_2,K_1 = k_1, K_2 = k_2, W^{[i]}=w^{[i]}, Y_i = y_i,X_{1i} = x_{1i}, X_{2i} = x_{2i}, J=i) \nonumber \\
&=\frac{P(M_1=m_1,M_2=m_2,K_1 = k_1, K_2 = k_2,W^{n}=w^{n},X_{1i}=x_{1i},X_{2i}=x_{2i},Y_i = y_i,J=i)}{P(M_1=m_1,M_2=m_2,K_1 = k_1, K_2 = k_2, W^{[i]}=w^{[i]}, Y_i = y_i,X_{1i} = x_{1i}, X_{2i} = x_{2i},J=i)} \nonumber \\
%%%%%%%%%%%%%%%%%%% new line %%%%%%%%%%%%%%%%%
&= \frac{\bigg[\sum_{x_1^{[i]},x_2^{[i]}} P(W^n=w^n,X_1^n=x^n,X_2^n=x_2^n, J=i)}{\bigg[\sum_{x_1^{[i]},x_2^{[i]}} P(W^{[i]}=w^{[i]},X_1^n=x^n,X_2^n=x_2^n, J=i)}\nonumber \\
& \hspace{1.6in} \times \frac{P(M_1=m_1|X_1^n=x^n, K_1 = k_1)P(M_2=m_2|X_2^n=x_2^n, K_2 = k_2)\bigg]} {P(M_1=m_1|X_1^n=x^n, K_1 = k_1)P(M_2=m_2|X_2^n=x_2^n, K_2 = k_2) \bigg]}\nonumber 
\end{align}
\begin{align}
%%%%%%%%%%%%%%%%%%% new line %%%%%%%%%%%%%%%%%
&= \frac{P(W_i = w_i, X_{1i} = x_{1i}, X_{2i} = x_{2i}, J=i)\bigg[\sum_{x_1^{[i]},x_2^{[i]}} P(X_1^{[i]}=x^{[i]},X_2^{[i]}=x_2^{[i]}|W^{[i]} = w^{[i]})}{P( X_{1i} = x_{1i}, X_{2i} = x_{2i}, J=i)\bigg[\sum_{x_1^{[i]},x_2^{[i]}} P(X_1^{[i]}=x^{[i]},X_2^{[i]}=x_2^{[i]}|W^{[i]}=w^{[i]})}\nonumber \\
& \hspace{1.6in} \times \frac{P(M_1=m_1|X_1^n=x^n, K_1 = k_1)P(M_2=m_2|X_2^n=x_2^n, K_2 = k_2)\bigg]} {P(M_1=m_1|X_1^n=x^n, K_1 = k_1)P(M_2=m_2|X_2^n=x_2^n, K_2 = k_2) \bigg]}\nonumber \\
&= \frac{P(W_i = w_i, X_{1i} = x_{1i}, X_{2i} = x_{2i}, J=i)}{P( X_{1i} = x_{1i}, X_{2i} = x_{2i},J=i)} =P_{W|X_1X_2}(w_i|x_{1i},x_{2i}).\nonumber
\end{align}
% \begin{align}
% &P(M_1=m_1,M_2=m_2,K_1 = k_1, K_2 = k_2, W^{[i]}=w^{[i]}, Y_i = y_i|X_{1i}=x_{1i},X_{2i}=x_{2i},W_i=w_i) \nonumber \\
% &=\frac{P(K_1 = k_1, K_2 = k_2)P(M_1=m_1,M_2=m_2,W^{[i]}=w^{[i]},X_{1i}=x_{1i},X_{2i}=x_{2i},W_i=w_i,Y_i = y_i)}{P(X_{1i}=x_{1i},X_{2i}=x_{2i},W_i=w_i)} \nonumber \\
% &= \frac{P(K_1 = k_1, K_2 = k_2)}{P(X_i=x_{1i},X_{2i}=x_{2i},W_i=w_i) } \sum_{x_1^{[i]},x_2^{[i]}}\bigg[ P(W^n=w^n,X_1^n=x^n,X_2^n=x_2^n)\nonumber \\
% & \hspace{1.2in} \times P(M_1=m_1|X_1^n=x^n, K_1 = k_1)
% P(M_2=m_2|X_2^n=x_2^n, K_2 = k_2) \bigg]\nonumber \\
% & \hspace{2.4in} \times\sum_{y^{[i]}} P(Y^n = y^n| M_1 = m_1, M_2 = m_2, K_1 = k_1, K_2= k_2) \nonumber \\
% %%%%%%%%%%%%%%%%%%% new line %%%%%%%%%%%%%%%%%
% &= {P(K_1 = k_1, K_2 = k_2)} \sum_{x_1^{[i]},x_2^{[i]}}\bigg[ P(W^{[i]}=w^{[i]},X^{[i]}=x^{[i]},X_2^{[i]}=x_2^{[i]})\nonumber \\
% & \hspace{1in} \times P(M_1=m_1|X_1^n=x^n, K_1 = k_1)
% P(M_2=m_2|X_2^n=x_2^n, K_2 = k_2) \bigg]\nonumber \\
% & \hspace{2.4in} \times \sum_{y^{[i]}} P(Y^n = y^n| M_1 = m_1, M_2 = m_2, K_1 = k_1, K_2= k_2). \nonumber
% % &=\sum_{x^{[i]},x_2^{[i]}} P(W^{[i]}=w^{[i]},X^{[i]}=x^{[i]},X_2^{[i]}=x_2^{[i]}) P(M_1=m_1|X_1^n=x^n)
% % P(M_2=m_2|X_2^n=x_2^n) \nonumber 
% \end{align}
We now have the right hand side of the above simplification independent of $ j, w^{[i]}, m_1, m_2, k_1, k_2,$ and $y_i$, which proves that the Markov Chain $(f)$ is satisfied.

We have shown that $(R_1,R_2,C_1,C_2)$ belongs to $\mathcal{R}_O(P_{\underline{X}Y},\epsilon)$ for all $\epsilon>0$, which is the desired proof of the outer bound.

% where 
% $\mathcal{R}_O(P_{\underline{X}Y},\epsilon)$ is defined similar to 
% $\mathcal{R}_O(P_{\underline{X}Y})$, where $R_i$ is replaced with $R_i + \epsilon$ and $ C_1 + C_2 $ is replaced with $C_1 + C_2 + g_c(\epsilon)$ and $\mathcal{P}_F$ is replaced with $\mathcal{P}_F(\epsilon)$ which is defined below:
% Let $\mathcal{P}_F(\epsilon)$ denote the collection of conditional PMFs $\tilde{P}_{JQUVY|X_1X_2}$ defined
% on $\CalJ \times \CalQ \times \CalU \times \CalV  \times {\CalY}$ such that the following conditions are satisfied: (a) $(Q,J)$ is independent of $(X_1,X_2)$,  (b) $U -(X_1, Q, J)-(X_2, Q, J)-V$,  (c) $(X_1,X_2,Q) - (J,U,V) - Y$, and (f) $\|P_{\underline{X}Y} - P_{\underline{X}}\tilde{P}_{Y|\underline{X}} \|_1 \leq \epsilon$, 
% where $\CalJ, \CalQ, \CalU$ and $\CalV$ are finite sets. 
% Using the continuity of $\mathcal{R}_O(P_{\underline{X}Y},\epsilon)$ as a function of $\epsilon$
% we get the desired proof of the outer bound.

\section{Proof of Propositions}
%%%%%%%%%%%%%%%%%%%%%%%%%%%%%%%%%%%%%%%%%%%%%%%%%

\subsection{Proof of Proposition \ref{propSI:S_tilde}}
\label{appx:propSI:S_tilde}
We begin by using the lower bound from (\ref{eq:chernoff1_Z}) given in Appendix \ref{AppSec:ProofOfLemma_PMF}.  If $\tilde{R} > I(X;W) + 4\delta_1$, we have
% an upper bound as
\begin{align}
\sum_{l =1}^{2^{n\tilde{R}}}  E^{(\mu_1)}_{L|X^n}(l|x^n) & =  \frac{1}{2^{n\tilde{R}}} \left(\frac{1-\epsilon}{1+\eta}\right) \sum_{\substack{w^n \in T_{\delta}(W|x^n)}} \sum_{l=1}^{2^{n\tilde{R}}} \mathbbm{1}_{\{\mathtt{w}^n(l,\mu_1) = w^n\}} \frac{P^{n}_{X|W}(x^n|w^n)}{P^{n}_X(x^n)} \nonumber \\
%%%%%%%%%%%%%%new line %%%%%%%%%%%%%
& =  \left(\frac{1}{1+\eta}\right)\frac{1}{P^{n}_X(x^n)}  \frac{1}{2^{n\tilde{R}}}\sum_{l=1}^{2^{n\tilde{R}}} Z_l^{(\mu)}(x^n)\nonumber \\
%%%%%%%%%%%%%%new line %%%%%%%%%%%%%
& \stackrel{w.h.p}\geq \left(\frac{1}{1+\eta}\right)\frac{1}{P^{n}_X(x^n)}(1-\eta)\EE[Z^{(\mu)}(x^n)] \nonumber \\
%%%%%%%%%%%%%%new line %%%%%%%%%%%%%
& \geq \left(\frac{1-\eta}{1+\eta}\right)(1-\epsilon_c) ,\label{eq:appxDA_result1}
\end{align}
where the second equality follows from the definition of $Z_l^{(\mu)}(x^n)$ as defined in \eqref{def:ZtZ}, the first inequality uses the lower bound  from (\ref{eq:chernoff1_Z}) which is true with probability greater than $1-\delta_{\tau}$, where $ \delta_{\tau} \deq 2\exp{\bigg(-\frac{\eta^2 2^{n(\tilde{R} - I(X,W)-4\delta_1)} }{4 \ln{2}}\bigg)} $, and the second inequality uses the fact that $\EE[Z^{(\mu)}(x^n)] = P^{n}_X(x^n)\sum_{\substack{w^n \in T_{\delta}(W|x^n)}}P^{n}_{W|X}(w^n|x^n) \geq P^{n}_X(x^n)(1-\epsilon_c)$, for sufficiently large $n$ and $\epsilon_c(\delta) \searrow 0$ as $\delta \searrow 0.$ 
% %%%%%%%%%%%%%%new line %%%%%%%%%%%%%
% & \geq 2^{-5n\delta_1}\left(\frac{1-\eta}{1+\eta}\right) \\ 
% %%%%%%%%%%%%%%new line %%%%%%%%%%%%%
% \geq \frac{1-\eta}{1+\eta} \mbox{ with high probability ,} \nonumber \\ \implies
%     1-&\sum_{l =1}^{2^{n\tilde{R}}}  E^{(\mu_1)}_{L|X^n}(l|x^n)  \leq \frac{2\eta}{1+\eta} \mbox{ whp.} \nonumber
% \end{align*}
% \begin{align}
% &\sum_{m=1}^{2^{nR}}p^{(\mu)}_{M|X^{n}}(m|x^{n}) \geq \frac{1-\eta}{1+\eta} \mbox{ with high probability (whp),} \nonumber \\ \implies
%     1-&\sum_{m=1}^{2^{nR}}p^{(\mu)}_{M|X^{n}}(m|x^{n}) \leq \frac{2\eta}{1+\eta} \mbox{ whp.} \nonumber
% \end{align}
Using this we get, with high probability,
\begin{align}
    \sum_{y^n,z^n} \widetilde{S}\cdot\11_{\mathtt{PMF}(\mathcal{C})} & \leq \frac{2\eta + \epsilon_c(1-\eta)}{1+\eta} \sum_{y^n,z^n }P^{n}_{Z|X}(z^n|x^n)P^{n}_{Y|WZ}(y^{n}|w_0,z^n) \nonumber \\
    & \leq \frac{2\eta + \epsilon_c(1-\eta)}{1+\eta} \sum_{y^n,z^n }P^{n}_{YZ|WX}(y^{n},z^n|w_0,x^n)  = \frac{2\eta + \epsilon_c(1-\eta)}{1+\eta}  \nonumber
\end{align}
where the first equality follows by using the Markov Chains $Z-X-W$ and $X - (W,Z)-Y$. Finally, since $\widetilde{S}\cdot\11_{\mathtt{PMF}(\mathcal{C})} \leq 1,$ using the above result, we have for sufficiently large $n$,
\begin{align}\label{eq:SI_ExpecSTilde_Small}
    \EE\left[\sum_{y^n,z^n} \widetilde{S}\cdot\11_{\mathtt{PMF}(\mathcal{C})} \right] \leq \frac{2\eta + \epsilon_c(1-\eta)}{1+\eta}(1-\delta_\tau) + \delta_\tau.
\end{align}
Therefore $\EE\left[\sum_{x^n \in T_\delta(X)}P^n_X(x^n)\sum_{y^n,z^n} \widetilde{S}\cdot\11_{\mathtt{PMF}(\mathcal{C})} \right]$ can be made arbitrarily small for all sufficiently large $n$. 
% Again, \cite{arxiv_CurrentPaper} contains the detailed bounding expressions used.

%%%%%%%%%%%%%%%%%%%%%%%%%%%%%%%%%%%%%%%%%%%%%%%%%

\subsection{Proof of Proposition \ref{propSI:Lemma for S_2}}
\label{appx:propSI:Lemma for S_2}
The term  $\Expectation\left[\sum_{x^n\in T_{\delta}(X)}P_X^n(x^n)\sum_{ y^{n},z^{n}}S_2\right]$ captures the binning error in terms of total variation.
% and can be bounded once we define $\tilde{w}^n =  f^{(\mu)}(b^{(\mu)}(\mathcal{I}_{\mathcal{C}}^{(\mu)}(l)),z^n )$ and bound
If we let $\tilde{w}^n =  f^{(\mu)}(b^{(\mu)}(l),z^n )$, we have
% Let $\tilde{w}^n = f^{(\mu)}(b^{(\mu)}(\mathcal{I}_{\mathcal{C}}^{(\mu)}(l)),z^n )$. This gives the second term in (\ref{eq:TVSimplify1}) as 
\begin{align}
\label{eq:PMFdiff}
    \sum_{y^n}\!\Bigg|&\left(P^{n}_{Y|WZ}(y^{n}|w^n,z^n) - P^{n}_{Y|WZ}(y^{n}|\tilde{w}^n,z^n)\right)\Bigg| 
    \leq 2 \cdot\mathbbm{1}_{\{w^n \neq \tilde{w}^n\}}.
    % =&  \sum_{\substack{\tilde{w}^n: \\ w^n \neq \tilde{w}^n}} \sum_{m=1}^{2^{nR}} \sum_{l'=1}^{2^n{\tilde{R}}} 
    % % \mathbbm{1}_{\{(w^n,z^n) \in T_{\delta}(W,Z)\}}
    % \mathbbm{1}_{\{(\tilde{w}^n,z^n) \in T_{\delta}(W,Z)\}} 
    % % & \leq 2 \cdot\mathbbm{1}_{\{w^n \neq \tilde{w}^n\}}  \!=\!\!\! \!\!\sum_{\substack{\tilde{w}^n: \\ w^n \neq \tilde{w}^n}}\!\!\! \sum_{m,l'} \!\mathbbm{1}_{\{(w^n,z^n) \in T_{\delta}(W,Z),(\tilde{w}^n,z^n) \in T_{\delta}(W,Z)\}} \nonumber \\
    %  \mathbbm{1}_{\{b^{(\mu)}(l)=m)\}}\mathbbm{1}_{\{b^{(\mu)}(l')=m)\}}\mathbbm{1}_{\{\mathtt{w}(l',\mu)=\tilde{w}^n\}} 
\end{align}
Substituting (\ref{eq:PMFdiff}) in $\Expectation[\sum_{x^n\in T_{\delta}(X)}P_X^n(x^n)\sum_{ y^{n},z^{n}}S_2], $ and using union bound, we obtain $\Expectation[\sum_{x^n\in T_{\delta}(X)}P_X^n(x^n)\sum_{ y^{n},z^{n}}S_2] \leq J_1 + J_2, $ where 
\begin{align*}
J_1 & \deq  2\cdot \Expectation\Bigg[\sum_{x^n\in T_{\delta}(X)} \sum_{z^{n}}\sum_{\substack{ \mu ,l}}  \sum_{\substack{w^n \in
\\T_{\delta}(W|x^n)}} \!\!\!\!\!
\frac{ (1-\epsilon) P^{n}_{Z|X}(z^{n}|x^{n}) 
P^{n}_{X|W}(x^n|w^n)}{2^{n(\tilde{R}+C)} (1+\eta)}\mathds{1}_{\left\{ 
\substack{w^n = \mathtt{w}^n(l,\mu)}\right\}}\11_{(w^n,z^n) \notin T_{\delta}(W,Z)}\Bigg]\\
%%%%%%%%%%%%%%%%%%%% new line %%%%%%%%%%%%%%%%%%%%%%%%
J_2 & \deq 2\cdot \Expectation\Bigg[\sum_{x^n\in T_{\delta}(X)} \sum_{z^{n}}\sum_{\substack{ \mu ,l}}  \sum_{\substack{w^n \in
\\T_{\delta}(W|x^n)}} \!\!\!\!\!
\frac{ (1-\epsilon) P^{n}_{Z|X}(z^{n}|x^{n}) 
P^{n}_{X|W}(x^n|w^n)}{2^{n(\tilde{R}+C)} (1+\eta)}\mathds{1}_{\left\{ 
\substack{w^n = \mathtt{w}^n(l,\mu)}\right\}} \\ &\hspace{1.5in}\sum_{m,l'}\sum_{\substack{\tilde{w}^n: \\ w^n \neq \tilde{w}^n}}   
% \mathbbm{1}_{\{(w^n,z^n) \in T_{\delta}(W,Z)\}}
\mathbbm{1}_{\{(\tilde{w}^n,z^n) \in T_{\delta}(W,Z)\}} 
 \mathbbm{1}_{\{b^{(\mu)}(l)=m)\}}\mathbbm{1}_{\{b^{(\mu)}(l')=m)\}}\mathbbm{1}_{\{\mathtt{w}(l',\mu)=\tilde{w}^n\}}\Bigg]
\end{align*}
We begin by showing $J_1 $ can be made arbitrarily small for sufficiently large $n$. Using the fact that $\EE[\mathds{1}_{\left\{ 
\substack{w^n = \mathtt{w}^n(l,\mu)}\right\}}] = \frac{P^n_W(w^n)}{(1-\epsilon)},$ for $w^n \in T_{\bar{\delta}}(W),$ we have
\begin{align*}
J_1 & = 2 \sum_{x^n\in T_{\delta}(X)}\sum_{z^{n}} \sum_{\substack{w^n: w^n \in
\\T_{\delta}(W|x^n) \cap T_{\bar{\delta}}(W), \\(w^n,z^n) \notin T_{\delta}(W,Z) }} \!\!\!\!\!
\frac{  P^{n}_{Z|X}(z^{n}|x^{n}) 
P^{n}_{XW}(x^n,w^n)}{ (1+\eta)} 
\end{align*}
\begin{align*}
%%%%%%%%%%%%%%%%%%% new line %%%%%%%%%%%%%%%%%%%%5
& \leq \frac{2}{(1+\eta)} \sum_{x^n\in T_{\delta}(X)}\sum_{z^{n}} \sum_{\substack{w^n: w^n \in
\\T_{\delta}(W|x^n) \cap T_{\bar{\delta}}(W), \\(w^n,z^n) \notin T_{\delta}(W,Z) }} \!\!\!\!\!{  
P^{n}_{XWZ}(x^n,w^n,z^n)} \\
& \leq \frac{2}{(1+\eta)} \sum_{\substack{(w^n,z^n) \notin T_{\delta}(W,Z) }} \!\!\!\!\!{  
P^{n}_{WZ}(w^n,z^n)} \leq \frac{2\epsilon_{J_1}}{(1+\eta)},
\end{align*}
where $\epsilon_{J_1} (\delta) \searrow 0$ as $\delta\searrow 0.$
Proceeding with $J_2$, we have
\begin{align}
    J_2  &\leq 2\cdot \Expectation\Bigg[\sum_{x^n\in T_{\delta}(X)} \sum_{z^{n}}\sum_{\substack{ \mu ,l}}  \sum_{\substack{w^n \in
\\T_{\delta}(W|x^n)}} \!\!\!\!\!
\frac{ (1-\epsilon) P^{n}_{XZ}(x^n,z^{n}) 
P^{n}_{X|W}(x^n|w^n)}{2^{n(\tilde{R}+C)} (1+\eta)}\mathds{1}_{\left\{ 
\substack{w^n = \mathtt{w}^n(l,\mu)}\right\}}\nonumber \\ &\hspace{1in}\sum_{m,l'}\sum_{\substack{\tilde{w}^n: \\ w^n \neq \tilde{w}^n}}   
% \mathbbm{1}_{\{(w^n,z^n) \in T_{\delta}(W,Z)\}}
\mathbbm{1}_{\{(\tilde{w}^n,z^n) \in T_{\delta}(W,Z)\}} 
     \mathbbm{1}_{\{b^{(\mu)}(l)=m)\}}\mathbbm{1}_{\{b^{(\mu)}(l')=m)\}}\mathbbm{1}_{\{\mathtt{w}(l',\mu)=\tilde{w}^n\}}\Bigg] \nonumber 
\end{align}

\begin{align}
     & = 2\sum_{x^n\in T_{\delta}(X)} \sum_{\substack{z^{n} \in T_{\delta}(Z)}}\sum_{m=1}^{2^{nR}}  \sum_{\substack{w^n \in 
\\T_{\delta}(W|x^n)}} \!\!\!\!\!
\frac{ (1-\epsilon)P^{n}_{XZ}(x^n,z^{n}) 
P^{n}_{X|W}(x^n|w^n)}{2^{n(\tilde{R}+C)} (1+\eta)}
\nonumber \\ 
%%%%%%%%%%%%%new line %%%%%%%%%%%
& \hspace{1in} \sum_{\substack{\tilde{w}^n: (\tilde{w}^n,z^n)\\\in T_{\delta}(W,Z), \\ w^n \neq \tilde{w}^n}}   
 \sum_{\substack{ \mu ,l,l'}}\Expectation\Bigg[\mathds{1}_{\left\{ \substack{ \mathtt{w}^n(l,\mu) = w^n}\right\}}  \mathbbm{1}_{\{\mathtt{w}(l',\mu)=\tilde{w}^n\}} \mathbbm{1}_{\{b^{(\mu)}(l)=m)\}} \mathbbm{1}_{\{b^{(\mu)}(l')=m)\}}\Bigg] \nonumber \\
 %%%%%%%%%%%%%new line %%%%%%%%%%%
& = 2 \sum_{x^n\in T_{\delta}(X)}\sum_{\substack{z^{n} \in T_{\delta}(Z)}}\sum_{m=1}^{2^{nR}}  \sum_{\substack{w^n \in 
\\T_{\delta}(W|x^n)}} \!\!\!\!\!
\frac{ (1-\epsilon) P^{n}_{XZ}(x^n,z^{n}) 
P^{n}_{X|W}(x^n|w^n)}{2^{n(\tilde{R}+C)} (1+\eta)} \nonumber \\
& \hspace{2in} \sum_{\substack{\tilde{w}^n: (\tilde{w}^n,z^n)\in T_{\delta}(W,Z), \\ w^n \neq \tilde{w}^n}}    \;
 \sum_{\substack{ \mu ,l,l'}}\Bigg[\frac{P_W^n(w^n)}{(1-\epsilon)}\frac{P_W^n(\tilde{w}^n)}{(1-\epsilon)}2^{-2nR}\Bigg] \nonumber \\
 %%%%%%%%%%%%%new line %%%%%%%%%%%
 & = 2 \cdot 2^{n(\tilde{R}-R)} \sum_{x^n\in T_{\delta}(X)}\sum_{\substack{z^{n} \in T_{\delta}(Z)}}  \sum_{\substack{w^n \in T_{\delta}(W|x^n)}} \!\!\!\!\!
\frac{ P^{n}_{XWZ}(x^n,w^n,z^{n}) 
}{ (1+\eta)}
\sum_{\substack{\tilde{w}^n: (\tilde{w}^n,z^n)\\\in T_{\delta}(W,Z), \\ w^n \neq \tilde{w}^n}}    
 \frac{P_W^n(\tilde{w}^n)}{(1-\epsilon)} \nonumber \\
 %%%%%%%%%%%%%new line %%%%%%%%%%%
 & \leq 2 \cdot 2^{n(\tilde{R}-R)}\sum_{x^n\in T_{\delta}(X)} \sum_{\substack{z^{n} \in T_{\delta}(Z)}}  \sum_{\substack{w^n \in T_{\delta}(W|x^n)}} \!\!\!\!\!
\frac{ P^{n}_{WZ|X}(w^n,z^{n}|x^{n}) 
}{ (1+\eta)(1-\epsilon)}
 2^{-n(I(W;Z)-\delta_I)} \nonumber \\
 & \leq  2^{n(\tilde{R}-R-I(W;Z)+\delta_I+\delta')}  \nonumber
\end{align}
where the second equality follows by using $\EE[\11_{\{b^{(\mu)}(l) = m\}}] = 2^{-nR}$ , the third equality follows from the Markov Chain $Z-X-W,$ the second inequality follows from the properties of $\delta$-typical sets where $\delta_I (\delta) \searrow 0$ as $\delta \searrow 0.$ Therefore, from above $\EE[\sum_{y^{n},z^{n}}S_2] $ can be made arbitrarily small, for sufficiently large $n$, if $\tilde{R} - R \leq I(W;Z) + \epsilon_1$, where $\epsilon_1 = \delta_I + \delta'. $

%%%%%%%%%%%%%%%%%%%%%%%%%%%%%%%%%%%%%%%%%%%%%%%%%

\subsection{Proof of Proposition \ref{prop:S_tilde}}
\label{appx:prop:S_tilde}
% Now, we are left to prove the terms $S_2, S_3$ and $S_4$ are small. For this, we use the constraints on the rates $\tilde{R}_1$ and $\tilde{R}_2$ provided in lemma (\ref{lem:2PMFWHP}), and as a result prove that these terms are small with high probability.
We begin by defining $\tilde{S}_i \deq |T_i|$ for $i \in \{2,3,4\}.$ Firstly, consider the following simplification of $\tilde{S_2}$.
\begin{align}
&\sum_{x_1^n,x_2^n \in T_{\delta}(\ulineX)}\sum_{y^n}P^{n}_{X_1 X_2}(x_1^{n},x_2^{n})\tilde{S}_{2} \nonumber \\
%%%%%%%%%%%%%%%%%%%%% new line %%%%%%%%%%%%%%%%
& \hspace{20pt}= \sum_{x_1^n,x_2^n \in T_{\delta}(\ulineX)}\sum_{y^n}\Bigg|\sum_{\substack{ \mu_1,\mu_2,l_2 }}\sum_{\substack{w_2^n \in \\T_{\delta}(W_2|x_2^n)}}\frac{P^{n}_{X_1 X_2}(x_1^{n},x_2^{n}) \left[1-\sum_{m_1=1}^{2^{n\tilde{R}_1}}E^{(\mu_1)}_{L_1|X_1^{n}}(l_1|x_1^{n}
)\right]}{2^{n(C_1+C_2)}}\nonumber\\
& \hspace{2.5in}\frac{(1-\epsilon_2) 
P^{n}_{X_2|W_2}(x_2^n|w_2^n)  }{2^{n\tilde{R}_2}(1+\eta)P^{n}_{X_2}(x_2^n)} \mathds{1}_{\left\{  w_2^n = \mathtt{w}_2^n(l_2,\mu_2) 
\right\}}P^{n}_{Y|W_1W_2}(y^{n}|\tilde{w}^n_1,\tilde{w}^n_2)\Bigg|\nonumber\\
%%%%%%%%%%%%%%%%%%%%% new line %%%%%%%%%%%%%%%%
&\hspace{20pt}= \sum_{x_1^n,x_2^n \in T_{\delta}(\ulineX)}\sum_{y^n}\sum_{\substack{ \mu_1,\mu_2,l_2 }}\sum_{\substack{w_2^n \in \\T_{\delta}(W_2|x_2^n)}}\frac{P^{n}_{X_1 X_2}(x_1^{n},x_2^{n}) \left|1-\sum_{m_1=1}^{2^{n\tilde{R}_1}}E^{(\mu_1)}_{L_1|X_1^{n}}(l_1|x_1^{n}
)\right|}{2^{n(C_1+C_2)}}\nonumber\\
&\hspace{2.5in}\frac{(1-\epsilon_2) 
P^{n}_{X_2|W_2}(x_2^n|w_2^n)  }{2^{n\tilde{R}_2}(1+\eta)P^{n}_{X_2}(x_2^n)} \mathds{1}_{\left\{  w_2^n = \mathtt{w}_2^n(l_2,\mu_2) \right\}} P^{n}_{Y|W_1W_2}(y^{n}|\tilde{w}^n_1,\tilde{w}^n_2)\nonumber
\end{align}

Taking expectation over the second encoder's codebook, we obtain
\begin{align}
    \EE_{\CalC_2}&\left[\sum_{x_1^n,x_2^n \in T_{\delta}(\ulineX)}\sum_{y^n}P^{n}_{X_1 X_2}(x_1^{n},x_2^{n})\tilde{S}_{2} \right] \nonumber \\
    %%%%%%%%%%%%%%%%%%%%% new line %%%%%%%%%%%%%%%%
    & \leq   \sum_{x_1^n,x_2^n \in T_{\delta}(\ulineX)}\sum_{y^n}\!\sum_{\mu_1,\mu_2}\!\!\!\sum_{\substack{w_2^n \in \\T_{\delta}(W_2|x_2^n)}}\!\!\!\frac{ \left|1-\sum_{l_1=1}^{2^{n\tilde{R}_1}}E^{(\mu_1)}_{L_1|X_1^{n}}(l_1|x_1^{n}
    )\right|}{2^{n(C_1+C_2)}(1+\eta)}P^{n}_{X_1,X_2,W_2}(x_1^n,x_2^n,w_2^n)  P^{n}_{Y|W_1W_2}(y^{n}|\tilde{w}^n_1,\tilde{w}^n_2)\nonumber\\
    %%%%%%%%%%%%%%%%%%%%% new line %%%%%%%%%%%%%%%%
    & \leq  \sum_{x_1^n,x_2^n \in T_{\delta}(\ulineX)}\sum_{w_2^n}\sum_{\mu_1,\mu_2}\frac{ \left|1-\sum_{l_1=1}^{2^{n\tilde{R}_1}}E^{(\mu_1)}_{L_1|X_1^{n}}(l_1|x_1^{n}
    )\right|}{2^{n(C_1+C_2)}(1+\eta)}P^{n}_{X_1,X_2,W_2}(x_1^n,x_2^n,w_2^n)  \nonumber\\
    %%%%%%%%%%%%%%%%%%%%% new line %%%%%%%%%%%%%%%%
    & \leq \frac{1}{2^{n(C_1+C_2)}}\sum_{\mu_1,\mu_2}\sum_{x_1^n \in T_{\delta}(X_1)}P^{n}_{X_1}(x_1^n)\frac{ \left|1-\sum_{l_1=1}^{2^{n\tilde{R}_1}}E^{(\mu_1)}_{L_1|X_1^{n}}(l_1|x_1^{n}
    )\right|}{(1+\eta)}.
\end{align}
Further, taking expectation over the first encoder's codebook and introducing the indicator $\11_{\mathtt{PMF}(\CalC_1,\CalC_2)}$, we get
\begin{align*}
 \EE\left[\sum_{x_1^n,x_2^n \in T_{\delta}(\ulineX)}\!\!\sum_{y^n}P^{n}_{X_1 X_2}(x_1^{n},x_2^{n})\tilde{S}_{2}\cdot \11_{\mathtt{PMF}(\CalC_1,\CalC_2)}\right] & \leq \EE_{\CalC_1}\!\!\left[\EE_{\CalC_2}\!\!\left[\sum_{x_1^n,x_2^n \in T_{\delta}(\ulineX)}\!\!\sum_{y^n}P^{n}_{X_1 X_2}(x_1^{n},x_2^{n})\tilde{S}_{2} \right]\!\!\11_{\mathtt{PMF}(\CalC_1)}\right] \nonumber \\
    & \leq \frac{2\eta + \epsilon_c(1-\eta)}{1+\eta}(1-\delta_\tau) + \delta_\tau. \nonumber
\end{align*}
where the last inequality follows using the result from \eqref{eq:SI_ExpecSTilde_Small} provided in Appendix \ref{appx:propSI:S_tilde}, and $\delta_\tau(\delta) \searrow 0$ as $\delta \searrow 0$ if $\tilde{R}_1 \geq I(X_1;W_1) +4\delta_1$,
% , then $\left|1-\sum_{l_1=1}^{2^{n\tilde{R}_1}}E^{(\mu_1)}_{L_1|X_1^{n}}(l_1|x_1^{n}  )\right| \leq \frac{2\eta + \epsilon_c(1-\eta)}{1+\eta} \mbox{ w.h.p.}$,
    for all sufficiently large $n$, where $\delta_1 \searrow 0$,  $\epsilon_c(\delta) \searrow 0$ as $\delta \searrow 0$. 
    
%     This implies
% \begin{align}
%     \EE\left[\sum_{x_1^n,x_2^n \in T_{\delta}(\ulineX)}\sum_{y^n}P^{n}_{X_1 X_2}(x_1^{n},x_2^{n})\tilde{S}_{2}\right] & \leq \frac{2\eta + \epsilon_c(1-\eta)}{1+\eta}\sum_{x_1^n \in T_{\delta}(X_1)}P^{n}_{X_1}(x_1^n) 
%     % + \frac{1}{1+\eta} \sum_{x_1^n\notin T_{\delta}(X_1)}P^{n}_{X_1}(x_1^n) 
%     \leq \frac{2\eta + \epsilon_c(1-\eta)}{1+\eta} \nonumber
% \end{align}
% Therefore, if $\tilde{R}_1 \geq I(X_1;W_1) +4\delta$, then $\EE\left[\sum_{x_1^n,x_2^n \in T_{\delta}(\ulineX)}\sum_{y^n}P^{n}_{X_1 X_2}(x_1^{n},x_2^{n})\tilde{S}_{2}\right]$ can be made arbitrarily small with high probability. 
Using very similar arguments as above, it can also be shown that if $\tilde{R}_2 \geq I(X_2;W_2) +4\delta_1$, then $ \EE\left[\sum_{x_1^n,x_2^n \in T_{\delta}(\ulineX)}\sum_{y^n}P^{n}_{X_1 X_2}(x_1^{n},x_2^{n})\tilde{S}_{3}\cdot \11_{\mathtt{PMF}(\CalC_1,\CalC_2)}\right] 
$ 
% and hence $\EE\left[\sum_{x_1^n,x_2^n \in T_{\delta}(\ulineX)}\sum_{y^n}P^{n}_{X_1 X_2}(x_1^{n},x_2^{n})\ S_3\right]$ 
can be made arbitrarily small  for all sufficiently large $n$. 

Similarly consider the final term corresponding to $\tilde{S}_4$. For $\tilde{R}_1$ and $\tilde{R}_2$ satisfying the above constraints, i.e., $\tilde{R}_1 \geq I(X_1;W_1) +4\delta$ and $\tilde{R}_2 \geq I(X_2;W_2) +4\delta$, we will have
\begin{align}
    &\EE\left[\sum_{x_1^n,x_2^n \in T_{\delta}(\ulineX)}\!\!\sum_{y^n}P^{n}_{X_1 X_2}(x_1^{n},x_2^{n})\tilde{S}_4\cdot \11_{\mathtt{PMF}(\CalC_1,\CalC_2)} \right] \nonumber \\
    %%%%%%%%%%%%%%%%%%%%% new line %%%%%%%%%%%%%%%%
    & \leq \EE\left[\sum_{x_1^n,x_2^n \in T_{\delta}(\ulineX)}\sum_{\substack{ \mu_1,\mu_2 }}\frac{P^{n}_{X_1 X_2}(x_1^{n},x_2^{n}) \left|1-\sum_{l_1=1}^{2^{n\tilde{R}_1}}E^{(\mu_1)}_{L_1|X_1^{n}}(l_1|x_1^{n}
    )\right|\left|1-\sum_{l_2=2}^{2^{n\tilde{R}_2}}E^{(\mu_2)}_{L_2|X_2^{n}}(l_2|x_2^{n})\right|}{2^{nC}}\11_{\mathtt{PMF}(\CalC_1,\CalC_2)}\right]\nonumber \\
    %%%%%%%%%%%%%%%%%%%%% new line %%%%%%%%%%%%%%%%
    % &\leq \sum_{\substack{x_1^n \in T_{\delta}(X_1)\\x^n_2\in T_{\delta}(X_2)}}\sum_{\substack{ \mu \in [2^{nC}] }}\frac{P^{n}_{X_1 X_2}(x_1^{n},x_2^{n}) }{2^{nC}} \left(\frac{2\eta + \epsilon_c(1-\eta)}{1+\eta}\right)^2 
    &\leq \left[\left(\frac{2\eta + \epsilon_c(1-\eta)}{1+\eta}\right)^2(1-2\delta_\tau) + 2\delta_\tau\right], \nonumber 
    % + \sum_{\substack{x_1^n \in T_{\delta}(X_1)\\x^n_2\notin T_{\delta}(X_2)}}\sum_{\substack{ \mu \in [2^{nC}] }}\frac{P^{n}_{X_1 X_2}(x_1^{n},x_2^{n}) 2\eta}{2^{nC}(1+\eta)} \nonumber \\
    % & \;\;\; + \sum_{\substack{x_1^n \notin T_{\delta}(X_1)\\x^n_2\in T_{\delta}(X_2)}}\sum_{\substack{ \mu \in [2^{nC}] }}\frac{P^{n}_{X_1 X_2}(x_1^{n},x_2^{n}) 2\eta}{2^{nC}(1+\eta)} + \sum_{\substack{x_1^n \notin T_{\delta}(X_1)\\x^n_2\notin T_{\delta}(X_2)}}\sum_{\substack{ \mu \in [2^{nC}] }}\frac{P^{n}_{X_1 X_2}(x_1^{n},x_2^{n})}{2^{nC}} \nonumber \\
    % &\leq \frac{ 4\eta^2}{(1+\eta)^2} + \sum_{x^n_2\notin T_{\delta}(X_2)}P^{n}_{X_2}(x^n_2)\sum_{x^n_1}\frac{P^{n}_{X_1| X_2}(x_1^{n}|x_2^{n}) 2\eta}{(1+\eta)} \nonumber \\
    % & \;\;\;  + \sum_{x^n_1\notin T_{\delta}(X_1)}P^{n}_{X_1}(x^n_1)\sum_{x^n_2}\frac{P^{n}_{X_2| X_1}(x_2^{n}|x_1^{n}) 2\eta}{(1+\eta)} + \sum_{\substack{x_1^n \notin T_{\delta}(X_1), x^n_2}}{P^{n}_{X_1 X_2}(x_1^{n},x_2^{n})} \nonumber\\
    %  &\leq \frac{ 4\eta^2}{(1+\eta)^2} + \frac{4\eta\epsilon}{(1+\eta)} + \epsilon\nonumber 
\end{align}
where the second inequality again uses the result from Appendix \ref{appx:propSI:S_tilde}. This completes the proof.

%%%%%%%%%%%%%%%%%%%%%%%%%%%%%%%%%%%%%%%%%%%%%%%%%%%%%%%%%%%%%%%%%%%%%%%%%%%%%%%%%%%%%%%%%%%%%%%%%%%%%%%%%%%%%%%%%%%%%%%%%%%%%%%%%%%%%%%%%%%%%%%%%%%%%%%%%%%%%%
\subsection{Proof of Proposition \ref{prop:Lemma for S_2}}
\label{appx:prop:Lemma for S_2}
Define $(\hat{w}^n_1,\hat{w}^n_2) = f^{(\mu)}(b^{(\mu_1)}(l_1),b^{(\mu_2)}(l_2))$. Consider,
\begin{align}
    \Bigg|&P^{n}_{Y|W_1W_2}(y^{n}|w_1^n,w^n_2) - P^{n}_{Y|W_1W_2}(y^{n}|f^{(\mu)}(b_1^{(\mu_1)}(l_1),b_2^{(\mu_2)}(l_2))) \Bigg| \leq 2 \cdot \mathbbm{1}_{\{(w_1^n,w^n_2) \neq (\hat{w}^n_1,\hat{w}^n_2)\}}. \nonumber 
% & =  2 \sum_{\substack{\hat{w}^n_1,\hat{w}^n_2: \\  (\hat{w}^n_1,\hat{w}^n_2) \neq (w_1^n,w_2^n) }} \sum_{m_1,m_2} \sum_{l'_1,l'_2} 
%     \mathbbm{1}_{\{(\hat{w}^n_1,\hat{w}^n_2) \in T_{\delta}(W_1,W_2)\}}      \mathbbm{1}_{\{b_1^{(\mu_1)}(\mathcal{I}_{\mathcal{C}_1}^{(\mu_1)}(l_1))=m_1)\}}\mathbbm{1}_{\{b_1^{(\mu_1)}(\mathcal{I}_{\mathcal{C}_1}^{(\mu_1)}(l'_1))=m_1)\}} \nonumber \\
%     & \hspace{40pt} \mathbbm{1}_{\{b_2^{(\mu_1)}(\mathcal{I}_{\mathcal{C}_2}^{(\mu_1)}(l_2))=m_2)\}}\mathbbm{1}_{\{b_2^{(\mu_1)}(\mathcal{I}_{\mathcal{C}_2}^{(\mu_1)}(l'_2))=m_2)\}}\mathbbm{1}_{\{W_1(l'_1,\mu_1) = w'_1\}}\mathbbm{1}_{\{W_2(l'_2) = w'_2\}} \label{eq:binIndicators}
\end{align}
Substituting the above bound in the $S_{12}$ term and using the union bound, we obtain $\EE\left[\sum_{\ulinex^n \in T_\delta(\ulineX)}P^n_{\ulineX}(\ulinex^n){\sum_{y_n}S_{2}}\right]  \leq J_1 + J_2, $ where 
\begin{align}
    J_1 & \deq 2 \sum_{\ulinex^n \in T_\delta(\ulineX)} \sum_{\substack{ \mu_1,\mu_2 }} \sum_{\substack{w_1^n \in \\T_{\delta}(W_1|x_1^n)}}\sum_{\substack{w_2^n \in \\T_{\delta}(W_2|x_2^n)}}
\!\!\!\!\frac{(1-\epsilon_1)(1-\epsilon_2) P^{n}_{X_1 X_2}(x_1^{n},x_2^{n}) 
P^{n}_{X_1|W_1}(x_1^n|w_1^n) P^{n}_{X_2|W_2}(x_2^n|w_2^n) }{2^{n(\tilde{R}_1+\tilde{R}_2+C_1+C_2)}(1+\eta)^2P^{n}_{X_1}(x_1^n) P^{n}_{X_2}(x_2^n) } \nonumber \\ &
\hspace{0.5in}\sum_{l_1,l_2} \sum_{m_1,m_2} \EE\Bigg[\mathds{1}_{\{ w_2^n = \mathtt{w}_2^n(l_2,\mu_2) \}}\mathds{1}_{\{  w_1^n = \mathtt{w}_1^n(l_1,\mu_1)\}}      \mathbbm{1}_{\{({w}^n_1,{w}^n_2) \notin T_{\delta}(W_1,W_2)\}} 
   \mathbbm{1}_{\{b_1^{(\mu_1)}(l_1)=m_1)\}}  \mathbbm{1}_{\{b_2^{(\mu_1)}(l_2)=m_2)\}}\Bigg],\nonumber \\
   %%%%%%%%%%%%%%%%%%%%%%%%% new line %%%%%%%%%%%%%%%%%%%%%%%%%%%%
    J_2 & \deq 2 \sum_{\ulinex^n \in T_\delta(\ulineX)} \sum_{\substack{ \mu_1,\mu_2 }} \sum_{\substack{w_1^n \in \\T_{\delta}(W_1|x_1^n)}}\sum_{\substack{w_2^n \in \\T_{\delta}(W_2|x_2^n)}}
    \!\!\!\!\frac{(1-\epsilon_1)(1-\epsilon_2) P^{n}_{X_1 X_2}(x_1^{n},x_2^{n}) 
    P^{n}_{X_1|W_1}(x_1^n|w_1^n) P^{n}_{X_2|W_2}(x_2^n|w_2^n) }{2^{n(\tilde{R}_1+\tilde{R}_2+C_1+C_2)}(1+\eta)^2P^{n}_{X_1}(x_1^n) P^{n}_{X_2}(x_2^n) } \nonumber \\ &
    \hspace{0.7in}\sum_{l_1,l_2}\sum_{\substack{\hat{w}^n_1,\hat{w}^n_2: \\  (\hat{w}^n_1,\hat{w}^n_2) \neq (w_1^n,w_2^n) }} \sum_{m_1,m_2} \sum_{l'_1,l'_2}\EE\Bigg[\mathds{1}_{\{ w_2^n = \mathtt{w}_2^n(l_2,\mu_2) \}}\mathds{1}_{\{  w_1^n = \mathtt{w}_1^n(l_1,\mu_1)\}}      \mathbbm{1}_{\{(\hat{w}^n_1,\hat{w}^n_2) \in T_{\delta}(W_1,W_2)\}} \nonumber \\
    & \hspace{1in}\mathbbm{1}_{\{b_1^{(\mu_1)}(l_1)=m_1)\}} \mathbbm{1}_{\{b_1^{(\mu_1)}(l'_1)=m_1)\}}  \mathbbm{1}_{\{b_2^{(\mu_1)}(l_2)=m_2)\}}\mathbbm{1}_{\{b_2^{(\mu_1)}(l'_2)=m_2)\}} \mathbbm{1}_{\{\mathtt{w}_1(l'_1,\mu_1) = w'_1\}}\mathbbm{1}_{\{\mathtt{w}_2(l'_2) = w'_2\}}\Bigg].\nonumber
\end{align}
Consider the term $J_1$. This can be bounded as
\begin{align*}
    %J_1 & \leq  2 \sum_{\ulinex^n \in T_\delta(\ulineX)}  \sum_{\substack{w_1^n \in \\T_{\delta}(W_1|x_1^n)}}\sum_{\substack{w_2^n \in \\T_{\delta}(W_2|x_2^n)}}\!\!\!\!\frac{ P^{n}_{X_1 X_2W_1W_2}(x_1^{n},x_2^{n},w_1^n,w_2^n) }{(1+\eta)^2 }  \EE\Bigg[      \mathbbm{1}_{\{({w}^n_1,{w}^n_2) \notin T_{\delta}(W_1,W_2)\}} \Bigg],\nonumber \\
   J_1 & \leq 2 \sum_{\ulinex^n \in T_\delta(\ulineX)}  \sum_{(w_1^n,w_2^n) \notin T_{\delta}(W_1,W_2)}
\!\!\!\!\frac{ P^{n}_{X_1 X_2W_1W_2}(x_1^{n},x_2^{n},w_1^n,w_2^n) }{(1+\eta)^2 } \leq 2 \sum_{\substack{(w_1^n,w_2^n) \notin\\ T_{\delta}(W_1,W_2)}}
\!\!\!\!\frac{ P^{n}_{W_1W_2}(w_1^n,w_2^n) }{(1+\eta)^2 } \leq \epsilon_{J_1},
\end{align*}
where $\epsilon_{J_1}(\delta) \searrow 0$ as $\delta \searrow 0$.
Now, consider the term corresponding to $J_2$.
\begin{align}
    J_2
%     &\leq 2 \sum_{x_1^n,x_2^n} \sum_{\substack{ \mu_1,\mu_2 }} \sum_{\substack{w_1^n \in \\T_{\delta}(W_1|x_1^n)}}\sum_{\substack{w_2^n \in \\T_{\delta}(W_2|x_2^n)}}
% \!\!\!\!\frac{(1-\epsilon_1)(1-\epsilon_2) P^{n}_{X_1 X_2}(x_1^{n},x_2^{n}) 
% P^{n}_{X_1|W_1}(x_1^n|w_1^n) P^{n}_{X_2|W_2}(x_2^n|w_2^n) }{2^{n(\tilde{R}_1+\tilde{R}_2+C_1+C_2)}(1+\eta)^2P^{n}_{X_1}(x_1^n) P^{n}_{X_2}(x_2^n) } \nonumber \\ &
% \hspace{25pt}\sum_{l_1,l_2}\sum_{\substack{\hat{w}^n_1,\hat{w}^n_2: \\  (\hat{w}^n_1,\hat{w}^n_2) \neq (w_1^n,w_2^n) }} \sum_{m_1,m_2} \sum_{l'_1,l'_2}\EE\Bigg[\mathds{1}_{\{ w_2^n = \mathtt{w}_2^n(l_2,\mu_2) \}}\mathds{1}_{\{  w_1^n = \mathtt{w}_1^n(l_1,\mu_1)\}}      \mathbbm{1}_{\{(\hat{w}^n_1,\hat{w}^n_2) \in T_{\delta}(W_1,W_2)\}} \nonumber \\
%     & \hspace{25pt}\mathbbm{1}_{\{b_1^{(\mu_1)}(l_1)=m_1)\}} \mathbbm{1}_{\{b_1^{(\mu_1)}(l'_1)=m_1)\}}  \mathbbm{1}_{\{b_2^{(\mu_1)}(l_2)=m_2)\}}\mathbbm{1}_{\{b_2^{(\mu_1)}(l'_2)=m_2)\}} \mathbbm{1}_{\{\mathtt{w}_1(l'_1,\mu_1) = w'_1\}}\mathbbm{1}_{\{\mathtt{w}_2(l'_2) = w'_2\}}\Bigg]
% \nonumber \\
%%%%%%%%%%%%%%%%%%% new line %%%%%%%%%%%%%%%%%%%%%%%
& \leq 2 \cdot 2^{n(\tilde{R}_1 + \tilde{R}_2 - R_1 -R_2)} \!\!\sum_{x_1^n,x_2^n} \!\!\sum_{\substack{w_1^n \in \\T_{\delta}(W_1|x_1^n)}}\sum_{\substack{w_2^n \in \\T_{\delta}(W_2|x_2^n)}} \!\!\!
\!\!\!\!\frac{(1-\epsilon_1)(1-\epsilon_2) P^{n}_{X_1 X_2}(x_1^{n},x_2^{n}) 
 }{(1+\eta)^2P^{n}_{X_1}(x_1^n) P^{n}_{X_2}(x_2^n) } \nonumber \\ &
\hspace{0.5in}P^{n}_{X_1|W_1}(x_1^n|w_1^n) P^{n}_{X_2|W_2}(x_2^n|w_2^n)\hspace{-0.3in}\sum_{\substack{\hat{w}^n_1,\hat{w}^n_2:   (\hat{w}^n_1,\hat{w}^n_2) \neq (w_1^n,w_2^n) \\ (\hat{w}^n_1,\hat{w}^n_2) \in T_{\delta}(W_1,W_2)}}\!\!\!\! \frac{P_{W_1}^n(w^n_1)}{(1-\epsilon)}\frac{P_{W_2}^n({w}^n_2)}{(1-\epsilon)}      \frac{P_{W_1}^n(\hat{w}_1)}{(1-\epsilon)}\frac{P_{W_2}^n(\hat{w}_2)}{(1-\epsilon)}
\nonumber \\
%%%%%%%%%%%%%%%%%%% new line %%%%%%%%%%%%%%%%%%%%%%%
& \leq  \frac{2 \cdot 2^{n(\tilde{R}_1 + \tilde{R}_2 - R_1 -R_2)}}{(1-\epsilon_1)(1-\epsilon_2)(1+\eta)^2}
\sum_{x_1^n,x_2^n}  \sum_{\substack{w_1^n \in \\T_{\delta}(W_1|x_1^n)}}\sum_{\substack{w_2^n \in \\T_{\delta}(W_2|x_2^n)}} 
\!\!\!\!{ P^{n}_{X_1 X_2}(x_1^{n},x_2^{n})
P^{n}_{W_1|X_1}(w_1^n|x_1^n) P^{n}_{W_2|X_2}(w_2^n|x_2^n) } \nonumber \\ &
\hspace{3.5in}\sum_{\substack{\hat{w}^n_1,\hat{w}^n_2:   (\hat{w}^n_1,\hat{w}^n_2) \neq (w_1^n,w_2^n) \\ (\hat{w}^n_1,\hat{w}^n_2) \in T_{\delta}(W_1,W_2)}}   {P_{W_1}^n(\hat{w}_1)}{P_{W_2}^n(\hat{w}_2)}\nonumber\\
%%%%%%%%%%%%%%%%%%% new line %%%%%%%%%%%%%%%%%%%%%%%
& \leq  \frac{2 \cdot 2^{n(\tilde{R}_1 + \tilde{R}_2 - R_1 -R_2)}}{(1-\epsilon_1)(1-\epsilon_2)(1+\eta)^2}   \sum_{\substack{\\ w_1^n,w^n_2}}\!{P^{n}_{ W_1 W_2}(w_1^{n},w_2^{n}) }  \sum_{\substack{\hat{w}^n_1,\hat{w}^n_2:  (\hat{w}^n_1,\hat{w}^n_2) \neq (w_1^n,w_2^n) \\ (\hat{w}^n_1,\hat{w}^n_2) \in T_{\delta}(W_1,W_2)}}   {P_{W_1}^n(\hat{w}_1)}{P_{W_2}^n(\hat{w}_2)}\nonumber\\
%%%%%%%%%%%%%%%%%%% new line %%%%%%%%%%%%%%%%%%%%%%%
& \leq  \frac{2 \cdot 2^{n(\tilde{R}_1 + \tilde{R}_2 - R_1 -R_2)}}{(1-\epsilon_1)(1-\epsilon_2)(1+\eta)^2} 2^{-n(I(W_1;W_2) + \delta_{J}')}. 
% \sum_{\substack{\hat{w}^n_1,\hat{w}^n_2: \\  (\hat{w}^n_1,\hat{w}^n_2) \neq (w_1^n,w_2^n) \\ (\hat{w}^n_1,\hat{w}^n_2) \in T_{\delta}(W_1,W_2)}}   2^{-n(H(W_1) + H(W_2) - 2\delta)}
% \nonumber\\
% & \leq  \frac{2 \cdot 2^{n(\tilde{R}_1 + \tilde{R}_2 - R_1 -R_2 - I(W_1;W_2)-3\delta)}}{(1-\epsilon_1)(1-\epsilon_2)(1+\eta)^2}   
\end{align}
Hence, from above if $\tilde{R}_1 + \tilde{R}_2 - R_1 -R_2 \leq  I(W_1;W_2)+\delta_J'' $, then the term $\EE\left[\sum_{\ulinex^n \in T_\delta(\ulineX)}P^n_{\ulineX}(\ulinex^n){\sum_{y_n}S_{2}}\right]$ goes to zero exponentially, where $\delta_J'(\delta), \delta_J''(\delta) \searrow 0$ as $\delta \searrow 0.$

%%%%%%%%%%%%%%%%%%%%%%%%%%%%%%%%%%%%%%%%%%%%%%%%%%%%%%%%%%%%%%%%%%%%%%%%%%%%%%%%%%%%%%%%%%%%%%%%%%%%%%%%%%%%%%%%%%%%%%%%%%%%%%%%%%%%%%%%%%%%%%%%%%%%%%%%%%%%%%%

\subsection{Proof of Proposition \ref{prop:Lemma for J}}
\label{appx:prop:Lemma for J}

We begin by considering the second term within the modulus of $P_{\underline{X}}^n(\underline{x}^n)J$, for $\underline{x}^n \in T_\delta(\underline{X})$, i.e.,
\begin{align}
    \frac{1}{2^{nC}}&\sum_{\substack{ \mu_1 \in [2^{nC_1}]  }} \sum_{\substack{w_1^n}}
\!P^{n}_{X_1 X_2}(x_1^{n},x_2^{n}) 
E^{(\mu_1)}_{W_1^n|X_1^n}(w_1^n|x^n_1)P^{n}_{W_2|X_2}(w_2^n|x^n_2)P^{n}_{Y|W_1W_2}(y^{n}|w_1^n,w^n_2)
)),  \nonumber \\ 
%%%%%%%%%%%%%%%%% new line %%%%%%%%%%%%%%%%%%%%%%%
= & \frac{1}{2^{nC_1}}\sum_{\substack{ \mu_1 \in [2^{nC_1}]}}\sum_{w_1^n}P^{n}_{X_1}(x^{n}_{1})E^{(\mu_1)}_{W_1^n|X_1^n}(w_1^n|x^n_1) \nonumber \\ &\left(P^{n}_{X_2|X_1W_1}(x^n_2|x_1^n,w_1^n)P^{n}_{W_2|X_1X_2W_1}(w^n_2|x_1^n,x_2^n,w_1^n)P^{n}_{Y|X_1X_2W_1W_2}(y^n|x^n_1,x_2^n,w_1^n,w_2^n)\right)\nonumber \\
%%%%%%%%%%%%%%%%% new line %%%%%%%%%%%%%%%%%%%%%%%
= & \frac{1}{2^{n(\tilde{R_1}+C_1)}} \frac{(1-\epsilon_1)}{(1+\eta)} \!\sum_{\mu_1,l_1} \!\sum_{\substack{w_1 \in \\ T_{\delta}(W_1|x_1^n)}}\!\!\!\!\!\!{P^{n}_{X_1|W_1}(x_1^n|w_1^n)}P_{X_2W_2Y|X_1W_1}^n (x_2^n,w_2^n,y^n|x_1^n,w_1^n) \mathbbm{1}_{\{\mathtt{w}_1^n(l_1,\mu_1) = w_1^n\}} \nonumber \\
 %%%%%%%%%%%%%%%%% new line %%%%%%%%%%%%%%%%%%%%%%%
 \end{align}
 \begin{align}
= & \frac{1}{2^{n(\tilde{R_1}+C_1)}} \frac{(1-\epsilon_1)}{(1+\eta)} \sum_{\mu_1,l_1} \sum_{\substack{w_1 \in \\ T_{\delta}(W_1|x_1^n)}}\!\!\!\!  P_{X_1X_2W_2Y|W_1}^n (x_1^n,x_2^n,w_2^n,y^n|w_1^n) \mathbbm{1}_{\{\mathtt{w}_1^n(l_1,\mu_1) = w_1^n\}}. \label{eq:J_term1Simplification}
\end{align}
We use the simplification from above and again using triangle inequality bound $\displaystyle \sum_{\underline{x}^n \in T_\delta(\underline{X})}\sum_{y^n,w_2^n}P_{\underline{X}}^n{(\underline{x}^n)}J$ by the following:
\begin{align}
    \sum_{\underline{x}^n \in T_\delta(\underline{X})}&\sum_{y^n,w_2^n}P_{\underline{X}}^n{(\underline{x}^n)}J \nonumber \\ \leq &\sum_{\underline{x}^n \in T_\delta(\underline{X})}\sum_{y^n,w_2^n} \Bigg| P_{X_1X_2W_2Y}^n(x_1^n,x_2^n,w_2^n,y^n) \nonumber \\  
    &\hspace{70pt}- \frac{1}{2^{n(\tilde{R_1}+C_1)}}  \sum_{\mu_1,l} \sum_{\substack{w_1 }} P_{X_1X_2W_2Y|W_1}^n (x_1^n,x_2^n,w_2^n,y^n|w_1^n) \mathbbm{1}_{\{\mathtt{w}_1^n(l_1,\mu_1) = w_1^n\}}  \Bigg| \nonumber \\ 
    & +    \sum_{\underline{x}^n \in T_\delta(\underline{X})}\sum_{y^n,w_2^n} \Bigg|  \frac{1}{2^{n(\tilde{R_1}+C_1)}}  \sum_{\mu_1,l} \sum_{\substack{w_1 }} P_{X_1X_2W_2Y|W_1}^n (x_1^n,x_2^n,w_2^n,y^n|w_1^n) \mathbbm{1}_{\{\mathtt{w}_1^n(l_1,\mu_1) = w_1^n\}} \nonumber \\ &  \hspace{20pt}- \frac{1}{2^{n(\tilde{R_1}+C_1)}} \frac{(1-\epsilon_1)}{(1+\eta)} \sum_{\mu_1,l} \sum_{\substack{w_1 \in \\ T_{\delta}(W_1|x_1^n)}}\!\!\!\!  P_{X_1X_2W_2Y|W_1}^n (x_1^n,x_2^n,w_2^n,y^n|w_1^n) \mathbbm{1}_{\{\mathtt{w}_1^n(l_1,\mu_1) = w_1^n\}} \Bigg| \label{eq:Q1}
\end{align}
The first term in (\ref{eq:Q1}) can be shown to be small in the expected sense using the Lemma \ref{lem:Change Measure Soft Covering Variance Based} given the constraint $\tilde{R_1}+C_1 \geq I(X_1,X_2,W_2,Y;W_1)$. 
Further, the second term in (\ref{eq:Q1}) can be bounded by first taking the expectation over the codebook of $W_1$ and then using a technique similar to that of bounding (\ref{eq:S12_sideInf}). 
% Further, since $\displaystyle\EE\left[\sum_{\underline{x}^n \in T_\delta(\underline{X})}\sum_{y^n,w_2^n}P_{\underline{X}}^n{(\underline{x}^n)}J\cdot\PMFIndicator{}\right] \leq \EE\left[\sum_{\underline{x}^n \in T_\delta(\underline{X})}\sum_{y^n,w_2^n}P_{\underline{X}}^n{(\underline{x}^n)}J\right],$ 
We therefore have $\EE\left[\sum_{\underline{x}^n \in T_\delta(\underline{X})}\sum_{y^n,w_2^n}P_{\underline{X}}^n{(\underline{x}^n)}J\right] \leq \epsilon_{\scriptscriptstyle J}$ for  $\tilde{R_1}+C \geq I(X_1,X_2,W_2,Y;W_1)$ and sufficiently large $n$.

\subsection{Proof of Proposition \ref{prop:Lemma for Q2}}
\label{appx:prop:Lemma for Q2}
\noindent\textit{Analysis of $\sum_{\underline{x}^n\in T_\delta(\underline{X}),y^n}P_{\underline{X}}^n{(\underline{x}^n)}Q_{2}$}: We recall $Q_{2}, 
E^{(\mu_1)}_{W_{1}^{n}|X_{1}^{n}}(\cdot|\cdot), 
E^{(\mu_2)}_{W_{2}^{n}|X_{2}^{n}}(\cdot|\cdot) $.
\begin{align}
Q_2 =  \Bigg|  &\frac{1}{2^{nC_1}}\sum_{\substack{ \mu_1 \in [2^{nC_1}]  }} 
\sum_{\substack{w_1^n,w_2^n }} 
\!\!E^{(\mu_1)}_{W_1^n|X_1^n}(w_1^n|x^n_1)P^{n}_{W_2|X_2}(w_2^n|x^n_2)P^{n}_{Y|W_1W_2}(y^{
n}|w_1^n,w^n_2)
  \nonumber \\
&\hspace{20pt}- \frac{1}{2^{nC_1}}\sum_{\substack{ \mu_1 \in [2^{nC_1}]  }} 
\sum_{\substack{w_1^n,w_2^n }} 
\!\!
E^{(\mu_1)}_{W_1^n|X_1^n}(w_1^n|x^n_1)E^{(\mu_2)}_{W_2^n|X_2^n}(w_2^n|x^n_2)P^{n}_{
Y|W_1W_2}(y^{n}|w_1^n,w^n_2) \Bigg| \end{align}
\begin{align} 
E_{W_{1}^{n}|X_{1}^{n}}^{(\mu_1)}(w_1^n|x_1^n) = \frac{1}{2^{n\tilde{R_1}}} 
\frac{1-\epsilon_1}{1+\eta} 
\sum_{l_1=1}^{2^{n\tilde{R_1}}}\frac{P^{n}_{X_1|W_1}(x_1^n|w_1^n)}{P^{n}_{X_1}
(x_1^n)}\mathbbm{1}_{\{\mathtt{w}_1^n(l_1,\mu_1) = w_1^n\}}\mathbbm{1}_{\{ w_1^n 
\in T_{\delta}(W_1|x_1^n)\}} \nonumber\\
E^{(\mu_2)}_{W_{2}^{n}|X_{2}^{n}}(w_2^n|x_2^n) = \frac{1}{2^{n\tilde{R_2}}} 
\frac{1-\epsilon_2}{1+\eta} 
\sum_{l_2=1}^{2^{n\tilde{R_2}}}\frac{P^{n}_{X_2|W_2}(x_2^n|w_2^n)}{P^{n}_{X_2}
(x_2^n)}\mathbbm{1}_{\{\mathtt{w}_2^n(l_2,\mu_2) = w_2^n\}}\mathbbm{1}_{\{ w_2^n 
\in T_{\delta}(W_2|x_2^n)\}}. \nonumber
\end{align}
Let us define the $\widetilde{P}_{\ulineX^n\ulineW^nY^n}(\ulinex^{n},\ulinew^{n},y^{n})$ on $\mathcal{X}_{1}^{n}\times \mathcal{X}_{2}^{n} 
\times \mathcal{W}_{1}^{n} \times \mathcal{W}_{2}^{n}\times \mathcal{Y}^{n}$ as 
% We remind the reader that $0\leq \sum_{w_1^n \in \mathcal{W}_1^n}E^{(\mu_1)}_{W_1^n|X_1^n}\leq 1$ since we only need to consider the case $\mathds{1}_{\{\mathtt{PMF}(C_1,C_2) \}}=1$. Refer to Lemma \ref{lem:2PMFWHP} for an upper bound on $\mathbb{P}(\mathds{1}_{\{\mathtt{PMF}(C_1,C_2) \}}=0).$
\begin{align}\label{def:s_PMF}
   \!\widetilde{P}_{\ulineX^n\ulineW^nY^n}(\ulinex^{n},\ulinew^{n},y^{n}) \!\deq 
\frac{1}{2^{nC_1}}\!\!\!\!\!\!\sum_{\mu_1 \in [2^{nC_1}]}\!\!\!\!\!
P_{\underline{X}}^n{(\underline{x}^n)}
{E}^{(\mu_1)}_{W_1^n|X_1^n}(w_1^n|x^n_1)P^{n}_{W_2|X_2}(w_2^n|x^n_2)P^{n}_{Y|W_1W_2}(y^{n}|w_1^n,w^n_2).
\end{align}
% where $\tilde{E}^{(\mu_1)}_{W_1^n|X_1^n}(w_1^n|x^n_1)$ is defined as the following
% \begin{align}
%     \tilde{E}^{(\mu_1)}_{W_1^n|X_1^n}(w_1^n|x^n_1) = \begin{cases} \displaystyle
%         {E}^{(\mu_1)}_{W_1^n|X_1^n}(w_1^n|x^n_1) &\text{ for } w_1^n \neq \Bar{w}_1   \\
%          1- \sum_{w_1^n \in \mathcal{W}_1^n}E^{(\mu_1)}_{W_1^n|X_1^n} &\text{ for } w_1^n = \Bar{w}_1^n  
%     \end{cases}.
% \nonumber
% \end{align} for some $\Bar{w}_1^n \notin T_{\delta}(W_1)$, which ensures  $\sum_{w_1^n \in \mathcal{W}_1^n}\tilde{E}^{(\mu_1)}_{W_1^n|X_1^n} = 1$ and hence $\widetilde{P}_{\ulineX^n\ulineW^nY^n}(\ulinex^{n},\ulinew^{n},y^{n})$ is a PMF.
We remind the reader that $0\leq \sum_{w_1^n \in \mathcal{W}_1^n}E^{(\mu_1)}_{W_1^n|X_1^n}\leq 1$, since we only need to consider the case $\mathds{1}_{\{\mathtt{PMF}(C_1,C_2) \}}=1$. Refer to Lemma \ref{lem:2PMFWHP} for an upper bound on $\mathbb{P}(\mathds{1}_{\{\mathtt{PMF}(C_1,C_2) \}}=0).$ 

% The first term in $P_{\underline{X}}^n{(\underline{x}^n)}Q_{2}$ is 
% $\widetilde{P}_{X_1^nX_2^nY^n}(x_{1}^{n},x_{2}^{n},y^{n})$. 
From the definition (\ref{def:s_PMF}), the first term in $P_{\underline{X}}^n{(\underline{x}^n)}Q_{2}$ is simply $\sum_{\ulinew^n}\widetilde{P}_{\ulineX^n\ulineW^nY^n}(\ulinex^n,\ulinew^n,y^n)$. Let us denote this expression by $\widetilde{P}_{\ulineX^nY^n}(\ulinex^n,y^n)$.
% , we also note the 
% marginal $\widetilde{P}_{W_{2}^{n}}(w_{2}^{n}) = P^{n}_{W_{2}}(w_{2}^{n})$. In 
% stating both 
% of these, we have again assumed that 
% $\widetilde{P}_{\ulineX^n\ulineW^nY^n}(\ulinex^{n},\ulinew^{n},y^{n})$ 
% is a PMF on 
% $\mathcal{X}_{1}^{n}\times \mathcal{X}_{2}^{n} \times \mathcal{W}_{1}^{n} 
% \times \mathcal{W}_{2}^{n}\times \mathcal{Y}^{n}$.
% \footnote{ It is not completely convincing to me that we should 
% write up the 
% argument like this - assuming things in the beginning that will be proved later. It will be more precise if we can use statements at every point that are already proved.} 
% Further, we can simplify the second term of $P_{\underline{X}}^n{(\underline{x}^n)}Q_{2}$ as
Further, its second term can be simplified as
% \begin{comment}{\begin{eqnarray}
% \frac{1}{2^{n(C_1+C_2)}}\sum_{\substack{\mu_1,\mu_2 }} 
% \sum_{\substack{w_1^n,w_2^n }} 
% \!\!P^{n}_{X_1 X_2}(x_1^{n},x_2^{n}) 
% E^{(\mu_1)}_{W_1^n|X_1^n}(w_1^n|x^n_1)E^{(\mu_2)}_{W_2^n|X_2^n}(w_2^n|x^n_2)P^{n}_{
% Y|W_1W_2}(y^{n}|w_1^n,w^n_2) = \nonumber\\
% \frac{1}{2^{n(C_2+\tilde{R_{2}})}}\sum_{\substack{ \mu \in [2^{nC}]  }} 
% \sum_{\substack{w_1^n,w_2^n }} 
% \!\!P^{n}_{X_1 X_2}(x_1^{n},x_2^{n}) 
% E^{(\mu_1)}_{W_1^n|X_1^n}(w_1^n|x^n_1)\nonumber\\
% \frac{1-\epsilon}{1+\eta} 
% \sum_{l=1}^{2^{n(\tilde{R_2}}}\frac{P^{n}_{X_2|W_2}(x_2^n|w_2^n)}{P^{n}_{X_2}
% (x_2^n)}\mathbbm{1}_{\{\mathtt{w}_2^n(l_2,\mu_2) = w_2^n\}}\mathbbm{1}_{\{ w_2^n 
% \in T_{\delta}(W_2|x_2^n)\}}P^{n}_{Y|W_1W_2}(y^{
% n}|w_1^n,w^n_2)
% \end{eqnarray}
% }\end{comment}
% In addition, the following simplification can be made to the second term in $Q_2$,
\begin{align}
    \frac{1}{2^{n(C_1+C_2)}}&\sum_{\substack{ \mu_1,\mu_2  }} \sum_{\substack{w_1^n,w_2^n }} 
\!\!P^{n}_{X_1 X_2}(x_1^{n},x_2^{n}) 
E^{(\mu_1)}_{W_1^n|X_1^n}(w_1^n|x^n_1)E^{(\mu_2)}_{W_2^n|X_2^n}(w_2^n|x^n_2)P^{n}_{
Y|W_1W_2}(y^{n}|w_1^n,w^n_2) \nonumber \\
=& \frac{1}{2^{n(\tilde{R_2}+C_1+C_2)}}\frac{(1-\epsilon_1)}{(1-\eta)}\sum_{\substack{\mu_1, \mu_2,l_2  }} \sum_{\substack{w_1^n,\\w_2^n\in T_{\delta}(W_2|x_2^n) }} 
\!\hspace{-10pt}P^{n}_{X_1 X_2}(x_1^{n},x_2^{n}) 
E^{(\mu_1)}_{W_1^n|X_1^n}(w_1^n|x^n_1)\nonumber \\
&\hspace{80pt}\frac{P^{n}_{W_2^n|X_2^n}(w_2^n|x^n_2)}{P^{n}_{W_2}(w_2^n)}P^{n}_{
Y|W_1W_2}(y^{n}|w_1^n,w^n_2)\mathbbm{1}_{\{\mathtt{w} _2^n(l_2,\mu_2) = w_2^n\}}   
\nonumber \\
=& \frac{1}{2^{n(\tilde{R}_2+C_2)}}\frac{(1-\epsilon_1)}{(1+\eta)}\sum_{\substack{ 
\mu_2,l_2 }} \sum_{\substack{w_1^n,\\w_2^n\in T_{\delta}(W_2|x_2^n) }} 
\!\hspace{-10pt}\frac{\widetilde{P}_{\ulineX^{n}\ulineW^{n}Y}(\ulinex^{n},
\ulinew^ {
n },y^{n})\mathbbm{1}_{\{\mathtt{w} _2^n(l_2,\mu_2) = w_2^n\}} } { P_{
W_2^{n}}(w_2^n)}  
 \nonumber 
%  =& 
% \frac{1}{2^{n(\tilde{R}_2+C_2)}}\frac{(1-\epsilon)}{(1-\eta)}\sum_{\substack{ 
% l_2  }} \sum_{\substack{w_2^n\in T_{\delta}(W_2|x_2^n) }} 
% \widetilde{P}_{X_1^nX_2^nY^n|W_2^n}(x_1^n,x_2^n,y^n|w_2^n)\mathbbm{1}_{\{
% \mathtt { w }
% _2^n(l_2,\mu_2) = w_2^n\}}   \nonumber
\end{align}
where the last equality follows by the definition from (\ref{def:s_PMF}). We therefore have
\begin{align}
 &\sum_{\ulinex^n \in T_{
\delta}(\ulineX),y^n}\hspace{-5pt}P_{\underline{X}}^n{(\underline{x}^n)}Q_2 = \nonumber \\& \hspace{20pt} \sum_{\ulinex^{n}\in T_{\delta}(\ulineX),y^{n}}\Bigg| \widetilde{P}_{\ulineX^nY^n}
(\ulinex^{n},y^{n}) -  
\frac{(1-\epsilon_1)}{(1+\eta)2^{n(\tilde{R_2}+C_2)}}\sum_{\substack{l_2,\mu_2  }} \sum_{\substack{w_2^n\in \\ T_{\delta}(W_2|x_2^n) }} \hspace{-10pt}
\frac{\widetilde{P}_{\ulineX^nY^nW_2^n}(\ulinex^n,y^n,w_2^n)}{P_{W_2}^n(w_2^n)}\mathbbm{1}_{\{\mathtt{w}_2^n(l_2,\mu_2) = w_2^n\}}
\Bigg|. \nonumber
\end{align}

% \begin{comment}{
% Substituting the above simplification in $Q_2$ and applying a triangle 
% inequality, similar to (\ref{eq:Q1}), we obtain
% \begin{align}
%      \sum_{x_1^n,x_2^n,y^n}Q_2 \leq \sum_{x_1^n,x_2^n,y^n}\Bigg|& \frac{1}{2^{nC}}\sum_{\mu}\bar{p}_{X_1^nX_2^nY^n}^{(\mu_1)} \nonumber \\ &- \frac{1}{2^{n(\tilde{R_2}+C)}}\sum_{\substack{ \mu,l_2  }} \sum_{\substack{w_2^n}} 
% \bar{p}_{X_1^nX_2^nY^n|W_2^n}^{(\mu_1)}(x_1^n,x_2^n,y^n|w_2^n)\mathbbm{1}_{\{\mathtt{w}_2^n(l_2,\mu_2) = w_2^n\}}   \Bigg| \nonumber \\
% + \sum_{x_1^n,x_2^n,y^n}\Bigg|& \frac{1}{2^{n(\tilde{R_2}+C)}}\sum_{\substack{ \mu,l_2  }} \sum_{\substack{w_2^n}} 
% \bar{p}_{X_1^nX_2^nY^n|W_2^n}^{(\mu_1)}(x_1^n,x_2^n,y^n|w_2^n)\mathbbm{1}_{\{\mathtt{w}_2^n(l_2,\mu_2) = w_2^n\}} \nonumber \\ &- \frac{1}{2^{n(\tilde{R_2}+C)}}\sum_{\substack{ \mu,l_2  }} \sum_{\substack{w_2^n\in T_{\delta}(W_2|x_2^n)}} 
% \bar{p}_{X_1^nX_2^nY^n|W_2^n}^{(\mu_1)}(x_1^n,x_2^n,y^n|w_2^n)\mathbbm{1}_{\{\mathtt{w}_2^n(l_2,\mu_2) = w_2^n\}}   \Bigg| \nonumber \\
% \end{align}
% }\end{comment}
To bound the above term, we add and subtract the following three terms within the modulus 
\begin{align}
    (i) \quad & P^{n}_{\ulineX Y}(\ulinex^n,y^n) \nonumber \\
    (ii) \quad & \frac{1}{2^{n(\tilde{R}_2+C_2)}}\sum_{\mu_2,l_2}\sum_{\substack{w_2^n\in T_{\delta}(W_2) }} P^{n}_{\ulineX Y|W_2}(\ulinex^n,y^n|w_2^n)\mathbbm{1}_{\{\mathtt{w}_2^n(l_2,\mu_2)=w_2^n\}} \nonumber \\
    (iii) \quad & \frac{1}{2^{n(\tilde{R}_2+C_2)}}\sum_{\substack{\mu_2
l_2  }} \sum_{\substack{w_2^n\in T_{\delta}(W_2) }} 
\frac{\widetilde{P}_{\ulineX^nY^nW_2^n}(\ulinex^n,y^n,w_2^n)}{P_{W_2}^n(w_2^n)}\mathbbm{1}_{\{\mathtt{w}_2^n(l_2,\mu_2) = w_2^n\}} \nonumber
\end{align}
Using triangle inequality on each pair of terms within the modulus, we obtain
\begin{align}
     \sum_{\ulinex^n \in T_\delta(\ulineX),y^n}\hspace{-5pt}&P_{\underline{X}}^n{(\underline{x}^n)}Q_2 \leq \sum_{\ulinex^n \in T_\delta(\ulineX),y^n}\hspace{-5pt}P_{\underline{X}}^n{(\underline{x}^n)}[Q_{21} + Q_{22} +Q_{23} + Q_{24}]
\end{align}
where, for all $\ulinex^n \in T_\delta(\ulineX)$, we define
\begin{align}
    P_{\underline{X}}^n{(\underline{x}^n)}Q_{21} = &  \Bigg|\widetilde{P}_{\ulineX^nY^n}
(\ulinex^{n},y^{n}) - P^{n}_{\ulineX Y}(\ulinex^n,y^n)\Bigg|, \nonumber \\
    P_{\underline{X}}^n{(\underline{x}^n)}Q_{22} = &  \Bigg|P^{n}_{\ulineX Y}(\ulinex^n,y^n) - \frac{1}{2^{n(\tilde{R}_2+C_2)}}\sum_{\mu_2,l_2} \sum_{\substack{w_2^n\in T_{\delta}(W_2) }} P^{n}_{\ulineX Y| W_2}(\ulinex^n,y^n|w_2^n)\mathbbm{1}_{\{\mathtt{w}_2^n(l_2,\mu_2)=w_2\}} \Bigg| \nonumber \\
    %%%%%%%%%%%%%%%%%%%%%%%% new line %%%%%%%%%%%%%%%%%%%%%
   P_{\underline{X}}^n{(\underline{x}^n)} Q_{23} = &     \frac{1}{2^{n(\tilde{R}_2+C_2)}}\sum_{\mu_2,l_2}\mathbbm{1}_{\{\mathtt{w}_2^n(l_2,\mu_2)=w_2\}}\Bigg|\!\! \sum_{\substack{w_2^n\in\\ T_{\delta}(W_2) }}\!\! P^{n}_{\ulineX Y| W_2}(\ulinex^n,y^n|w_2^n) -  \!\!\!\!\sum_{\substack{w_2^n\in \\T_{\delta}(W_2) }}\!\!\! \frac{\widetilde{P}_{\ulineX^nY^nW_2^n}(\ulinex^n,y^n,w_2^n)}{P_{W_2}^n(w_2^n)}\Bigg| \nonumber \\
   %%%%%%%%%%%%%%%%%%%%%% new line %%%%%%%%%%%%%%%%%%%%%%%
    P_{\underline{X}}^n{(\underline{x}^n)}Q_{24} = &  \Bigg|\frac{1}{2^{n(\tilde{R}_2+C_2)}}\sum_{\substack{\mu_2,l_2  }} \sum_{\substack{w_2^n\in \\ T_{\delta}(W_2) }}\frac{\widetilde{P}_{\ulineX^nY^nW_2^n}(\ulinex^n,y^n,w_2^n)}{P_{W_2}^n(w_2^n)}\mathbbm{1}_{\{\mathtt{w}_2^n(l_2,\mu_2) = w_2^n\}} \nonumber \\ & \hspace{50pt}- \frac{(1-\epsilon_1)}{(1+\eta)2^{n(\tilde{R_2}+C_2)}}\sum_{\substack{\mu_2,l_2  }} \sum_{\substack{w_2^n\in \\ T_{\delta}(W_2|x_2^n) }}\hspace{-10pt} \frac{\widetilde{P}_{\ulineX^nY^nW_2^n}(\ulinex^n,y^n,w_2^n)}{P_{W_2}^n(w_2^n)}\mathbbm{1}_{\{\mathtt{w}_2^n(l_2,\mu_2) = w_2^n\}}\Bigg|
\end{align}

Now we look at bounding each of these four terms, starting with the term corresponding to $Q_{21}$. Since
    $\sum_{\ulinex^n \in T_\delta(\ulineX),y^n} P_{\underline{X}}^n{(\underline{x}^n)}Q_{21} \leq \sum_{\ulinex^n \in T_\delta(\ulineX),y^n,w_2^n}P_{\underline{X}}^n{(\underline{x}^n)} J$, the result from Proposition \ref{prop:Lemma for J}  implies if $\tilde{R_1}+C_1 \geq I(X_1,X_2,W_2,Y;W_1)$ then, for sufficiently large n, the term corresponding to $Q_{21}$ can be made arbitrarily small in expected sense.
    
    \noindent Secondly, we look at $\sum_{\ulinex^n \in T_\delta(\ulineX),y^n}P^{n}_{\ulineX}(\ulinex^n)Q_{22}$. Using Lemma \ref{lem:Change Measure Soft Covering Variance Based}, we get, if $\tilde{R}_2+C_2 \geq I(X_1X_2Y;W_2) + \delta_{\scriptscriptstyle Q_{22}}$, then for sufficiently large n, $ \EE\left[\sum_{\ulinex^n \in T_\delta(\ulineX),y^n}P^{n}_{\ulineX}(\ulinex^n)Q_{22}\right] \leq \epsilon_{\scriptscriptstyle Q_{22}}$, where $\epsilon_{\scriptscriptstyle Q_{22}}, \delta{\scriptscriptstyle Q_{22}} \searrow 0 $ as $\delta \searrow 0$.
    
    \noindent Thirdly, consider $\sum_{\ulinex^n\in T_\delta(\ulineX),y^n}P_{\ulineX}^n(\ulinex^n)Q_{23}.$ Applying expectation over the second codebook followed by the first gives
    \begin{align}
        \EE&\left[\sum_{\ulinex^n\in T_\delta(\ulineX),y^n}P_{\ulineX}^n(\ulinex^n)Q_{23}\right] \nonumber \\
        & = \EE_{\mathcal{C}_1}\left[\sum_{\ulinex^n\in T_\delta(\ulineX),y^n}\frac{P_{W_2}^n(w_2^n)}{(1-\epsilon_2)}\Bigg| \sum_{w_2^n\in  T_{\delta}(W_2)}P^{n}_{\ulineX Y| W_2}(\ulinex^n,y^n|w_2^n) -  \sum_{\substack{w_2^n\in  T_{\delta}(W_2) }} \!\!\!\!\!\!\frac{\widetilde{P}_{\ulineX^nY^nW_2^n}(\ulinex^n,y^n,w_2^n)}{P_{W_2}^n(w_2^n)}\Bigg|\right] \nonumber \\
        & \leq \EE_{\mathcal{C}_1}\left[\frac{1}{(1-\epsilon_2)}\sum_{\ulinex^n\in T_\delta(\ulineX),y^n}\sum_{w_2^n}P_{\ulineX}^n(\ulinex^n)J\right]
    \end{align}
    where the first equality follows by expectation of the indicator function over the second codebook, and the subsequent inequality follows from using the triangle inequality and using the definition of $J$ (\ref{eq:strongerResult}). Again using Proposition \ref{prop:Lemma for J} proves $\EE\left[\sum_{\ulinex^n\in T_\delta(\ulineX),y^n}P_{\ulineX}^n(\ulinex^n)Q_{23}\right]$ can be made arbitrarily small.
    
    \noindent Finally, we remain with $\sum_{\ulinex^n\in T_\delta(\ulineX),y^n}P_{\underline{X}}^n{(\underline{x}^n)}Q_{24} $. This term can be split into two terms such that $Q_{24} = Q'_{24} + Q''_{24}$ where
    \begin{align}
        P_{ \underline{X}}^n{(\underline{x}^n)}Q'_{24} & = 2^{-n(\tilde{R}_2+C_2)}\Bigg|\left(1-\frac{1-\epsilon_1}{1+\eta}\right)\sum_{\mu_2,l_2}\sum_{\substack{w_2^n\in T_{\delta}(W_2|x_2^n) }}\hspace{-10pt} \frac{\widetilde{P}_{\ulineX^nY^nW_2^n}(\ulinex^n,y^n,w_2^n)}{P_{W_2}^n(w_2^n)}\mathbbm{1}_{\{\mathtt{w}_2^n(l_2,\mu_2) = w_2^n\}}\Bigg|\nonumber \\
        P_{\underline{X}}^n{(\underline{x}^n)}Q''_{24} & = 2^{-n(\tilde{R}_2+C_2)}\left(\frac{1-\epsilon_1}{1+\eta}\right)\Bigg|\sum_{\mu_2,l_2}\sum_{\substack{w_2^n\notin T_{\delta}(W_2|x_2^n) \\ w_2^n\in T_{\delta}(W_2)  }}\hspace{-10pt} \frac{\widetilde{P}_{\ulineX^nY^nW_2^n}(\ulinex^n,y^n,w_2^n)}{P_{W_2}^n(w_2^n)}\mathbbm{1}_{\{\mathtt{w}_2^n(l_2,\mu_2) = w_2^n\}}\Bigg|\nonumber 
    \end{align}
    
    Consider $\EE\left[\sum_{\ulinex^n\in T_\delta(\ulineX),y^n}P_{\underline{X}}^n{(\underline{x}^n)}Q'_{24}\right]$,
    \begin{align}
        & = \EE\Bigg[\frac{(\eta-\epsilon_1)(1-\epsilon_1)}{(1+\eta)^2}\frac{1}{2^{n(\tilde{R_1}+\tilde{R_2}+C_1+C_2)}} \nonumber \\
        & \hspace{50pt} \sum_{\ulinex^n\in T_\delta(\ulineX),y^n} \sum_{\mu_1,\mu_2,l_1,l_2} \sum_{\substack{w^n_1 \in  T_{\delta}(W_1|x_1^n) \\ w_2^n\in T_{\delta}(W_2|x_2^n)}}\!\! \frac{P_{\ulineX \ulineW Y}^n (\ulinex^n,\ulinew^n,y^n)}{P_{W_1}^n(w_1^n)P_{W_2}^n(w_2^n)} \mathbbm{1}_{\{\mathtt{w}_1^n(l_1,\mu_1) = w_1^n\}}\mathbbm{1}_{\{\mathtt{w}_2^n(l_2,\mu_2) = w_2^n\}}\Bigg]
        \nonumber \\
        & =\frac{(\eta-\epsilon_1)}{(1+\eta)^2(1-\epsilon_1)} \sum_{\ulinex^n\in T_\delta(\ulineX),y^n}  \sum_{\substack{w^n_1 \in  T_{\delta}(W_1|x_1^n) \\ w_2^n\in T_{\delta}(W_2|x_2^n)}}\!\! {P_{\ulineX \ulineW Y}^n (\ulinex^n,\ulinew^n,y^n)} \nonumber \\
        & \leq \frac{(\eta-\epsilon_1)}{(1+\eta)^2(1-\epsilon_1)} \sum_{\ulinex^n\in T_\delta(\ulineX),y^n}  \sum_{\substack{w_1^n, w_2^n}} {P_{\ulineX \ulineW Y}^n (\ulinex^n,\ulinew^n,y^n)} = \frac{(\eta-\epsilon)}{(1+\eta)^2(1-\epsilon)} \nonumber 
    \end{align}
    where the first equality above is obtained by substituting the definition of $\widetilde{P}_{\ulineX^n Y^n\ulineW^n}(\ulinex^n,y^n,\ulinew^n)$ followed by using the simplification from (\ref{eq:J_term1Simplification}), and the second equality is followed by using the fact that
    \begin{align}
        \EE\left[\mathbbm{1}_{\{\mathtt{w}_1^n(l_1,\mu_1) = w_1^n\}}\mathbbm{1}_{\{\mathtt{w}_2^n(l_2,\mu_2) = w_2^n\}}\right] = \frac{P_{W_1}^n(w_1^n)P_{W_2}^n(w_2^n)}{(1-\epsilon_1)(1-\epsilon_2)}
    \end{align}
    
    \noindent Similarly, consider $\EE\left[\sum_{\ulinex^n\in T_\delta(\ulineX),y^n}P_{\underline{X}}^n{(\underline{x}^n)}Q''_{24}\right]$,
    \begin{align}
        & = \EE\left[\frac{1-\epsilon_1}{1+\eta}\frac{1}{2^{n(\tilde{R}_1+\tilde{R}_2+C_1+C_2)}}\Bigg|\sum_{\substack{\mu_1,\mu_2,\\l_1,l_2}}\sum_{\substack{\ulinex^n\in T_\delta(\ulineX),\\y^n}} \sum_{\substack{\left\{w_1^n\in T_{\delta}(W_1|x_1^n)\right\}\\ \left\{w_2^n\notin T_{\delta}(W_2|x_2^n)\right. \\ \left.w_2^n\in T_{\delta}(W_2)\right\}  }}\!\!\! \frac{P_{\ulineX \ulineW Y}^n (\ulinex^n,\ulinew^n,y^n)}{P_{W_1}^n(w_1^n)P_{W_2}^n(w_2^n)} \right.\nonumber \\
        & \hspace{4in} \left.\mathbbm{1}_{\{\mathtt{w}_1^n(l_1,\mu_1) = w_1^n\}}\mathbbm{1}_{\{\mathtt{w}_2^n(l_2,\mu_2) = w_2^n\}}\Bigg|\right] \nonumber
        \end{align}
        \begin{align}
        & \leq \frac{1}{(1+\eta)(1-\epsilon_2)}\!\!\!\sum_{\substack{x_2\in T_{\delta}(X_2) \\ w_2 \notin T_{\delta}(W_2|x_2^n)}}\hspace{-15pt}P^{n}_{X_2W_2}(x_2^n,w_2^n)\sum_{x_1,w_1} P^{n}_{X_1W_1|X_2}(x_1^n,w_1^n|x_2^n) \sum_{y^n}P^{n}_{Y|W_1W_2}(y^n|w_1^n,w_2^n) \nonumber \\
        & \leq \frac{\epsilon'}{(1+\eta)(1-\epsilon_2)}
    \end{align}
This completes the analysis of all the terms corresponding to $Q_2$.

%\textcolor{red}{
% Change epsilons to epsilon1 and epslion2, and varepsilon and epsilon carefully \\ 
% Contribution: typicality, relaxation: if we restrict to normalization it doesn't work in the quantum case, not aware of any proof in the distributed case, one shot, one shot covering lemma, extended to distributed case, also obtained an outer bound.}

\bibliographystyle{IEEEtran}
\bibliography{SoftCovering}

\end{document}